\documentclass[11pt]{article}

\usepackage{geometry}
\usepackage{longtable}
\usepackage{booktabs}
\usepackage{rotating} 
\usepackage{cite}
\usepackage{graphbox, subfig} 
\usepackage{url}
\usepackage{textcomp}
\usepackage[dvipsnames]{xcolor}
\usepackage{amsmath}
\usepackage{amsmath,amssymb, enumerate, fullpage}
\usepackage{appendix}
\usepackage{enumitem}
\usepackage{amssymb}
\usepackage{amsthm}
\usepackage{graphicx,color}
\usepackage{epstopdf}
\usepackage{epsfig} 
\usepackage{tikz-cd}
\usepackage{fullpage,algorithm}
\usepackage{algorithmic}
\usepackage{framed}
\usepackage[utf8]{inputenc}
\usepackage[english]{babel}
\usepackage{titlesec}
\usepackage{xcolor, soul}
\usepackage{multirow}
\usepackage{hyperref}
\sethlcolor{green}
\titleformat{\chapter}[hang] 
{\normalfont\huge\bfseries}{\chaptertitlename\ \thechapter:}{1em}{}

\begin{document}
\title{StemP: A fast and deterministic Stem-graph
approach \\ for RNA and protein folding prediction}
\author{Mengyi Tang and Kumbit Hwang and Sung Ha Kang\footnote{Mengy~Tang (tangmengyi@gatech.edu), Kumbit Hwang (kb\_hwang@gatech.edu) and  Sung Ha~Kang (kang@math.gatech.edu) are with School of Mathematics, Georgia Institute of Technology, 686 Cherry Street, Atlanta, 30032, GA, USA. }}
\date{}

\maketitle
\begin{abstract}
We propose a new deterministic methodology to predict RNA sequence and protein folding.   \textit{Is  stem enough for structure prediction?}   The main idea is to  consider all possible stem formation in the given sequence.   With the stem loop energy and the strength of stem, we explore how to deterministically utilize stem information for RNA sequence and protein folding structure prediction. 
We use graph notation, where all possible stems are represented as vertices, and co-existence as edges.  This full Stem-graph  presents all possible folding structure, and we pick  sub-graph(s) which give the best matching energy for folding structure prediction.  
We introduce a Stem-Loop score to add structure information and to speed up the computation.  
The proposed method can handle  secondary structure prediction as well as protein folding with pseudo knots.  
Numerical experiments are done using a laptop and results take only a few minutes or seconds.  
One of the strengths of this approach is in the simplicity and flexibility of the algorithm, and it gives deterministic answer.  
We explore protein sequences from Protein Data Bank, rRNA 5S sequences,  and tRNA sequences from the Gutell Lab.  Various experiments and comparisons are included to validate the propose method.
\end{abstract}

\textbf{Remark} The code for the proposed method is available at https://github.com/sunghakang/StemP. 

\section{Introduction}

Ribonucleic acid (RNA) plays an indispensable role in biological process  and contributes to the cellular life. It carries out the instructions encoded in DNA, while DNA stores the information for protein synthesis, and most biological activities are carried out by proteins. 
The primary structure of RNA is dominated by the interaction between nucleotide bases connected by phosphate groups and polysaccharide molecules. This base-paired RNA's secondary structure provides a bridge to predict the 3D structure while being more stable to analyze \cite{tinoco1999rna}.  Next generation sequencing made it possible to determine the whole genomic sequence, utilizing RNA sequencing to discover RNA variants and quantify mRNAs for gene expression analysis.  

There have been a wide range of literature on prediction of the secondary structure of a RNA sequence. Minimizing Free Energy (MFE) \cite{mathews2006prediction}  is applicable for large molecules in terms of efficiency and is widely extended.  For an efficient computation, dynamic programming is suggested in \cite{lorenz2016predicting}, which conquers numerous possible folding structures by dividing them into smaller sub-problems.  A practical dynamic algorithm can be found in \cite{zuker_stiegler_1981}.   
In \cite{bellaousov2010probknot}, an optimal structure is assembled based on Maximum Energy Accuracy (MEA) without using  dynamic programming algorithm.  A novel method based on both MFE and MEA is proposed in \cite{wu2015improved}  for experimental probing data, and structural probing data is incorporated in related work such as \cite{lorenz2016shape} as supplementary information to enhance accuracy.

In this paper, we explore single sequence approach to focus on the effect of stems in each sequence.  Single sequence analysis approach includes, CONTRAfold\cite{zakov2011rich} which uses fully-automated statistical learning algorithms to evolve model parameters instead of relying on thermodynamics, and CyloFold\cite{bindewald2010cylofold}, an approach based on reproducing the folding procedure in a coarse-grained manner by choosing folding structures based on free energy contribution.  Related work on single sequence analysis includes \cite{perriquet2003finding, swenson2012gtfold, Sato01072011, tsang2008sarna}, and we refer to \cite{wiki_rna} for more details in various categories.  
Alternatively, comparative methods can reduce the space of possible folding structure by using evolutionary approaches \cite{hamada2011improving, hamada2009centroidalign, sato2012dafs}, and recently,  various machine learning techniques are developed, e.g. \cite{li2016protein, singh2019rna, zhang2016deep}.

To represent the structure, we construct a Stem-graph, where all possible stems are represented as vertices of the graph, and edges represents all possible co-existences among the vertices.   This setting is similar to  \cite{ji2004graph}, where a maximum clique finding algorithm is implemented to assemble compatible conserved stems.  Multiple possible optimal structures are assembled in topological order according to their compatibility among $k$ sequences.  In \cite{fera2004rag}, the author employs the idea of tree graphs   for archiving RNA tree motifs and dual graphs for general RNA motifs. 
A graph mining algorithm is proposed in \cite{hamada2006mining} to  detects all the possible motifs exhaustively.
One important advantage of these graph methods is that it allows the existence of pseudo-knots in plausible folding structures  \cite{ji2004graph,fera2004rag,gaspin1995interactive}.  In general, pseudo-knots are particularly difficult to identify efficiently since it leads to a structure with at least two helical stems. 
With the help of a graphic representation, a pseudo-knot can be efficiently considered as a possible folding structure without further annotation.

In this paper,  we propose Stem-graph based Prediction (StemP).  We first build the graph which represents all possible folding structure of the sequence, which we refer to as the full Stem-graph.   Secondly, we extract a sub-graph or multiple sub-graphs which have the best matching energy for folding structure prediction.   We introduce the Stem-Loop score to give structure information in vertex construction, and to make algorithm computationally more efficient.

The main contributions of this paper is to propose a simple, flexible, efficient and deterministic method for folding structure prediction. 
\begin{itemize}
\item{This method can handle pseudo-knots and 3D structure naturally without any other modification. }
\item{The proposed method is  computationally efficient that for sequences of length smaller than 200, results can be computed within one second. }  
\item{The proposed method is flexible, it can be applied to protein structure prediction, tRNA and rRNA 5S, which we explore in details in following sections. }
\end{itemize}   

This paper is organized as follows.  In Section \ref{sec: spa}, we give a general outline and details of Stem-graph approach, including how to construct the vertices and edges utilizing Stem-Loop score.  In  Section \ref{sec: pdb}, \ref{sec: trna} and \ref{sec: 5S}, we present details and comparison results for Protein Data Base (PDB), tRNA and 5S rRNA sequences respectively.  We conclude the paper with some discussions in Section \ref{sec:conclu}.

\section{StemP Methodology} 
\label{sec: spa}

\begin{figure*}
    \centering
    \includegraphics[width=0.95 \textwidth]{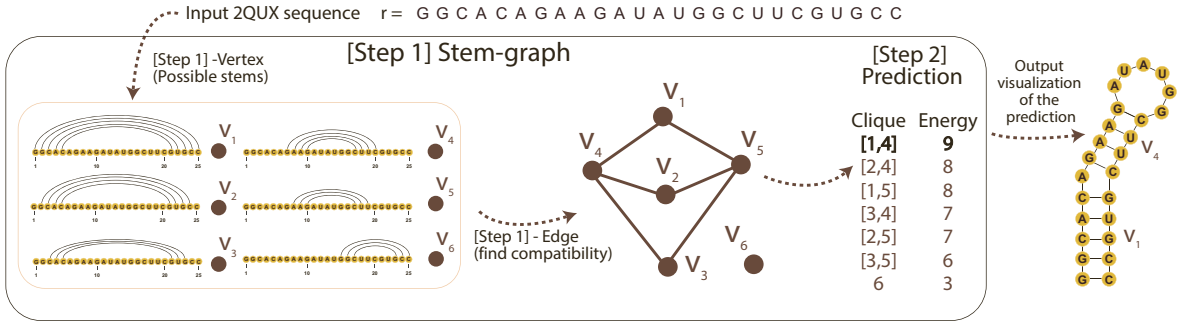}
    \caption{Outline of StemP with an example of protein 2QUX, From the input sequence, in [Step 1] the Stem-graph  is constructed where vertex represents stem and edges represent co-existences.  [Step 2] cliques from the full Stem-graph are ordered, then maximum energy one is picked as a prediction.  }
    \label{fig:2qux}
\end{figure*}

We briefly review some definitions.  \textit{A complete graph} is a simple undirected graph in which every pair of distinct vertices is connected by a unique edge.  \textit{A clique} in an undirected graph is a subset of vertices, such that every two distinct vertices are adjacent.  It's induced subgraph is complete.  \textit{Maximal Clique}  is a clique which cannot be extended by including any more adjacent vertices. 

\vspace{0.2cm}
The proposed Stem-graph based Prediction algorithm has two steps: [Step1] Stem-graph construction which presents all possible folding structure from the given sequence, and [Step2] Prediction, where we pick an optimal folding structure among maximal cliques.  Figure \ref{fig:2qux} represents the outline of StemP, using an example of protein 2QUX sequence:
\begin{equation} \label{E:seq 2qux}
\verb|r = GGCAC AGAAG AUAUG GCUUC GUGCC|.
\end{equation}

\textbf{[Step1]} Stem-graph construction.  
First, all possible stems are assigned as  vertices. Starting from the first base \texttt{G} in (\ref{E:seq 2qux}), we search for the first base \texttt{C} after \texttt{G} which matches it.  We consider the canonical base-pair matching, unless mentioned otherwise.  
A stem is constructed if there are at least $L$ consecutive base pairs starting from this \texttt{G}-\texttt{C} pair as starting and ending bases.  Here the integer $L\geq 2$ represents the minimum stem length to consider for each vertex.  We consider the longest stem which can be constructed from these starting and ending bases.  
If the first base \texttt{G}  forms  different stems, they are assigned as separate vertices.   Once all possible stems starting with the first base \texttt{G}  are found and assigned as vertices,  we move  to the second base \texttt{G}  and repeat the process. Figure \ref{fig:2qux} [Step 1]-Vertex shows all vertex for 2QUX.  We note that the stem formed from the first base \texttt{G}  may include the stem formed from the second base, as a sub-stem, e.g. $v_1, v_2$ in Figure \ref{fig:2qux}.   We consider these to be two different vertices.

Each vertex stores it's corresponding property, for example:
\begin{equation}
v_1 
= (i_1,j_1,l_1,d_1) = (1,25,5,24)\\
\label{eqn:v_def}
\end{equation}
here $i_1$ is the starting base number, $j_1$ the ending base number, $l_1$ the length of the stem (the number of consecutively matched bases), and $d_1=j_1-i_1$ the distance between the starting and the ending bases.  We use the ratio between $l_1$ and $d_1$ to define Stem-Loop score in subsection \ref{sec: sindex}.

After all possible stems are found and represented as vertices, in [Step1]-Edges, edges are constructed based on the co-existing possibilities between vertices.  For example, $v_2$ and $v_3$ are sub-stems of $v_1$ that they cannot co-exists, but $v_1$ and co-exits with $v_4$. Figure \ref{fig:2qux} [Step 1]-Edge considers all such possibility and $e_{14}$, $e_{24}$, $e_{34}$, $e_{15}$, $e_{25}$, $e_{35}$ are constructed in the full Stem-graph. 

Note that in this full Stem-graph (i) every subgraph represents different folding structure, and that (ii) pseudo-knots and 3D prediction comes naturally without any extra consideration.

\textbf{[Step2]} Prediction. With all possible structure presented in the full Stem-graph, folding prediction is to choose a subgraph satisfying certain energy.  Since plausible structure requires compatibility between every two vertices, for all subgraphs with multiple vertices, only a complete subgraph can be recognized as a possible folding structure.

\begin{quote}
We consider \textit{maximal cliques} of the full Stem-graph as  possible folding structures.   We sort all maximal clique by the total number of  \text{basepair matching} as a simple energy.
\end{quote}

The principle idea of Stem-graph Prediction is to find the maximum matching among the stable structures, which is maximal clique among complete sub-graphs since it represents the most complete structure where at least one vertex is a part of.   

For 2QUX (\ref{E:seq 2qux}), as shown in Figure \ref{fig:2qux} [Step2], there are 7 cliques in total.  The full Stem-graph of 2QUX only have 2-vertex cliques (six of them), and no 3-vertex cliques, and one 1-vertex clique.   The maximal clique constructed by  $v_1$ and $v_4$ is the maximum base-pair matching with maximum energy 9.  This is picked as the result of StemP, which matches with the true folding.  

For numerical computation for finding maximal cliques, we employ  a modified Bron-Kerbosch algorithm with pivoting \cite{Tomita2006} to find maximal cliques of the full Stem-graph.  In this algorithm, depth-first search algorithm with pruning methods are implemented based on Bron and Kerbosch algorithm \cite{Bron_Kerbosch}, which is a recursive backtracking algorithm. The time complexity of the worst-case time is $O(3^{\frac{n}{3}})$ for an undirected graph with $n$ vertex. The problem of finding cliques is also studied in many research area such as social network, bioinformatics, computer vision and computational topology.    Another improved methods based on Bron-Kerbosch algorithm can be found in  \cite{Eppstein2010}.

StemP gives a deterministic folding prediction, while showing all possible folding structure in a compact full Stem-graph.  In practice, to reduce the number of vertices for more efficient computation, we use the minimum stem length $L$, and Stem-Loop score which we introduce in the following subsection.

\subsection{Stem-Loop score and Generalized Stem-Loop score} \label{sec: sindex}

In  \textbf{[Step1]} Stem-graph construction, we further utilize structure information to reduce the number of vertices.  We introduce Stem-Loop score for each vertex $v_k$: 
\begin{equation}\label{E:sindex}
  SL(v_k) = \displaystyle{\frac{\mbox{the total length of vertex } v_k}{\mbox{the stem length of vertex } v_k}
= \frac{d_k}{l_k} }.
\end{equation}
The main idea behind this score is to explore the energy balance between the stem length $l_k$ and the size of the loop the stem encloses.   The stem length $l_k$ is not too small compared to the size of the loop $d_k$, since there will not be enough binding force to hold the loop stable.  The  stem length $l_k$ is not too large compared to the size of the loop $d_k$, since it is somewhat unnatural.  We observed that $SL$ values among the same type of sequences are similar, and we can learn the range of this value from known sequences. This is in spirit similar to recent trend of active learning methodology such as  Neural Network \cite{li2016protein, zhang2016deep} or Motif analysis\cite{tanner2003q, cordin2004newly}, that we learn Stem-Loop score from a known sequence, and biological property can be added.  

For a long sequence like rRNA 5S, Stem-Loop score itself may not be enough to capture the complex structure of the folding, such as stems enclosing another stems.  We generalize $SL$ to include such structures.  We define a set of vertices, $V_k$, which is understood as a structure starting from a base stem leading up to one hairpin loop, including all internal loops and  bulges in between. 
This set $V_k =\{ v_{k_1},...v_{k_m} \}$ represents a set of several stems in which one encloses the rest of the stems.   For this set $V_k$, we consider the following Generalized-Stem-Loop score:
\begin{equation} \label{E:gsl}
  GSL(V_k)   = GSL(v_{k_1},...v_{k_m}) 
 = \frac{d_{k_1}}{ l_{k_1}+...+l_{k_m}}.
 \end{equation}
The first vertex $v_{k_1}$ encloses all the other vertices $v_{k_2},...v_{k_m}$ (i.e.  $a_{k_1}<a_{k_j}$, $b_{k_1}>b_{k_j}$ for all $j>1$).  Here $d_{k_1}$ denotes  the total length of $V_k$, and $l_{k_1}+...+l_{k_m}$ denotes the sum of stem length within $V_k$.  $GSL$ represents the total strength of base-pair matching over the total length of the sequence enclosed in $V_k$.
We discuss details in Section \ref{sec: 5S} for 5s rRNA. 

The case such as tRNA Acceptor, the sequences have a self closing form, that is there is a stem connecting starting and ending bases of the entire sequence.  When the total length $d_k$ of the stem-loop is sufficiently large to enclose the whole sequence, we  consider the Acceptor-Stem-Loop score:
\begin{equation}\label{E:msl}
ASL (v_k) 
= \frac{\hat{l} - d_k + 2l_k - 2}{l_k}, 
\end{equation}
where $\hat{l}$ is the size of the given sequence.
This is a way to account for the open-end closing stems.  We consider the open-end loop to be closed and compute the Stem-Loop score in the opposite direction of a normal stem. 
We found (\ref{E:msl}) to be simpler to find the range of values. 


\subsection{Comparison Measure for Numerical Experiments}
\label{ss: mcc}

In section \ref{sec: pdb}, \ref{sec: trna} and \ref{sec: 5S}, we present StemP results and compare with existing methods.   To evaluate the prediction, we use Specificity (Spec)/ Positive Prediction Value (PPV), Sensitivity (Sens), the {Matthews Correlation Coefficient} (MCC) \cite{wu2015improved, parisien2008mc,hamada2006mining}, and $F_1$-score.  
Following \cite{baldi2000assessing} and \cite{wiese2008rnapredict}, the specificity utilized in our work is identical to the positive predictive value: 
\begin{align}
 \text{ Sensitivity  } = \frac{TP}{TP+FN}, \label{eq_sens}\\
\text{ Specifity / Positive  Prediction Value }  = \frac{TP}{TP + FP},  \label{eq_ppv}\\
\text{ Matthews  Correlation Coefficient  }   = \sqrt{\text{Sens} \times \text{PPV}}, 
\label{eq_mcc}\\
F_1\text{-score }
= \frac{2 \text{PPV} \cdot \text{Sens}}{\text{PPV} + \text{Sens}}, \label{eq_f1}
 \end{align}  
where  TP, TN, FP, FN refer to True Positive, True Negative, False Positive, and  False Negative base pairs.  Both MCC and $F_1$ takes values in between [0,1] and 1 represents 100 \% matching without any additional nor missing  base-pair matching.  Both MCC and $F_1$-score can give an overall justification of the prediction  with respect to the False prediction. 

In practice, StemP prediction of the maximum matching maximal clique may not be unique.  We use Standard Competition Ranking (SCR)(``1224" ranking) to present the results.  If there are multiple ranked 1 folding, we report with the value $m$ in the parenthesis to indicates that there are multiple folding with the same maximum number of base pairs matched.  For example, SCR($m$) = 1(2) represents that there are total of 2 sequences in the top rank, the same maximum matching, and 3(4) represents that there are 4 sequences in the third rank.  For some cases, we also present Dense Ranking (DR) (``1223" ranking).

\section{StemP for Protein Folding Prediction}\label{sec: pdb}

We present the details and results for protein folding prediction.  We use data from 
the Protein Data Bank (PDB) \cite{bernstein1977protein}, which preserves  structure information of a large number of biological molecules including proteins and nucleic acids.   
There are various experimental work on protein structure prediction.  In \cite{parisien2008mc}, Nucleotide Cyclic Motif (NCM) is introduced to represent nucleotide relationships in structured RNA. Experiments were performed on 182 PDBs of sizes from 8 to 35 and the corresponding result  reached  0.87 of MCC on average. 
From the point of view of graph structure, base triples were explored in \cite{muller2015combinatorics} to represent RNA secondary structure.  There are other approaches utilizing  Direct-Coupling Analysis\cite{de2015direct}, orientation and twisting of $\beta$-sheets \cite{micsonai2018bestsel} and multiple threading alignment approaches \cite{yang2015tasser}.

\subsection{Parameters} We experiment with protein data of length up to 50, and provide the accuracy of our results based on the true folding structures retrieved from Nucleic Acid Database \cite{berman1992nucleic, coimbatore2013nucleic}.  
Table \ref{T:PDBe} shows the parameters we chose for StemP.  We set the minimum stem length to be $L=3$, and set  Stem-Loop score to be $2 \leq SL \leq 20$.  
These two mild conditions enhance the computing speed by reducing the number of vertices in constructing the full Stem-graph.  
For example, for sequence 1KXK, MCC 0.96 is obtained in 11 seconds with $2 \leq SL \leq 20$, while it took 89 seconds to obtain the same accuracy without the conditions. 
We found that for a short sequence, especially for a protein sequence, the correct folding can be given by a single vertex when there is no clique of size bigger than 2.
%
\begin{table}
\centering
\begin{tabular}{l|c}
\hline
\multicolumn{2}{c}{ StemP parameters for Protein folding} \\
\hline
Basepair matching & Canonical base pair \\
Minimum Stem Length  & $L=3$ (or 2) \\
Stem-Loop score     & $2 \leq SL \leq 20$ (optional) \\
Optimal Structure & Maximum base-pair matching (often) \\ 
\hline
\end{tabular}
\caption{StemP parameters for proteins from Protein Data Bank\cite{bernstein1977protein}.   Wobble and \texttt {UU} base pairing are only considered when it is known to have such pairing.  This is indicated by superscript $w$ and $u$ respectively in Table \ref{T:pdb_experiment_1_50}. }\label{T:PDBe}
\end{table}


\begin{table}
\small
\begin{center}
\begin{tabular}{lrlrlcccccccc}
\hline
PDB &
  \multicolumn{1}{r}{length} &
  \multicolumn{1}{r}{StemP} &
  \multicolumn{1}{c}{SCR(m)} &
  \multicolumn{1}{c}{DR} &
  \multicolumn{1}{r}{CPU(s)} &
  FOLD &
  MaxExpect &
  ProbKnot &
  {MC} &
  \multicolumn{1}{l}{NAST} \\
  \hline
1RNG   $^{w}$ & 12 & \textbf {1.00}      & 1 (1) & 1 & 0.00  & \textbf {1.00}  & \textbf {1.00} & \textbf {1.00}  &   &   \\
2F8K          & 16 & \textbf {0.91}      & 1 (1) & 1 & 0.00  & \textbf {0.91}  & 0.82           & \textbf {0.91}  & O & X \\
2KVN          & 17 & \textbf {1.00}      & 1 (1) & 1 & 0.00  & \textbf {1.00}  & \textbf {1.00} & 0.91            &   &   \\
2AB4          & 20 & \textbf {1.00}$ ^S$ & 1 (1) & 1 & 0.00  & \textbf {1.00}  & \textbf {1.00} & 0.93            & O & X \\
361D          & 20 & \textbf {1.00}      & 1 (1) & 1 & 0.00  & 0.79            & 0.79           & 0.79            & O & O \\
2ANN          & 25 & \textbf {1.00}      & 4 (4) & 3 & 0.01  & 0.65            & 0.71           & 0.77            & X & O \\
1RLG          & 25 & \textbf{0.91}$^L$   & 1 (3) & 1 & 0.03  & 0.79            & 0.79           & 0.79            & X & O \\
2QUX          & 25 & \textbf {1.00}      & 1 (1) & 1 & 0.01  & \textbf{1.00}   & \textbf{1.00}  & \textbf{1.00}   & O & X \\
387D          & 26 & \textbf {0.77}      & 4 (3) & 4 & 0.01  & 0               & 0              & 0.42            & X & X \\
2L5Z $^w$     & 26 & \textbf {0.95}      & 1 (1) & 1 & 0.01  & \textbf {0.95}  & \textbf {0.95} & \textbf {0.95}  &   &   \\
1MSY $^{w}$   & 27 & \textbf{0.91}$^S$   & 4 (1) & 3 & 0.02  & 0.77            & 0.77           & 0.83            & X & O \\
1L2X $^p$     & 28 & \textbf {0.94}$^S$  & 1 (1) & 1 & 0.02  & 0.79            & 0.79           & 0.72            & X & O \\
2AP5 $^{p}$   & 28 & \textbf {1.00}$^S$  & 1 (1) & 1 & 0.01  & 0.79            & 0.79           & 0.79            & X & X \\
1JID $^w$     & 29 & \textbf{0.80}       & 1 (2) & 1 & 0.02  & \textbf {0.80}  & \textbf {0.80} & \textbf {0.80}  & O & X \\
1OOA $^w$     & 29 & \textbf{1.00}       & 3 (3) & 3 & 0.02  & \textbf {1.00}  & \textbf {1.00} & 0.87            & X & O \\
430D          & 29 & \textbf{0.83}       & 1 (3) & 1 & 0.00  & \textbf {0.83 } & \textbf {0.83} & \textbf {0.83 } & X & O \\
2OZB $^{w}$   & 33 & \textbf{1.00}$^L$   & 293(512) & 5 & 0.98  & \textbf {1.00}  & {0.95}         & 0.89             & O & O \\
1MJI          & 34 & \textbf{0.95}       & 1 (1) & 1 & 0.01  & \textbf{0.95}   & \textbf {0.95} & \textbf {0.95}  & X & X \\
1ET4 $^p$     & 35 & \textbf{0.47}       & 6 (1) & 2 & 0.01  & 0.13            & 0.13           & 0.15            & X & O \\
2HW8 $^w$     & 36 & 0.96                & 9 (11) & 3 & 0.07  & \textbf{1.00}   & \textbf{1.00}  & 0.89            & O & O \\
1I6U $^w$     & 37 & \textbf{0.87}                & 3 (13) & 2 & 0.12  & \textbf {0.87}  & \textbf {0.87} & \textbf {0.87}  & O & O \\
1F1T          & 38 & \textbf{0.88}$^{L}$ & 2 (3) & 2 & 0.07  & \textbf{0.88}   & \textbf {0.88} & 0.73            & O & O \\
1ZHO          & 38 & \textbf{1.00}       & 2 (4) & 2 & 0.02  & \textbf {1.00}  & \textbf {1.00} & 0.9             & O & O \\
5NZ3 $^p$     & 41 & \textbf{0.82}       & 8 (4) & 3 & 0.02  & 0.55             & 0.55            & 0.68            &   &   \\
1SO3          & 47 & \textbf{1.00}$^L$   & 1 (18) & 1 & 2.43  & 0.89             & 0.89            & 0.92            & O & O \\
1XJR $^w$     & 47 & \textbf{0.94}$^{L}$ & 6 (34) & 2 & 26.31 & \textbf{0.94}   & {0.90}         & 0.79            & O & O \\
1U63          & 49 & \textbf{0.97}       & 2 (3) & 2 & 0.12  & \textbf {0.97}  & \textbf {0.97} & \textbf {0.97}  & O & X \\
2PXB $^w$     & 49 & \textbf{0.97}       & 12(42) & 4 & 0.28  & \textbf {0.97}  & \textbf {0.97} & \textbf {0.97}  & O & O \\
 \\
\hline
\end{tabular}
\end{center}
\caption{StemP for Protein sequence (length up to 50).  In the protein list (first column), superscript $p$ indicates pseudo knots, superscript $w$ indicates including wobble base pairs, and $u$ includes \texttt{UU} base pairs. Superscript $L$ indicates using $L=2$ otherwise $L=3$, and $S$ indicates when $SL$ is used. The second column shows the length of the sequences. The third column shows the best MCC value among all maximal clique, the forth column shows Standard Competition Ranking (SCR) of this best MCC in the form of SCR($m$), the fifth column shows the Dense Ranking (DS) of the best MCC, and the sixth column shows the CPU time.  $293^*$ denotes SCR($m$)= 293(512) for 2OZB.
} \label{T:pdb_experiment_1_50}
\end{table}

\subsection{StemP Results and Comparison for Protein Folding} Table \ref{T:pdb_experiment_1_50} presents protein folding results for sequences of length up to 50. 
The comparisons are presented between the proposed method StemP, FOLD\cite{mathews2004incorporating}, MaxExpert\cite{lu2009improved},  ProbKnot\cite{bellaousov2010probknot}, MC \cite{parisien2008mc} and NAST \cite{jonikas2009coarse} based on MCC value in (\ref{eq_mcc}).  The experimental results of those methods were performed on RNAstructure web server\cite{reuter2010rnastructure} with default parameters\footnote[1]{Temperature = 310.15(K), Maximum Loop Size = 30, Maximum \% Energy Difference = 10, Maximum Number of Structures = 20, Window Size = 3, Gamma = 1, Iterations = 1, Minimum Helix Length = 3, SHAPE Intercept = -0.6, SHAPE Slope = 1.8, Maximum probabilities to show = 2.  }.
For the columns MC \cite{parisien2008mc} and NAST \cite{jonikas2009coarse}, this is based on the table in \cite{laing2010computational}, where  O and X indicate success or failure of these methods. and if empty, experiments were not known. 

Some variations are presented as superscripts in Table \ref{T:pdb_experiment_1_50}.  The superscript $p$ indicates existence of  pseudo-knots.  For StemP,  pseudo-knots are naturally predicted without any a priori information, but for some methods it is important to indicate.  The superscripts $w, u$ indicate wobble base \texttt{G-U} pairs and \texttt{U-U} respectively.  We included this  when constructing vertex.  If a sequence folding is known to have a wobble base pair, it helps to add this condition in vertex construction to identify the true folding.  
For protein folding, typically protein sequences don't have a strong structure known a priori, that using Stem-Loop score was not necessary.  We indicated with superscript $S$ to denote the PDBs that Stem-Loop score  are imposed.   For other sequences, the same result can be obtained with or without $SL$ condition.

StemP is computationally efficient. In Table \ref{T:pdb_experiment_1_50} forth column, we present the CPU time of  predicting folding with StemP for each sequence.  The average CPU times in Table \ref{T:pdb_experiment_1_50} is  1.18 seconds.  We used MATLAB with Intel\textregistered Core i5-9600K processor with 3.7GHz 6 Core CPU and 16 GB of RAM. 
%

In Table \ref{T:pdb_experiment_1_50}, StemP results in second column show the best MCC values among all maximal cliques.  These best MCC values show that they are more accurate compared to other methods for all sequences except for one 2HW8. 
The SCR($m$) in the forth column shows, 10 out of 28 results are SCR($m$)=1(1), that the unique top choice (maximum matching) is the folding prediction, and the matching accuracy is of MCC 1 or above 0.91 (i.e. 100\% or above 91\% matching).  Another 4 is the top choice, but with a multiple possibilities.   We list the rank 1, MCC values for the case, when the best folding prediction is not ranked 1: in Table \ref{T:pdb_experiment_1_50}, 2ANN 0.77, 387D 0.00, 1MSY 0.00 , 1OOA 0.50, 2OZB 0.68, IET4 0.22, 2HW8 0.67, 1I6U 0.84, 1F1T 0.76, 1ZHO 0.74,  5NZ3 0.67, 1XJR 0.45, 1U63 0.95, 2PXB 0.17.

\begin{figure}
\begin{center} 
\begin{tabular}{ccc}
\includegraphics[height=1in]{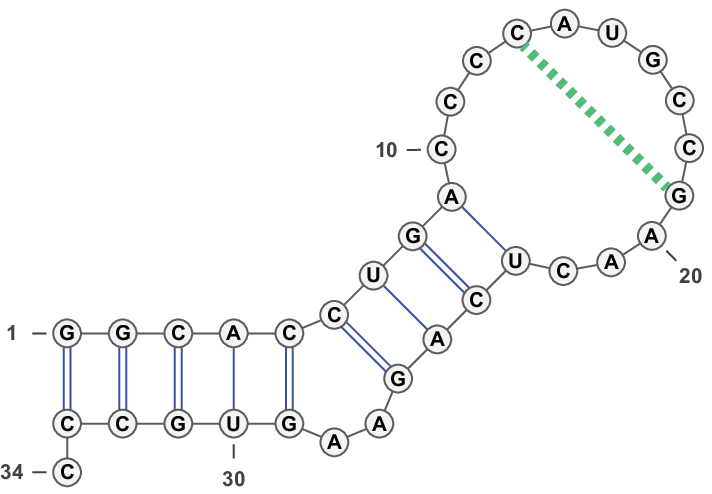} &
\includegraphics[height=0.5in]{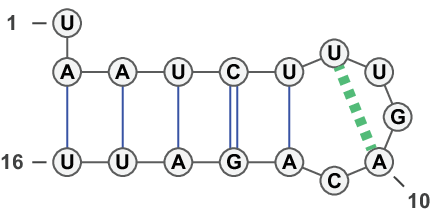}&
\includegraphics[height=1in]{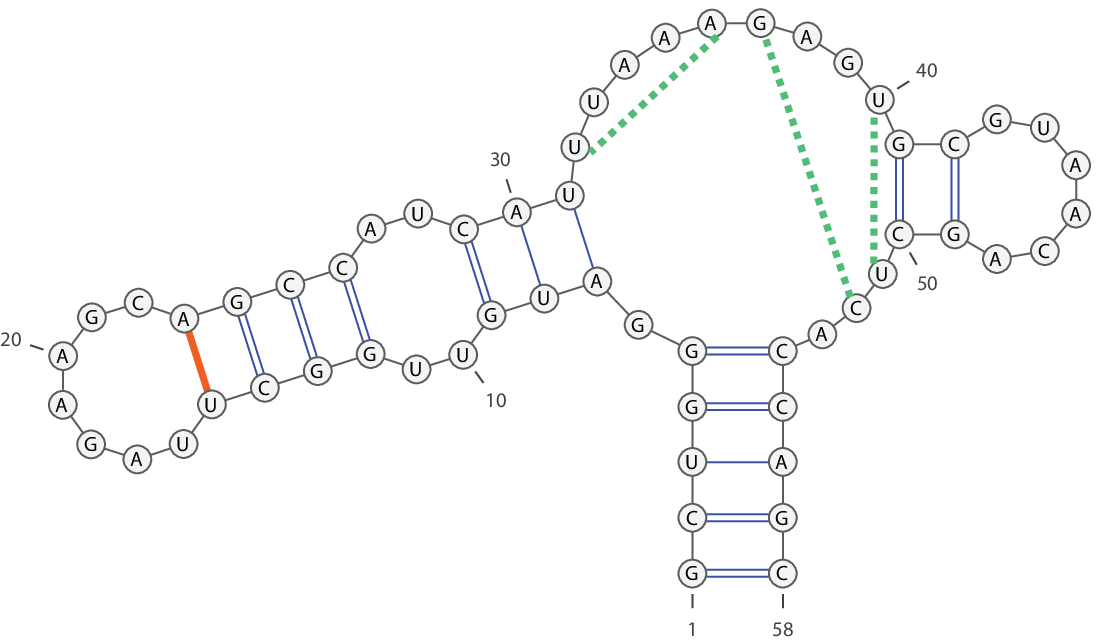} \\
(a) MCC 0.98 & (b) MCC 0.92 & (c) MCC 0.95 
\end{tabular}
\end{center}
\caption{Examples of StemP of MCC above 0.9.  (a) StemP of 1MJI (length 34).  (b) StemP of 2F8K (length 16).  (c) StemP of 1MMS (length 58).  
The green lines is missing basepair matching (False Negative), which are either non-canonical pairing or a single canonical pairing. }
\label{F:pdb_not100}
\end{figure}

Figure \ref{F:pdb_not100} shows examples when the best matching is not  MCC 1 (100 \% matching).   Typical examples are shown for MCC 0.95, 0.91 and 0.85.  These mismatches are caused by (i) non-canonical base-pair matching such as \texttt{G-G}, \texttt{A-G}, \texttt{C-U}, or  (ii) a singe canonical base pair with length 1.   In our current setting, we only consider the stem of length $L \geq 2$.  

\begin{figure}
\begin{center} 
\begin{tabular}{ccccc}
\includegraphics[width = 1.2in]{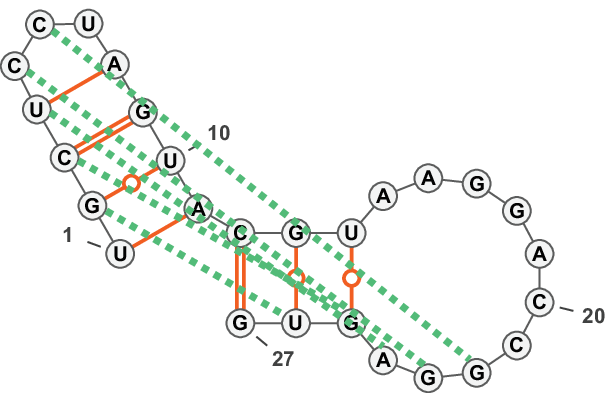}  
&\includegraphics[width = 0.8in]{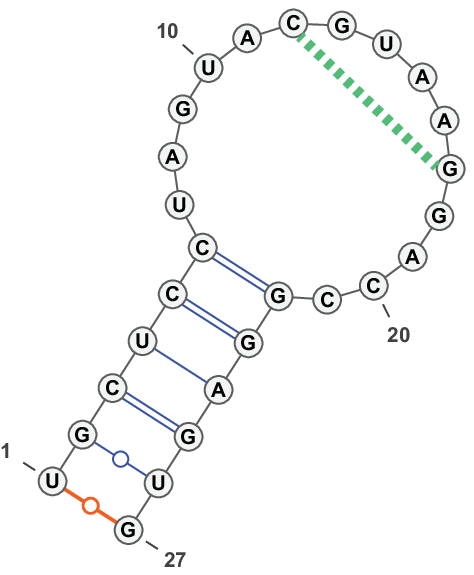} 
& \includegraphics[width = 1in]{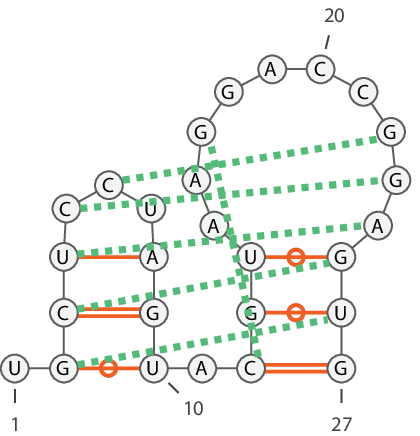}
& \includegraphics[width = 0.7in]{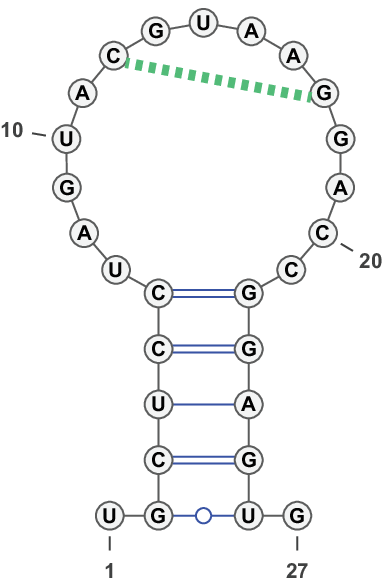}
& \includegraphics[width = 0.7in]{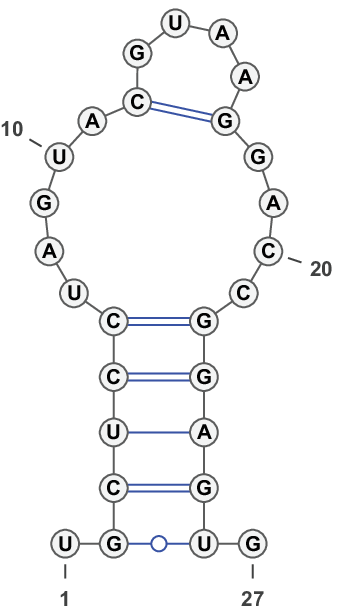} \\
(a) SCR =1&  (b)SCR =2(2) & (c) SCR=2(2) & (c) SCR=4 & (d) True folding \\
MCC 0 & MCC 0.83 & MCC 0 & MCC 0.91 & 
\end{tabular}
\end{center}
\caption{An example of StemP when the best MCC is not SCR=1.  For 1MSY (length 27), the true folding is not maximum matching.  (a) StemP result rank 1 with MCC 0. (b) StemP result SCR($m$)=2(2) and MCC 0.83. (c) Another SCR($m$)=2(2) and MCC 0.  (d) Next StemP result SCR=4 with most matching to the true folding structure (MCC 0.91). The green lines show the missing base pairs, and the red line wrong matching. One base pair \texttt{C(12) - G(16)} is missing in this example, since we use  $ L \geq 2$.  (VARNA\cite{darty2009varna} is adopted for visualization of secondary structures.)}
\label{F:pdb_not1_2} 
\end{figure}

For the cases when the best MCC is no SCR=1, we give an example in Figure  \ref{F:pdb_not1_2} with 1MSY (length 27).  The best folding result is found in rank 4.  The true folding is similar to (b) rank 2 and (d) rank 3, with one more base-pair in (b), yet the true folding is closer to (d).   This example shows, although there are similar folding structures with more matching, the true folding does not connect all the base-pair matching.  

\section{StemP for tRNA folding prediction}\label{sec: trna}
\begin{figure}
\begin{center}
\begin{tabular}{ccc}
    \includegraphics[width = 1.4 in]{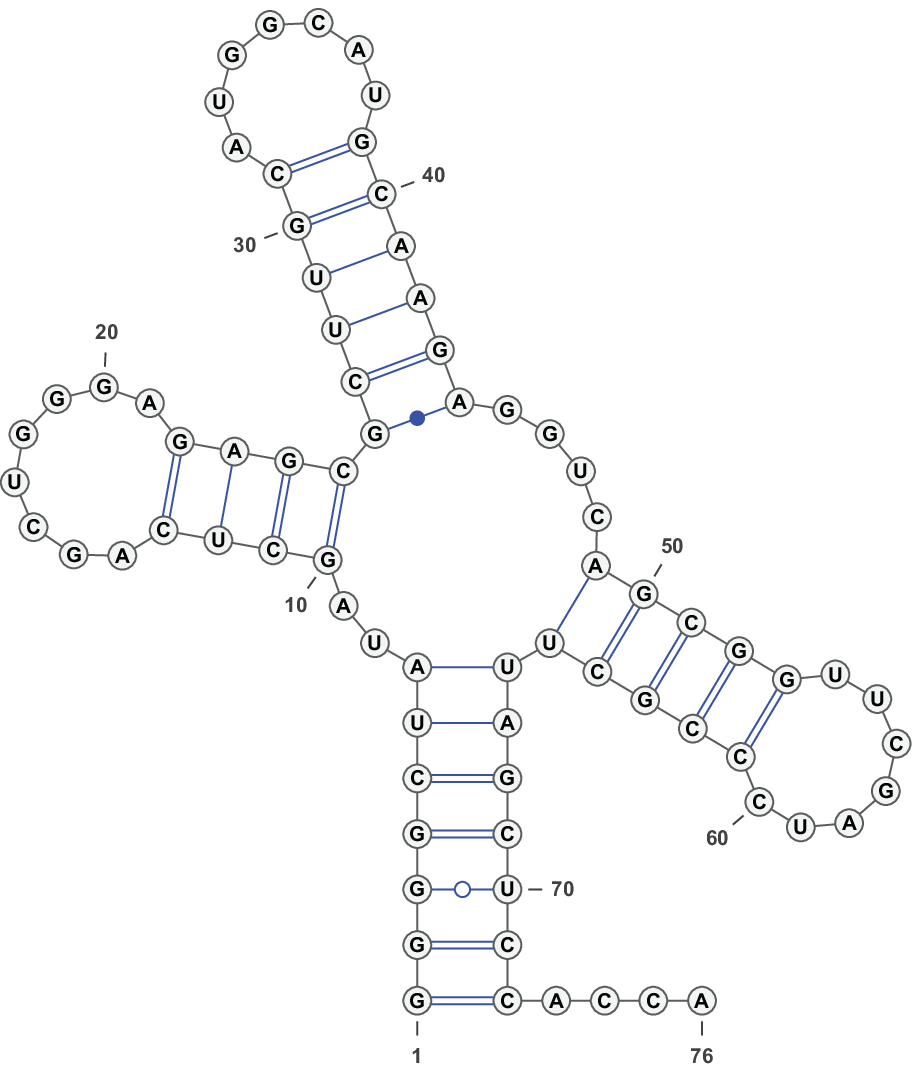} &
    \includegraphics[width = 1.4 in]{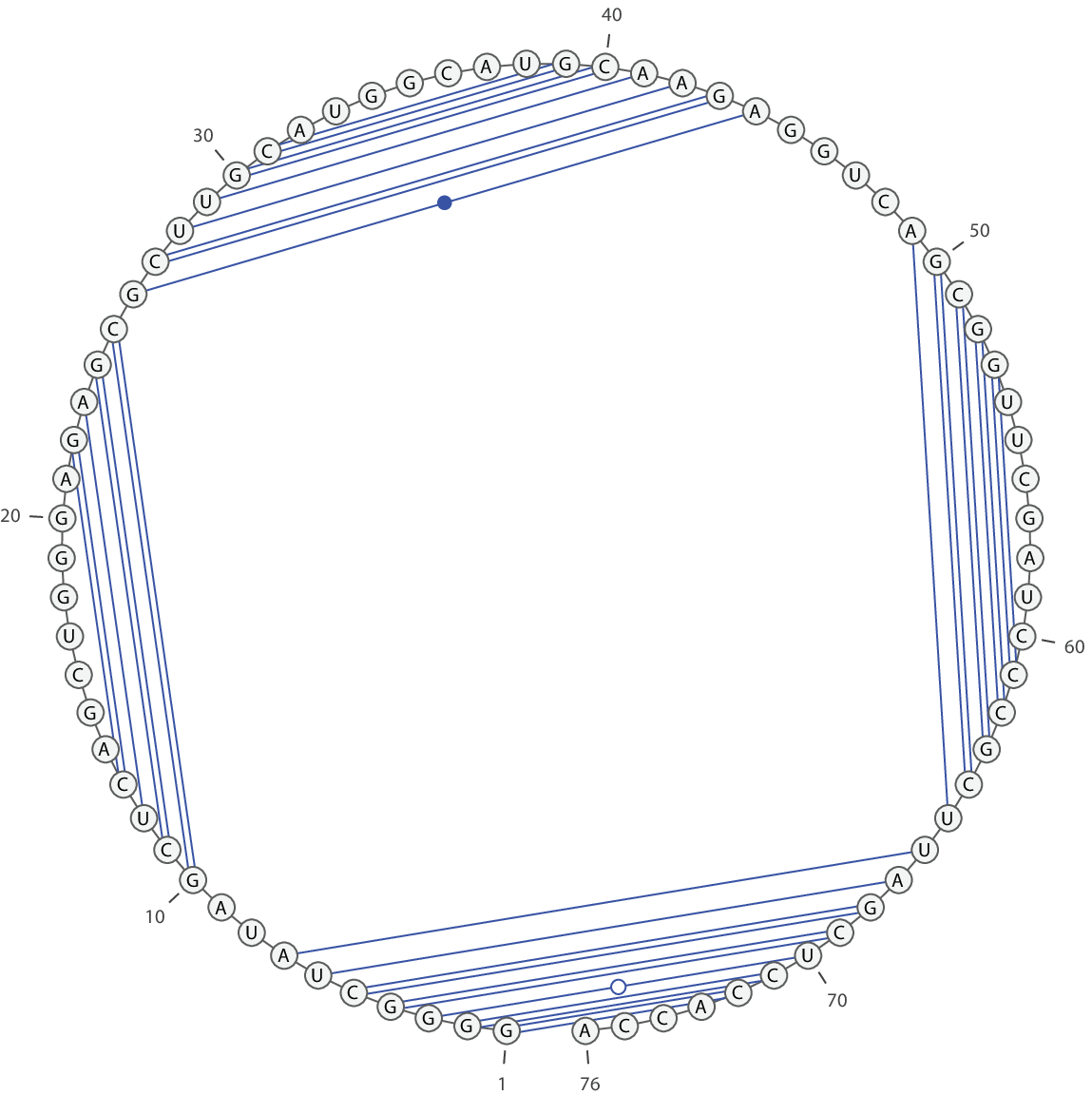}\\
    (a) Radiate view & (b) Circular view\\
\end{tabular}
\end{center}
    \caption{True folding structure of a tRNA sequence (Accession Number AE009341) , from The Gutell Lab \cite{cannone2002comparative}. tRNA has relatively standard form of folding. }
    \label{F: true tRNA example}
\end{figure}

The folding structure of tRNA is typically standard as shown in Figure \ref{F: true tRNA example}.  There are 4 distinct regions, which are Acceptor, D loop, Anticodon loop and TC loop. 
The structure of transfer RNA (tRNA) was determined through comparative analyses of RNA structure, and various methods are developed \cite{bevilacqua2016genome}.  
Different types of RNA including tRNA were  studied in \cite{tinoco1999rna} regarding secondary structure and tertiary structure. It was shown that the secondary structure is more significant and stable than the latter one.  MC-fold \cite{parisien2008mc} is a classical model to unify all basepair matching energetic contributions. Minimum Free Energy model was adopted by \cite{washietl2005fast}, \cite{lorenz2016predicting} and \cite{ren2005hotknots} to discover tRNA sequences, where probing data is utilized in \cite{lorenz2016predicting}.  In \cite{hamada2006mining}, a possible RNA alignment sequence was represented by finding the most frequent stem patterns from a database, and experiments were performed  on a variety of RNA sequences including tRNA. From the geometric point of view, the author of \cite{fera2004rag} employs the idea of tree graphs and dual graphs to represent RNA tree motifs and general RNA motifs which makes it possible to characterize the typologies of RNA structure.   There are probability based methods\cite{bellaousov2010probknot,tan2017turbofold} which measures the probability of the structure of a nucleic or base pair based on a large learning group. Some methods provide multiple sequences alignments with probabilities and can be extended to different types of RNA sequences.

\subsection{Parameters} 
\begin{figure}
\begin{center}
\begin{tabular}{ccc}
\includegraphics[width = 1.3in]{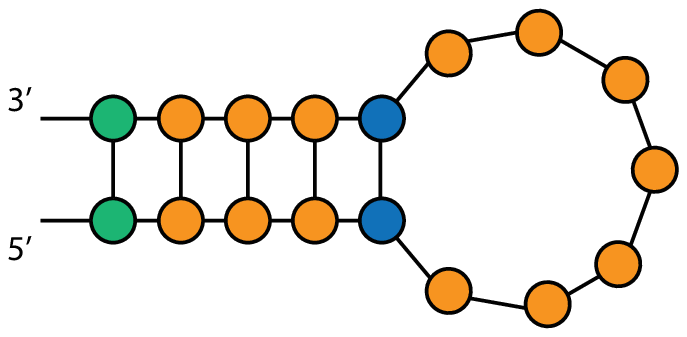} &
\includegraphics[width = 1.3in]{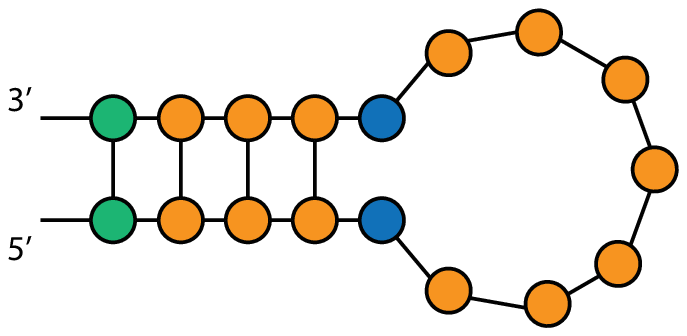}&
\includegraphics[width = 1.3in]{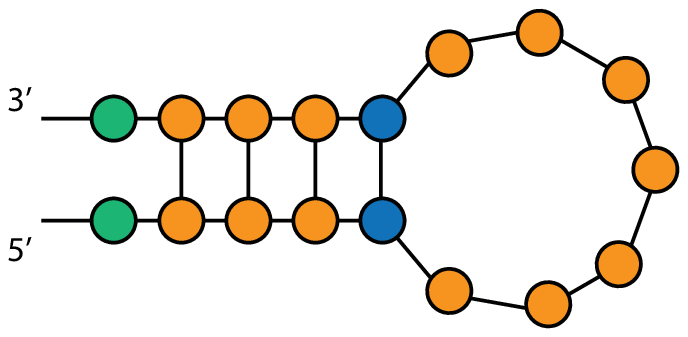} \\
(a) & (b) & (c) 
\end{tabular}
\caption{Partial stem considered for tRNA folding.  (a) and (c) are typical vertex constructed. 
(b) a new type of partial stem considered for tRNA in addition.  Notice one interior basepair is not connected. } 
\label{F:partialSum}
\end{center}
\end{figure}

For StemP for tRNA, (i) Table \ref{T:trna parameters} shows all the parameters used for the tRNA folding prediction.  In addition, to account for the standard shape of tRNA, we make a couple of following modifications: (ii) We use Acceptor-Stem-Loop score in (\ref{E:msl}) to find the Acceptor stem, and   consider (iii) Partial Stem.  For tRNA, it is common that some of matching base pairs at the end of the stem do not pair.  We consider these Partial Stems, which is a sub-stem of a typical vertex. Figure  \ref{F:partialSum} (b) shows the new partial stem we consider in addition.  

\begin{table}
\centering
\begin{tabular}{l|c}
\hline
 \multicolumn{2}{c}{StemP parameters for tRNA folding}\\
\hline
Basepair matching & Canonical and  Wobble matching \\
Stem type & Partial Stems included\\
Minimum Stem Length & $L= 3 $ \\
Stem-Loop Distance & $12 \leq d_i \leq 18 $  \\
Stem-Loop score    &  $3 < SL \leq 4.7$ or $ 5.4$  \\
Stem-Loop Distance (Acceptor) &  $ \hat{l}/2 < d_i$ \\
Acceptor-Stem-Loop score    & $ ASL \leq 3$\\
Optimal Structure & Maximum matching Maximal clique \\
\hline
\end{tabular}
\caption{StemP parameters for tRNA. For Cysteine, Glutamic Acid, Glutamine, Histidine, $SL$ upper bound 4.7 is used.  For Alanine, Asparagine, Aspartic Acid, Glutamic Acid, Glutamine, Glycine, Histidine, Isoleucine, Lysine, Methionine,  Phenylalanine,  Proline, Tryptophan, Tyrosine, $SL$ upper bound 5.4 is used. $\hat{l}$ is the total length of the sequence.  }\label{T:trna parameters}
\end{table}

\subsection{StemP Results for tRNA} 
\begin{figure}
\centering
\begin{tabular}{cccc}
\includegraphics[width = 1.5 in]{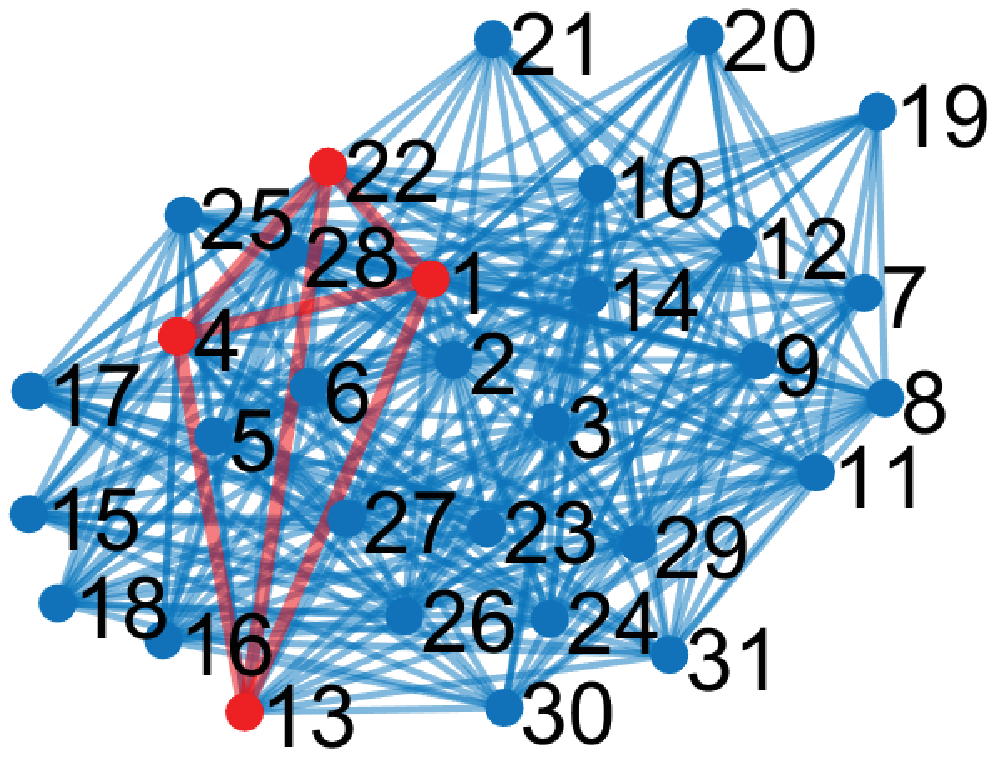} &
\includegraphics[width = 1 in]{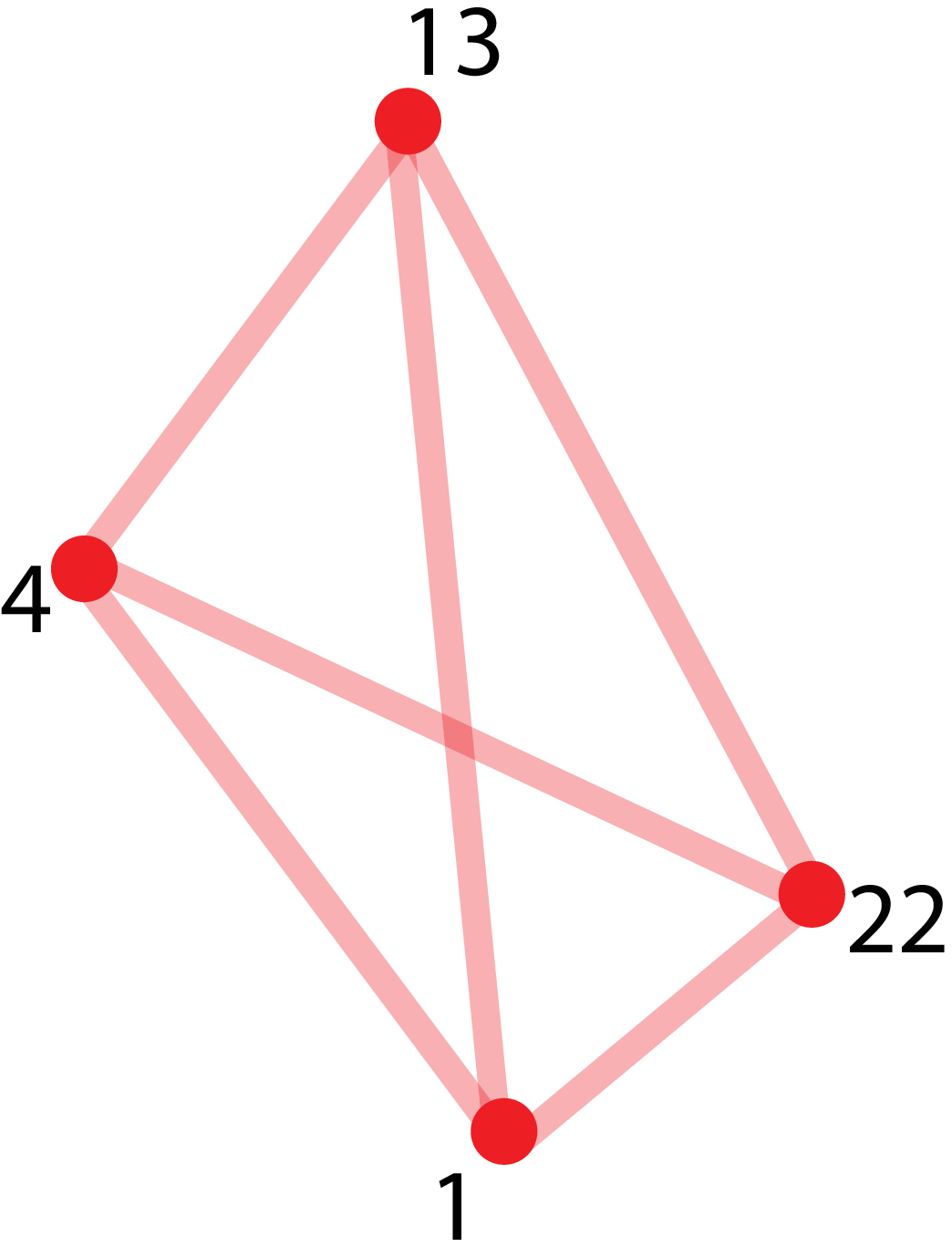} &
\includegraphics[width = 1.5 in]{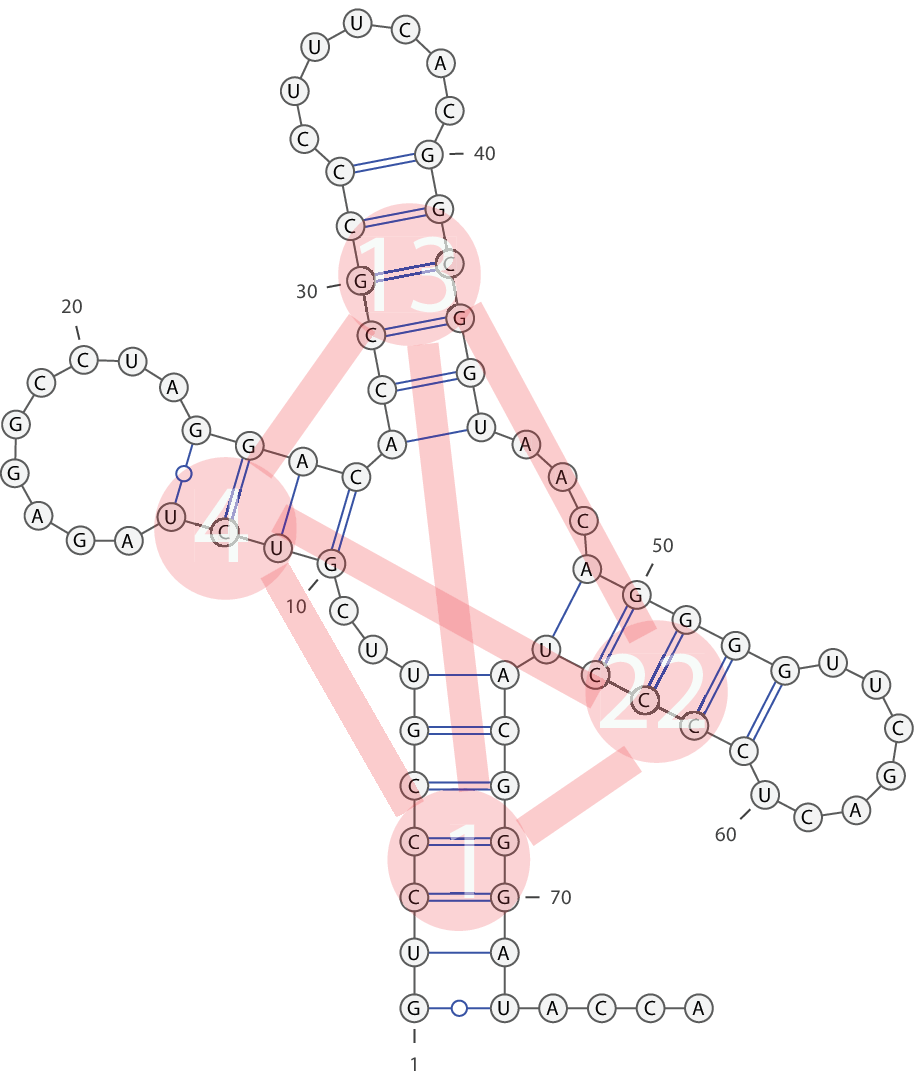} &
\includegraphics[width = 1.5 in]{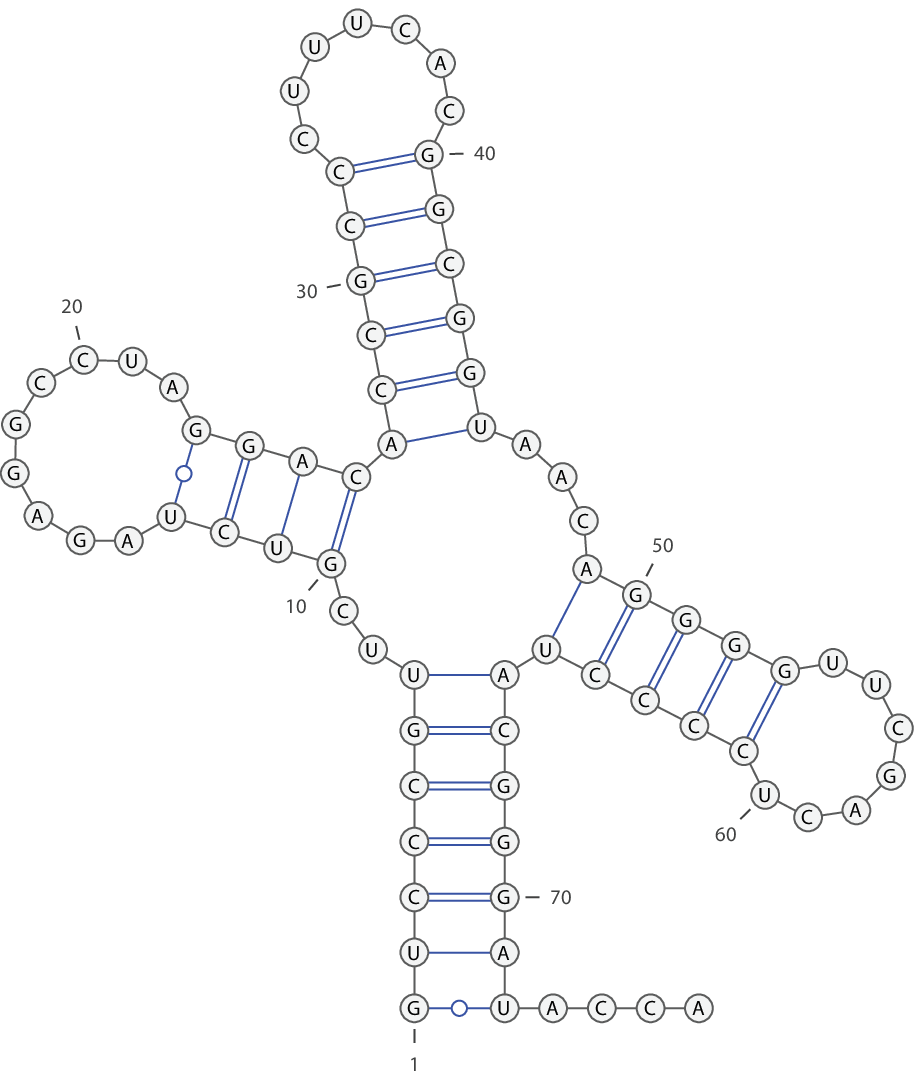} \\
(a) Stem-graph   & (b) Maximal Clique & (c) Prediction & (d) True
\end{tabular}
\caption{StemP for tRNA (MCC = 1.00). (a) Stem-graph.  (b) The maximum matching maximal clique (rank 1).  (c) StemP with Accession Number AB041850, superposed with the clique (vertices indicates the stems).  (d) The true folding structure, which is identical to the prediction}
\label{F:tRNAexample}
\end{figure}

Figure \ref{F:tRNAexample} shows an outline and typical result of StemP for tRNA, with an example (Accession Number AB041850) from organism Alanine.  In Figure \ref{F:tRNAexample},  (a) is the full Stem-graph,  (b) one of the identified maximal clique, and (c) what the clique (b) is representing.   In (c), red Stem-graph is superposed over the folding structure to show the clique. 
Figure \ref{F:tRNAexample}(c) shows StemP result with 100\% matching in the all stems, as shown in the true folding structure in Figure \ref{F:tRNAexample}(d).  StemP gives 100\% matching in the all stems.

\begin{figure}
\centering
\subfloat[SCR ranking]{\includegraphics[width=0.47 \textwidth]{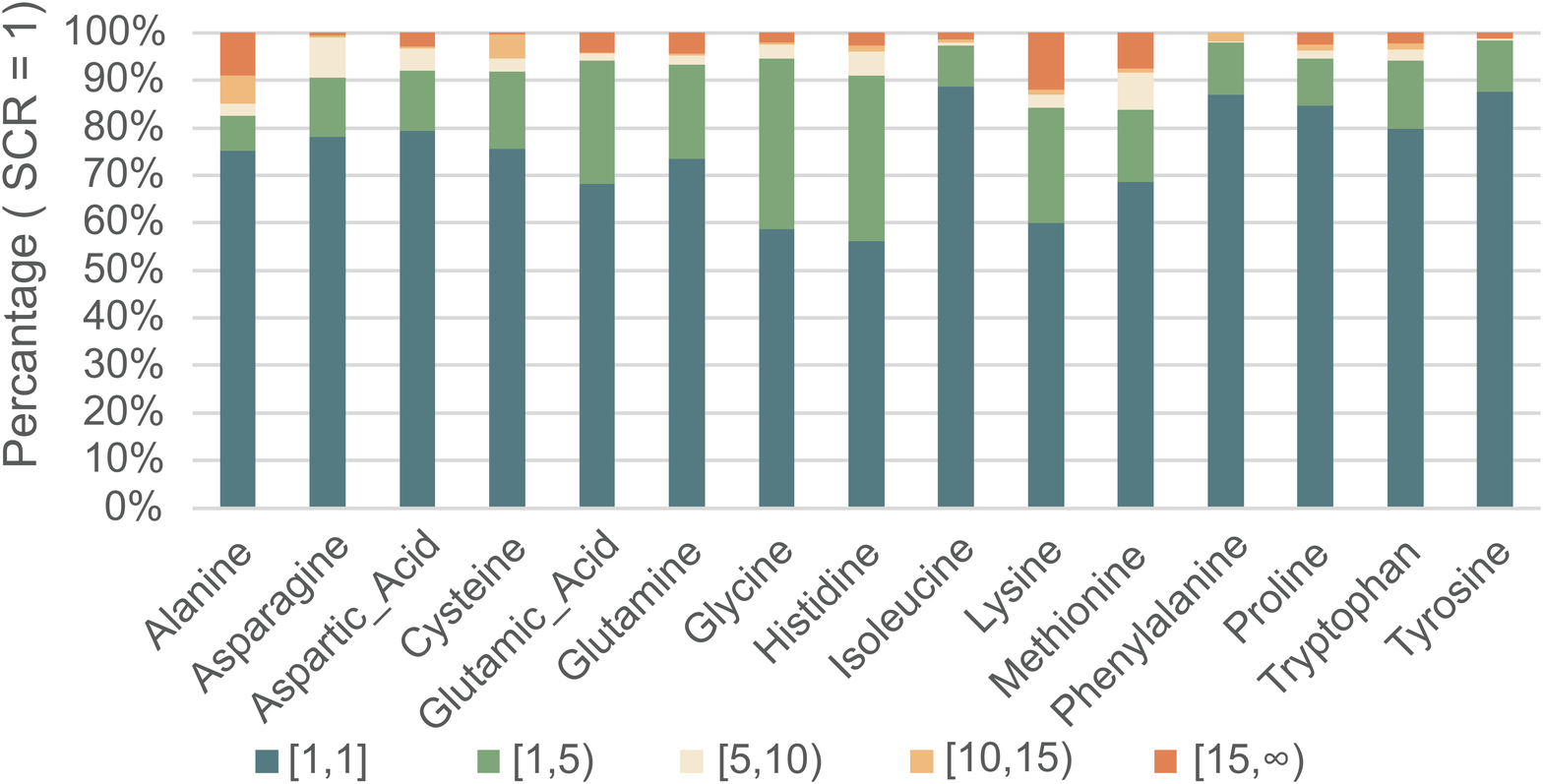}
\label{fig_second_case}} 
\subfloat[MCC values of SCR=1]{\includegraphics[width=0.47 \textwidth]{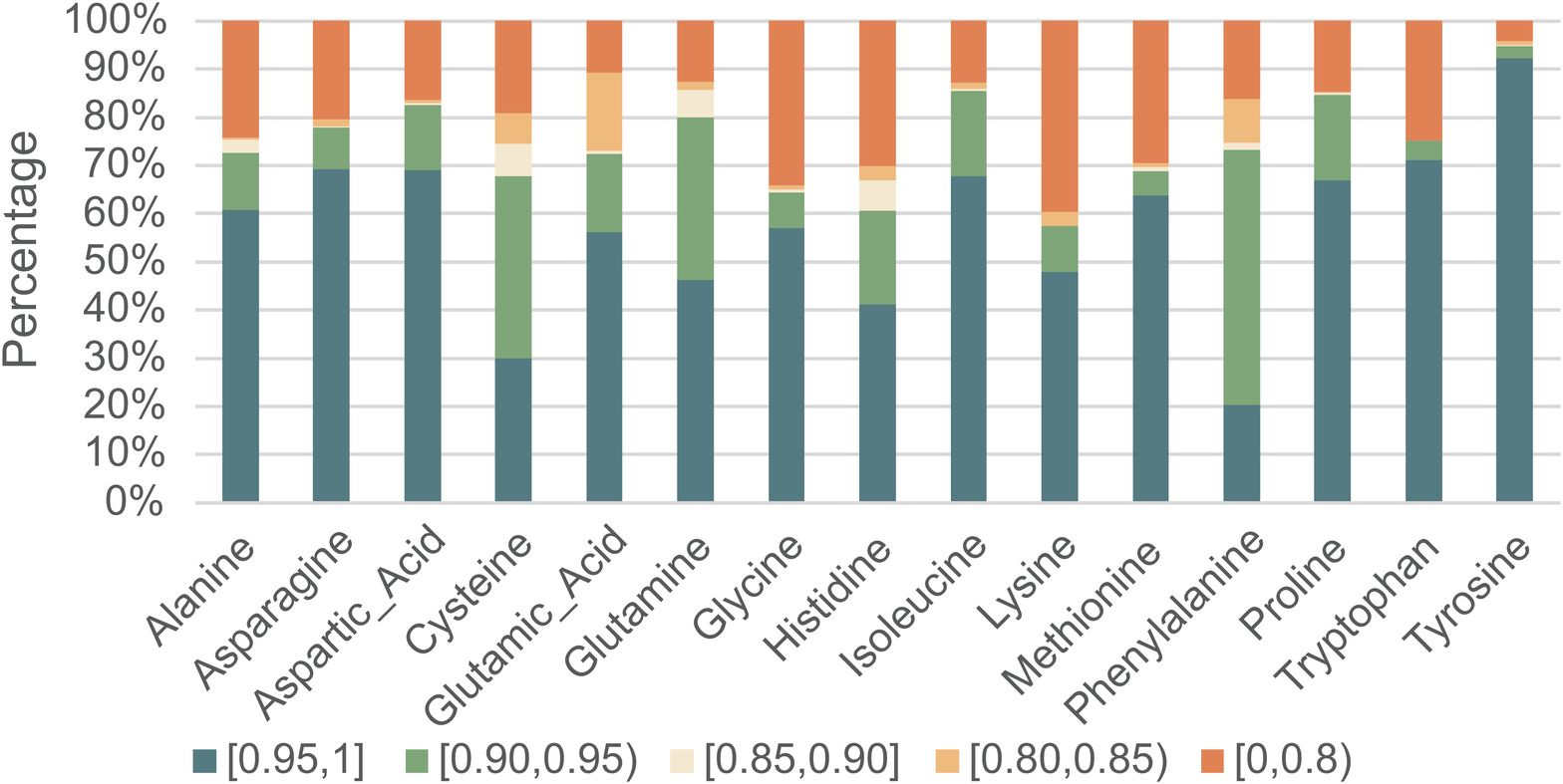}
\label{fig_top_choice}}\\
\subfloat[MCC values (Best)]{\includegraphics[width=0.47 \textwidth]{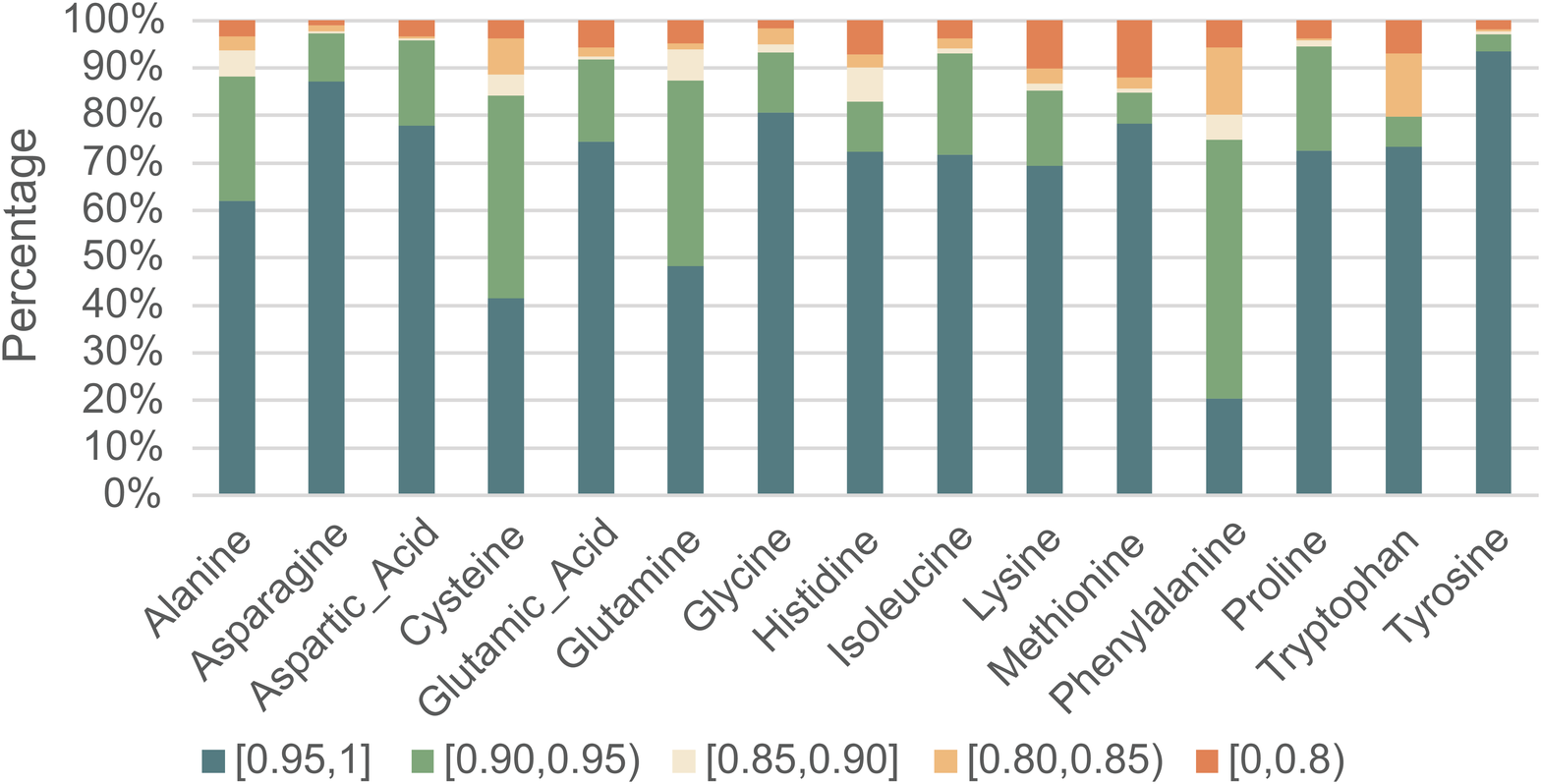}
\label{fig_first_case}}
\caption{StemP for 27,010 different tRNA sequences.   (a) The percentage of sequences that have SCR in $[1,1], (1,5],(5,10],(10,15],(15, \infty)$ for the highest MCC prediction. (b) The percentage of sequences with SCR = 1  that hat the prediction score (MCC) to be in $[0.95,1], [0.90, 0.05), [0.85,0.90),[0.80, 0.85),[0, 0.8)$ in each organism.  (c) The percentage of sequences that have the best prediction score (MCC) to be in $[0.95,1], [0.90, 0.05), [0.85,0.90),[0.80, 0.85),[0, 0.8)$ for each organism. }
\label{fig_tRNA_stat}
\end{figure}

In Table \ref{T:tRNA_results} and Figure \ref{fig_tRNA_stat}, we present the results of StemP for tRNA folding prediction for 27,010 number of tRNA sequences containing 15 subsets of tRNA from The Gutell Lab \cite{cannone2002comparative}.  We consider sequences that satisfies the following two criterias as valid inputs: (i) The length larger than or equal to 50; (ii) There exists at least one base pair in the true folding structure.  
The typical number of vertices for tRNA ranged from 20 to 40, the number of edges ranged from 120 to 150, and the number of cliques ranged from 180 to 2000.  
 In Table \ref{T:tRNA_results}  (a) shows the SCR values of the best prediction of StemP,  (b) MCC values of the top predictions, and (c) MCC values of the best predictions.   The second column shows the total number of experiments for each organism.   The third column shows the total number of results with the percentage in the parenthesis.  The forth column shows additional number of folding results, and the  combined percentage in parenthesis.  The last column shows the CPU time in seconds computed in average.   The bold numbers show when it is near 90\%. Notice majority of sequences (many over 85\% of the sequence) matched MCC 0.9 or higher.  It is shown that there are more than 90\% of the sequences that have maximum MCC within top 5 SCR.  
\begin{table}
\centering
\small
(a) The SCR values for the best StemP prediction on tRNA 
\begin{tabular}{lrrrrrrrr}
\hline
\textbf{Organism} &
  \textbf{\#} &
  \multicolumn{1}{r}{\textbf{$\leq 1$}} &
  \multicolumn{1}{r}{\textbf{$\leq 5$}} &
  \multicolumn{1}{r}{\textbf{$\leq 10$}} &
  \multicolumn{1}{r}{\textbf{$\leq 15$}} &
  \multicolumn{1}{r}{\textbf{$< 15$}} \\
  \hline
Alanine        & 4344 & 3261(75.1\%) & 325(82.6\%) & 108(85.0\%)  & 257(\textbf{91.0}\%)  & 393(9.0\%)  \\
Asparagine     & 1250 & 977(78.1\%)  & 156(\textbf{90.6}\%) & 106 (99.1\%) & 5(99.5\%)    & 6(0.4\%)  \\
Aspartic\_Acid & 1399 & 1111(79.4\%) & 177(\textbf{92.1}\%) & 66 (96.8\%)  & 5(97.1\%)  & 40(2.9\%)  \\
Cysteine       & 596  & 450(75.5\%) & 98(\textbf{91.9}\%) & 16(94.6\%)  & 30(99.7\%)  & 2(0.3\%)  \\
Glutamic\_Acid & 1895 & 1291(68.1\%) & 492(\textbf{94.1}\%) & 31(95.7\%)  & 2(95.8\%)  & 79(4.2\%)  \\
Glutamine      & 1151 & 846(73.5\%) & 229(\textbf{93.4}\%) & 22(95.3\%)  & 4(95.7\%)  & 50(4.3\%)  \\
Glycine        & 2493 & 1465(58.8\%) & 892(\textbf{94.5}\%) & 75(97.6\%)  & 12(98.0\%)  & 49(2.0\%)  \\
Histidine      & 987  & 554(56.1\%) & 345(\textbf{91.1}\%) & 49(96.1\%)  & 13(97.4\%)  & 26(2.7\%)  \\
Isoleucine     & 4558 & 4046(88.8\%) & 393(\textbf{97.4}\%) & 23(97.9\%)  & 31(98.6\%)  & 65(1.4\%)  \\
Lysine         & 1566 & 938(59.9\%) & 381(84.2\%) & 42(86.9\%)  & 17(\textbf{88.0}\%)  & 188(12.0\%)  \\
Methionine     & 1789 & 1228(68.6\%) & 273(83.9\%) & 137(\textbf{91.6}\%)  & 15(92.4\%)  & 136(7.6\%)  \\
Phenylalanine  & 2621 & 2279(87.0\%) & 286(\textbf{97.9}\%) & 11(98.3\%)  & 45(100.0\%)  & 0(0.0\%)  \\
Proline        & 1411 & 1197(84.7\%) & 139(\textbf{94.5}\%) & 25(96.3\%)  & 16(97.4\%)  & 36(2.6\%)  \\
Tryptophan     & 173  & 138(79.8\%) & 25(\textbf{94.2}\%) & 4 (96.5\%)  & 2(97.7\%)  & 4(2.3\%)  \\
Tyrosine       & 777  & 681(87.7\%) & 84(\textbf{98.5}\%) & 3(98.8\%)  & 0(98.8\%)  & 9(1.2\%) \\
\hline
\end{tabular}

\vspace{0.5cm}
(b) The MCC values of the top SCR = 1 StemP prediction.
\begin{tabular}{lrrrrrrrr}
\hline
\textbf{Organism} &
  \textbf{\#} &
  \multicolumn{1}{r}{\textbf{$\geq 0.95$}} &
  \multicolumn{1}{r}{\textbf{$\geq 0.90$}} &
  \multicolumn{1}{r}{\textbf{$\geq 0.85$}} &
  \multicolumn{1}{r}{\textbf{$\geq 0.80$}} &
  \multicolumn{1}{r}{\textbf{$< 0.80$}} \\
  \hline
Alanine        & 4344 & 2645(60.9\%)& 508(72.6\%)& 123(75.4\%)& 20(75.9\%)& 1048(24.1\%)\\
Asparagine     & 1250 & 865(69.2\%)& 109(77.9\%)& 4 (78.2\%)& 16(79.5\%)& 256(20.5\%)\\
Aspartic\_Acid & 1399 & 966(69.0\%)& 190(82.6\%)& 4 (82.9\%)& 11(83.7\%)& 228(16.3\%)\\
Cysteine       & 596  & 179(30.0\%)& 225(67.8\%)& 40(74.5\%)& 38(80.9\%)& 114(19.1\%)\\
Glutamic\_Acid & 1895 & 1065(56.2\%)& 309(72.5\%)& 11(73.1\%)& 307(89.3\%)& 203(10.7\%)\\
Glutamine      & 1151 & 533(46.3\%)& 389(80.1\%)& 66(85.8\%)& 18(87.4\%)& 145(12.6\%)\\
Glycine        & 2493 & 1422(57.0\%)& 184(64.4\%)& 15(65.0\%)& 20(65.8\%)& 852(34.2\%)\\
Histidine      & 987  & 407(41.2\%)& 192(60.7\%)& 62(67.0\%)& 29(69.9\%)& 297(30.1\%)\\
Isoleucine     & 4558 & 3093(67.9\%)& 807(85.6\%)& 21(86.0\%)& 58(87.3\%)& 579(12.7\%)\\
Lysine         & 1566 & 751(48.0\%)& 148(57.4\%)& 0 (57.4\%)& 47(60.4\%)& 620(39.6\%)\\
Methionine     & 1789 & 1140(63.7\%)& 93 (68.9\%)& 13(69.6\%)& 17(70.6\%)& 526(29.4\%)\\
Phenylalanine  & 2621 & 533(20.3\%)& 1386(73.2\%)& 39(74.7\%)& 241(83.9\%)& 422(16.1\%)\\
Proline        & 1411 & 944(66.9\%)& 252(84.8\%)& 4 (85.0\%)& 5 (85.4\%)& 206(14.6\%)\\
Tryptophan     & 173  & 123(71.1\%)& 7  (75.1\%)& 0 (75.1\%)& 0 (75.1\%)& 43 (24.9\%)\\
Tyrosine       & 777  & 717(92.3\%)& 20 (94.9\%)& 2 (95.1\%)& 6 (95.9\%)& 32 (4.1\%)\\
\hline
\end{tabular}

\vspace{0.5cm}
(c) The MCC values of the best StemP prediction.
\begin{tabular}{lrrrrrrrr}
\hline
\textbf{Organism} &
  \textbf{\#} &
  \multicolumn{1}{r}{\textbf{$\geq 0.95$}} &
  \multicolumn{1}{r}{\textbf{$\geq 0.90$}} &
  \multicolumn{1}{r}{\textbf{$\geq 0.85$}} &
  \multicolumn{1}{r}{\textbf{$\geq 0.80$}} &
  \multicolumn{1}{r}{\textbf{$< 0.80$}} &
  \textbf{cpu} \\
  \hline
Alanine        & 4344 & 2693(62.0\%)  & 1141(88.3\%)  & 233(\textbf{93.6}\%) & 134(86.7\%) & 134 (3.4\%)   & 0.24 \\
Asparagine     & 1250 & 1090(87.2\%)  & 127 (\textbf{97.4}\%)  & 5  (97.8\%) & 16 (99.0\%) & 12  (1.0\%)   & 0.23 \\
Aspartic\_Acid & 1399 & 1090(77.9\%)  & 249 (\textbf{95.7}\%)  & 6  (96.1\%) & 7  (96.6\%) & 47  (3.4\%)   & 0.22 \\
Cysteine       & 596  & 247 (41.4\%)  & 255 (84.2\%)  & 26 (88.6\%) & 45 (\textbf{96.1}\%) & 23  (3.9\%)   & 0.09 \\
Glutamic\_Acid & 1895 & 1412(74.5\%)  & 328 (\textbf{91.8}\%)  & 12 (92.5\%) & 35 (94.3\%) & 108  (5.7\%)   & 0.19 \\
Glutamine      & 1151 & 556 (48.3\%)  & 449 (87.3\%)  & 75 (\textbf{93.8}\%) & 14 (95.0\%) & 57  (5.0\%)   & 0.11 \\
Glycine        & 2493 & 2008(80.5\%)  & 314 (\textbf{93.1}\%)  & 46 (95.0\%) & 82 (98.3\%) & 43  (1.7\%)   & 0.14 \\
Histidine      & 987  & 713 (72.2\%)  & 105 (82.9\%)  & 71 (\textbf{90.1}\%) & 27 (92.8\%) & 71  (7.2\%)   & 0.19 \\
Isoleucine     & 4558 & 3267(71.7\%)  & 970 (\textbf{93.0}\%)  & 53 (94.1\%) & 94 (96.2\%) & 174 (3.8\%)   & 0.18 \\
Lysine         & 1566 & 1087(69.4\%)  & 247 (86.8\%)  & 25 (86.8\%) & 49 (\textbf{89.9}\%) & 158 (10.1\%)  & 0.25 \\
Methionine     & 1789 & 1399(78.2\%)  & 118 (84.8\%)  & 17 (85.7\%) & 38 (\textbf{87.9}\%) & 217 (12.1\%) & 0.17 \\
Phenylalanine  & 2621 & 533 (20.3\%)  & 1430(74.9\%)  & 139(80.2\%) & 371(\textbf{94.4}\%) & 148 (5.6\%)   & 0.04 \\
Proline        & 1411 & 1023(72.5\%)  & 310 (\textbf{94.5}\%)  & 19 (95.8\%) & 5  (96.2\%) & 54  (3.8\%)   & 0.19 \\
Tryptophan     & 173  & 127 (73.4\%)  & 11  (79.8\%)  & 0  (79.8\%) & 23 (\textbf{93.1}\%) & 12   (6.9\%)  & 0.16 \\
Tyrosine       & 777  & 726 (\textbf{93.4}\%)  & 28  (97.0\%)  & 5  (97.7\%) & 4  (98.2\%) & 14  (1.8\%)   & 0.15\\
\hline
\end{tabular}
\caption{StemP for 27,010 different tRNA sequences.  (a) The SCR values for the best StemP prediction on tRNA.  (b) The MCC values of the top SCR = 1 StemP prediction.  (c) The MCC values of the best StemP prediction.}
 \label{T:tRNA_results}
\end{table}

In Figure \ref{fig_tRNA_stat}, we present results of Table \ref{T:tRNA_results} in statistical plot.  %
In (a), we show the percentage of SCR in $[1,1], (1,5],(5,10],(10,15],(15, \infty)$ for the best MCC prediction.
In (b), we show the percentage of MCC values in $[0.95,1], [0.90, 0.05), [0.85,0.90),[0.80, 0.85),[0, 0.8)$ for the top SCR=1 prediction in each organism respectively . 
In (c), we show the percentage of the best prediction score (MCC)   for each organism respectively. 

StemP finds the shape of the tRNA folding well with a short computation time, in general in less than 0.3 seconds.  Figure \ref{fig_tRNA_stat} shows that the majority of the sequences, 20,463 among 27,010 (75.8\%), reach the maximum MCC as the top ranking (SCR=1) using StemP, i.e., maximum base-pair matching.  
The second graph shows that for SCR=1, the majority of the sequences (74.8\%),  20,202 among 27,010, give MCC above 0.9. 
The majority of the sequences (89.1\%), 24,053 among 27,010, reaches MCC value in $[0.90,1]$ in the best prediction, and 24,398 (90.3\%) gave the maximum MCC within top 5 SCR. 

\begin{figure}
\centering
\subfloat[MCC 0.95]{\includegraphics[width = 1.3 in]{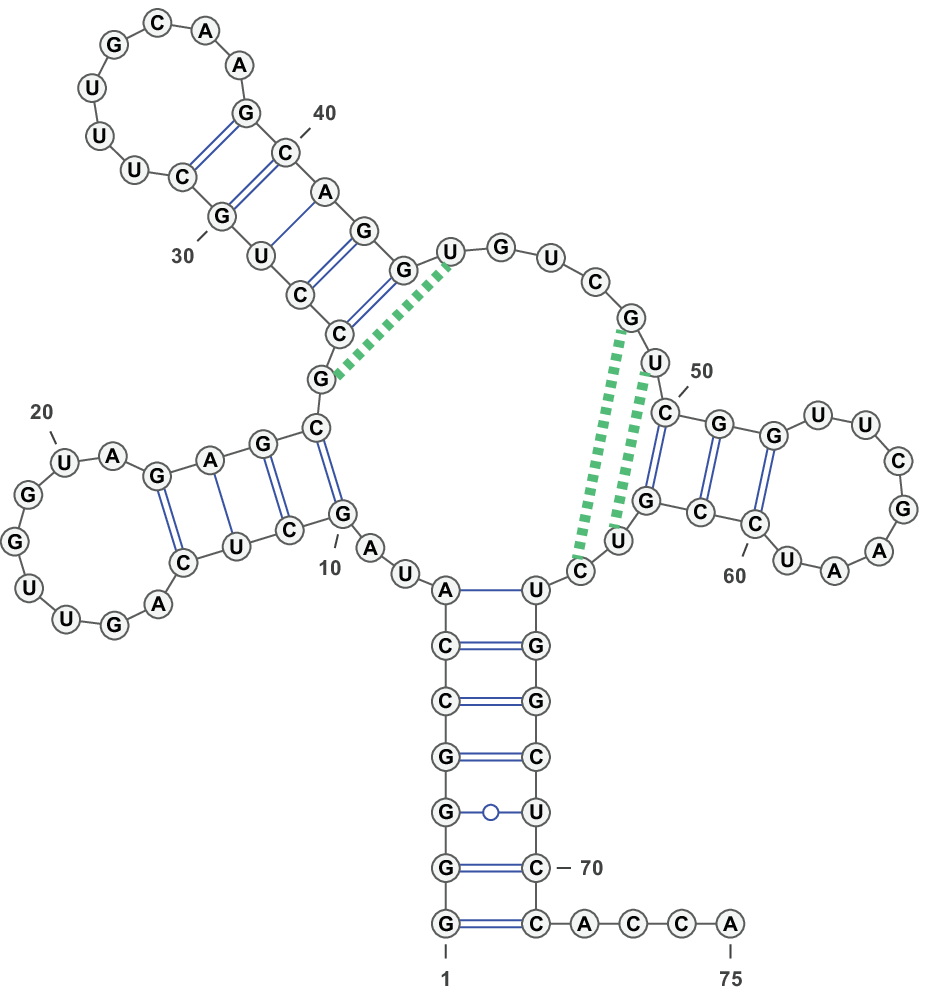}} \hfil
\subfloat[MCC 0.90]{\includegraphics[width = 1.2 in]{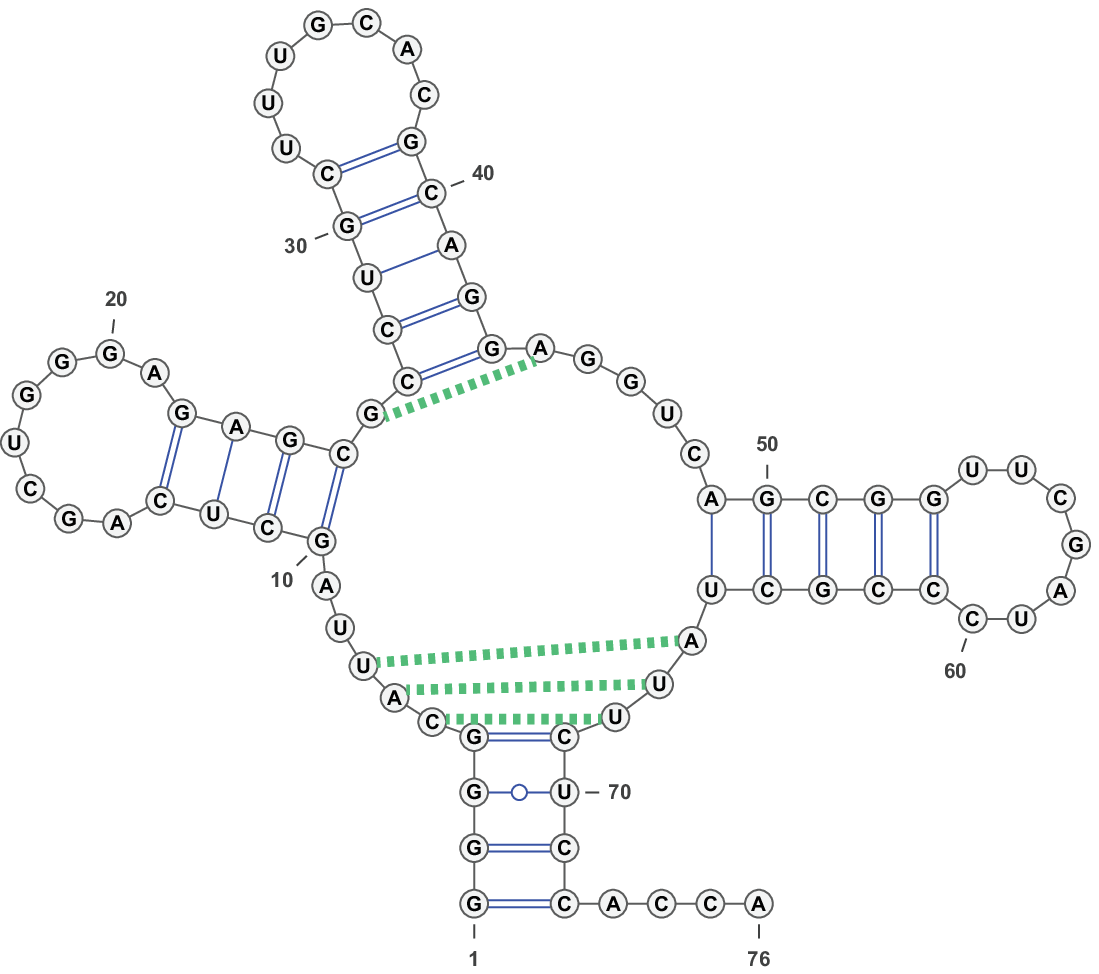}} \hfil
\subfloat[MCC 0.85]{\includegraphics[width = 1.3 in]{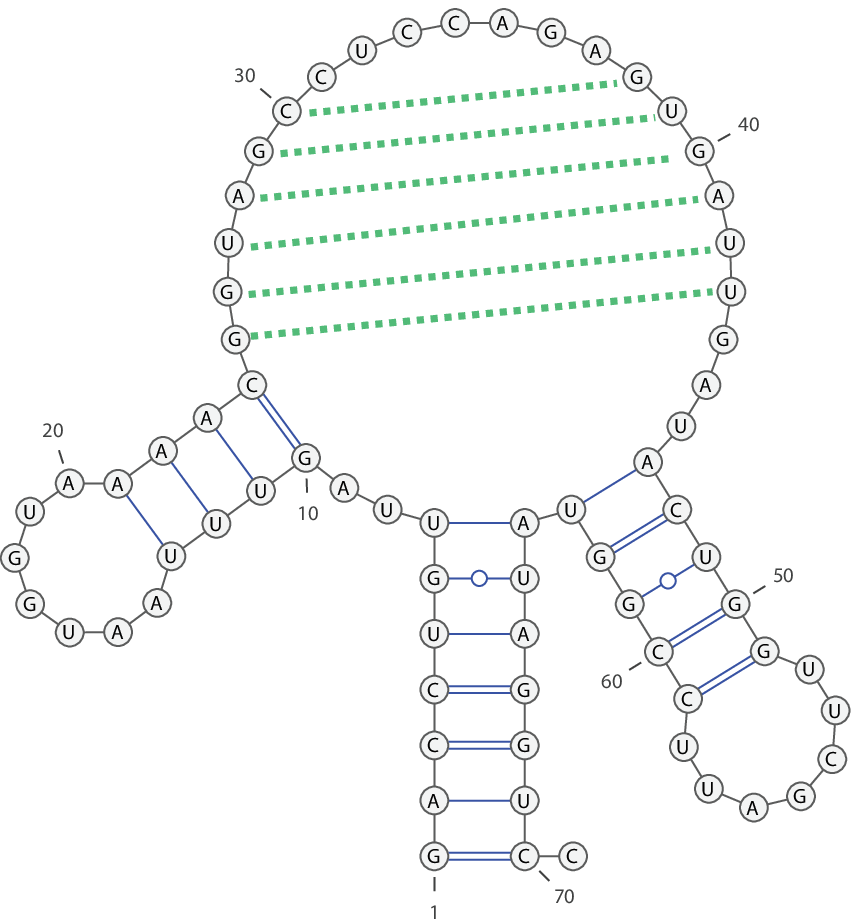}}
\caption{Examples of StemP results with MCC above 0.85 tRNA.  (a) StemP (rank 1) and true folding of (L00194) superposed.  Green dotted line shows the missing pairs (False Negatives), while all other base-pairs matching are correct.  
(b) StemP, SCR=1(9), and true folding of (AF146727.1) superposed.  
(c) AY653733. StemP, SCR = 7(3), with true folding.   }
\label{F:tRNAmis}
\end{figure}

When MCC is around 0.95, the prediction may miss base-pair matching, typically non-canonical pair like \texttt{U-U}, \texttt{G-A} or \texttt{C-U}.   Figure \ref{F:tRNAmis} shows example of typical folding with MCC 0.95, MCC 0.90 and MCC 0.85.   (a) with Accession Number  L00194  and  MCC 0.95,  (b) with Accession Number AF146727.1 and MCC 0.90. (c) with Accession Number AY653733 and MCC 0.85.
In (a), the mismatch (False Positive) is  non-canonical pair \texttt{U(49)-U(63)}. Similarly, in (b), the mismatches are non-canonical pair \texttt{G(26)-A(44)} and \texttt{C(5)-U(68)}. In (c), the missing stem starting with \texttt{G(25)-U(43)} and ending with \texttt{C(30)-G(38)} is broken into two sub-stems of length 3 and 2 due to the non-canonical pair \texttt{A(28)-G(40)}, which fall out of the Stem-Loop score range and the stem length range. It is shown that these miss-match does not affect the general shape of the folding structure.

\subsection{Comparison results for tRNA} 
\begin{figure}
\centering
\begin{tabular}{ccc}
\includegraphics[width = 1.3 in]{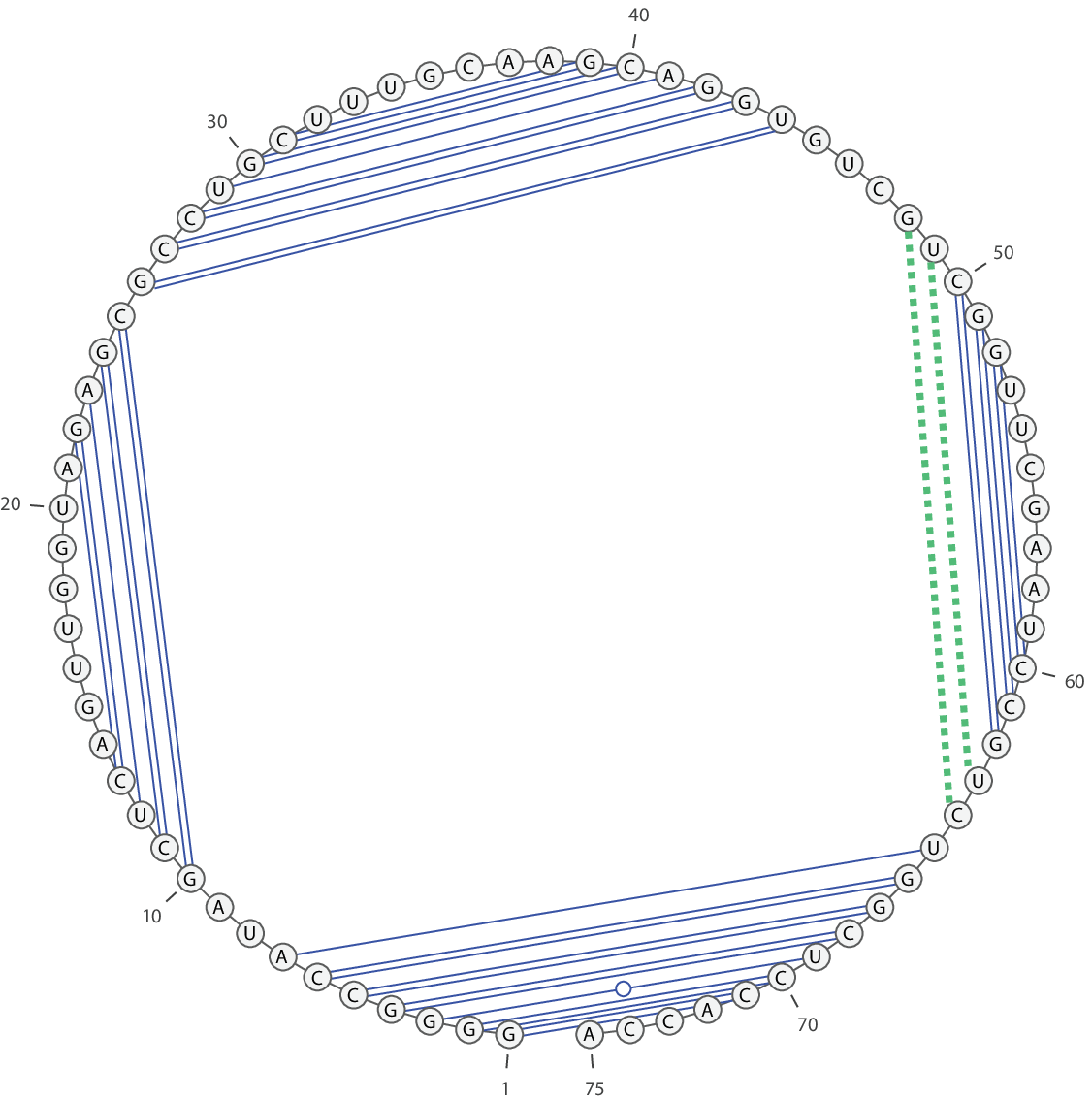} &
\includegraphics[width = 1.3 in]{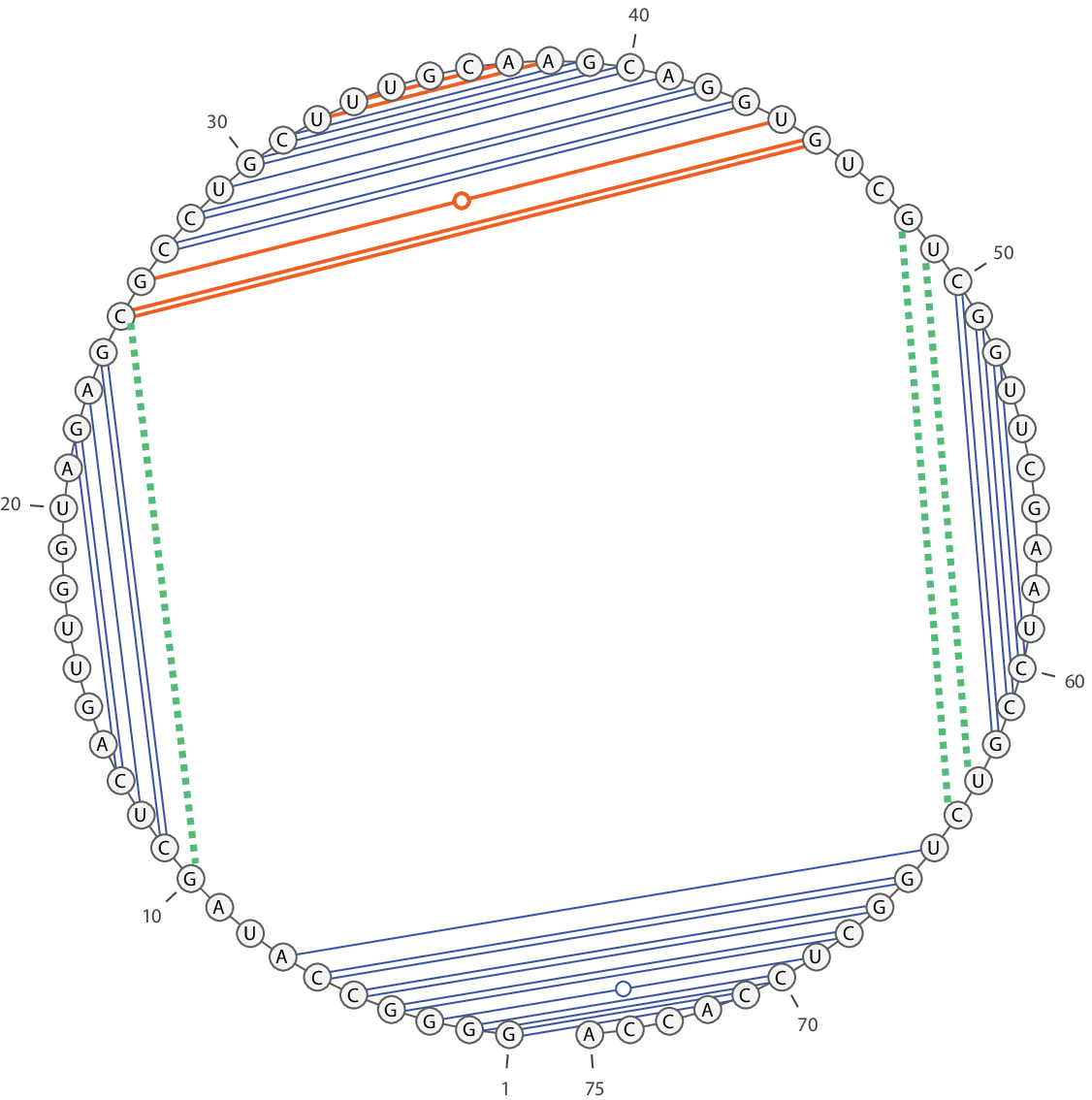}  &
\includegraphics[width = 1.3 in]{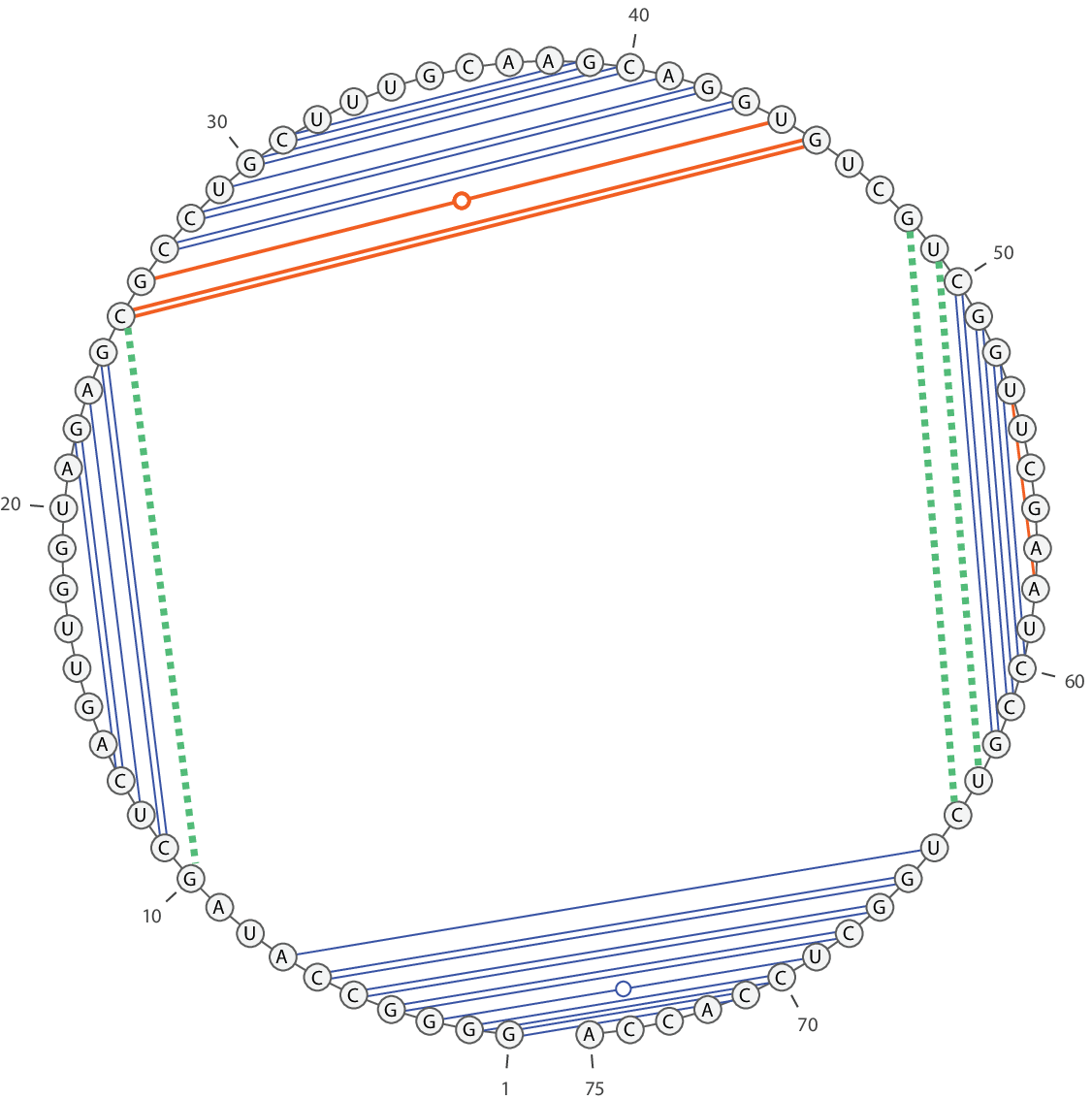} \\
(a) StemP& (b) MaxExpect & (c) ProbKont \\
\includegraphics[width = 1.3 in]{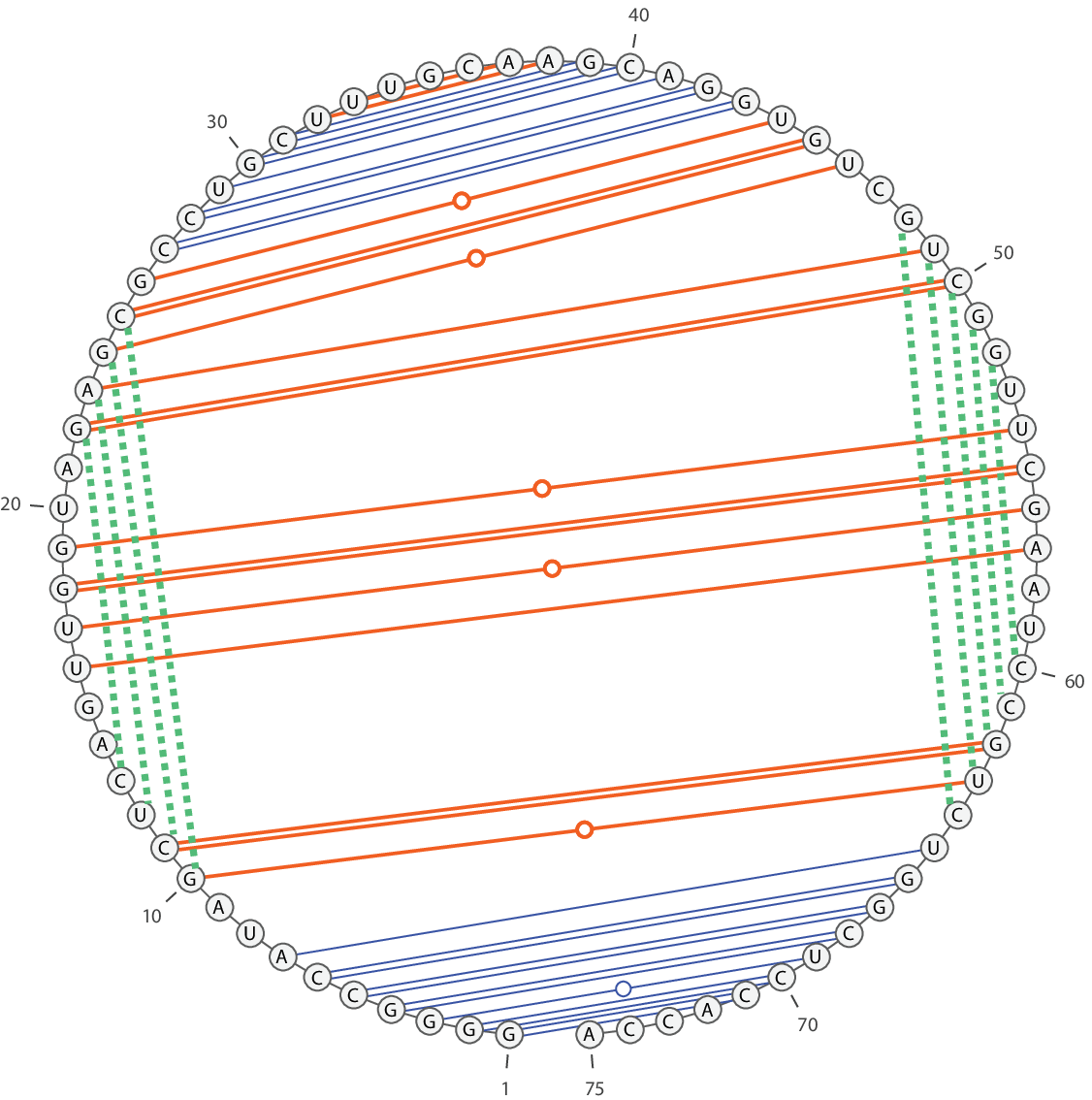}  &
\includegraphics[width = 1.3 in]{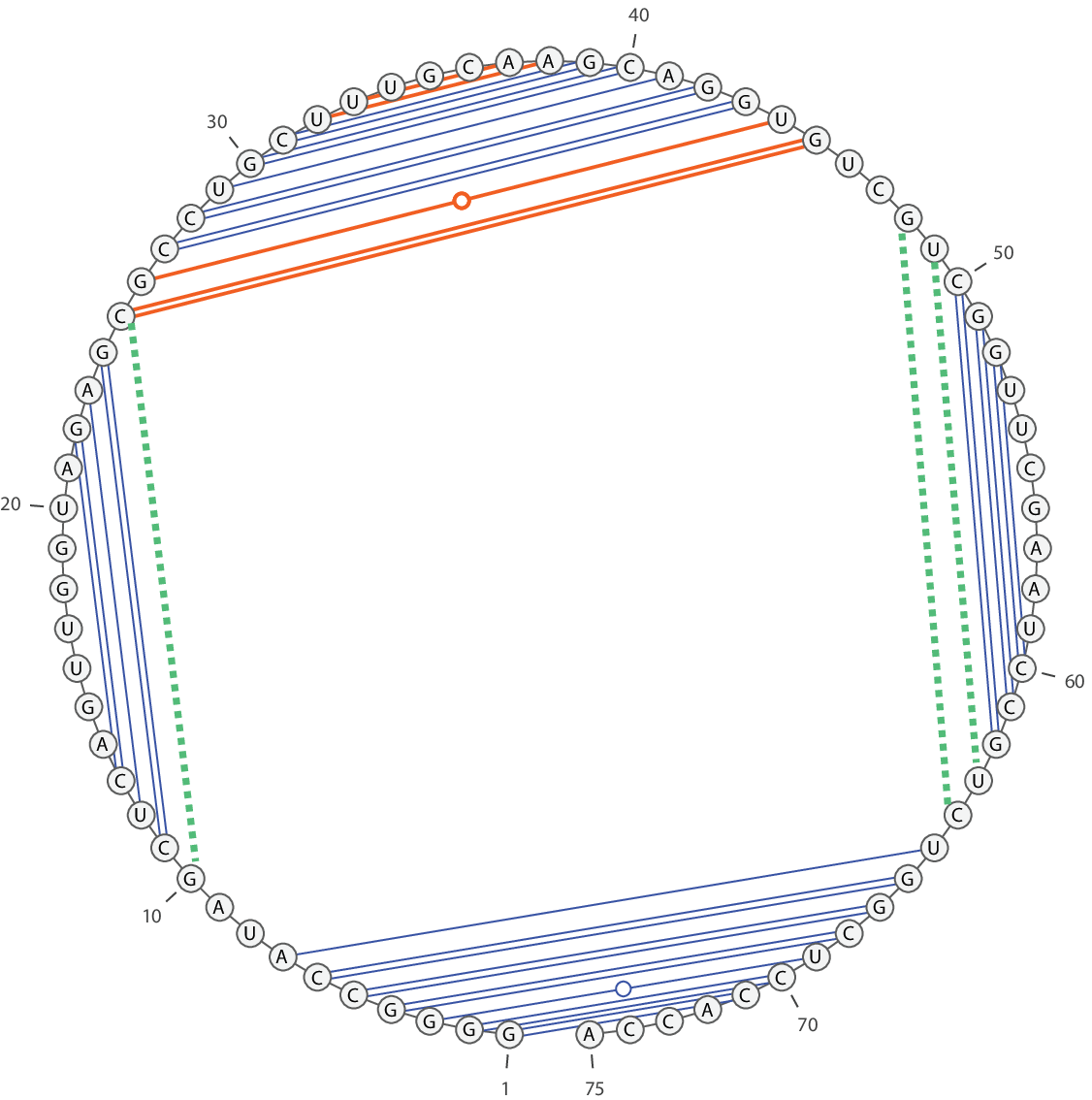}  &
\includegraphics[width = 1.3 in]{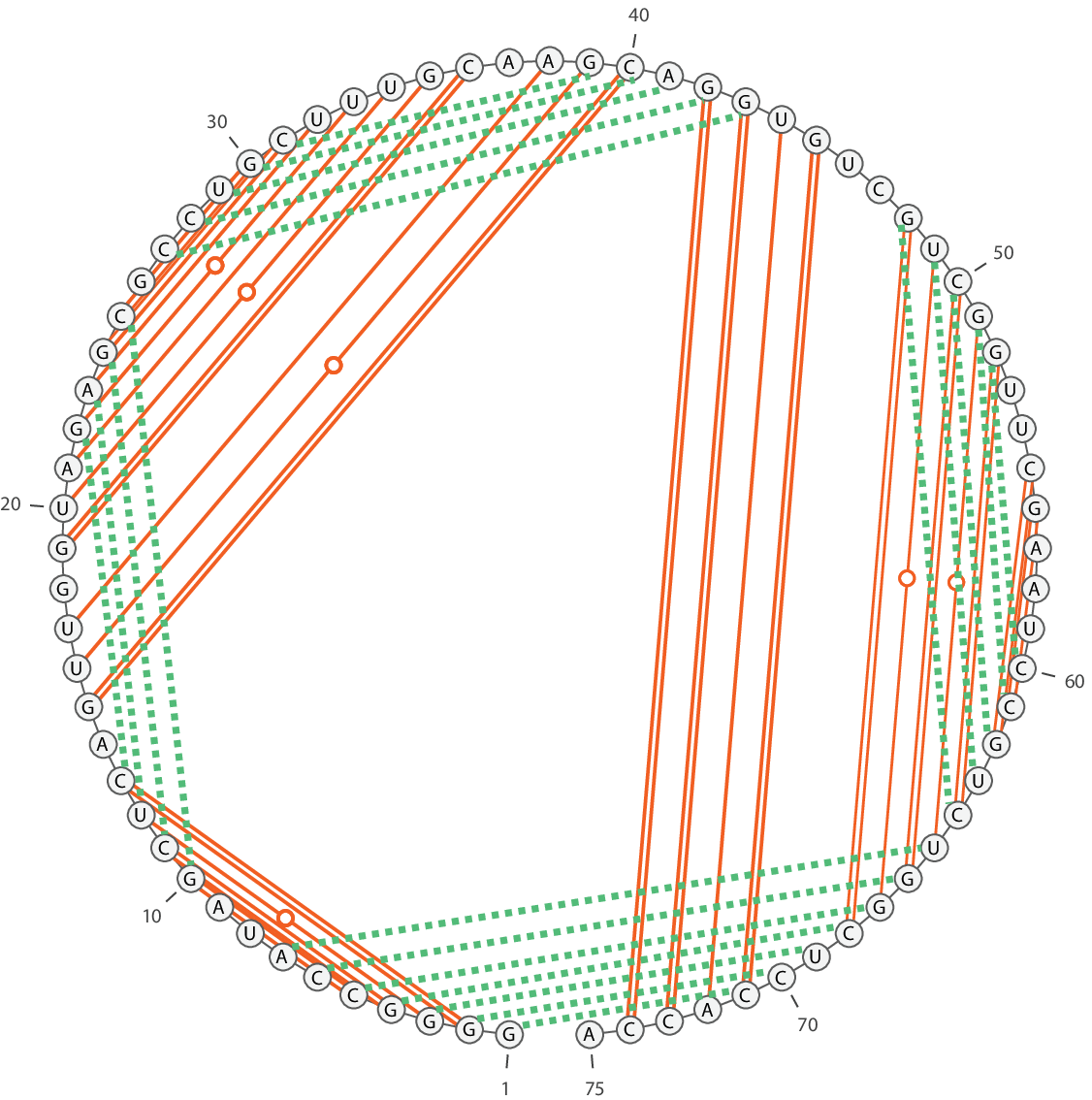}  \\
(d) FOLD1& (e) FOLD2& (f) FOLD3
\end{tabular}
\caption{ tRNA (Accession Number L00194) prediction comparison.  False-Positive (Green), and False-Negative (Red).  (a) StempP result of SCR=1(1) (MCC 0.95) (b) MaxExpect (MCC 0.86), (c) ProbKont (MCC 0.84), and (d)-(f) 3 possible FOLD predictions (MCC 0 - 0.86).  }
\label{F:tRNAcompare3}
\end{figure}

Figure \ref{F:tRNAcompare3} shows comparison results using StemP, FOLD, MaxExpect, and Probknot for a tRNA sequence with Accession Number L00194.  

In Figure \ref{F:tRNAcompare3}, (a) StemP has only 2 base pairs missing (False Negative) and no extra matching (False Positive).  (b) MaxExpect and (c) ProbKont have only acceptor stem completely correctly identified, and in (c), there are 3 base pairs missing (False Negative) and 4 extra pairs (False Positive).   (d)-(f) are multiple structures provided by FOLD, there are some base pairs missing (False Negative), and extra pairs (False Positive). 

In Table \ref{T:comparison tRNA 50 seqs}, we present comparisons on 47 different tRNA sequences between StemP and \cite{poznanovic2020challenge}.  We show the highest $F_1$ among the predictions with SCR=1 by Top, the highest $F_1$ among all possible clique structure predictions by Best.
StemP has higher top $F_1$ and best $F_1$ for 46 sequences out of 47.  There are 40 sequences reach the highest $F_1$-score when  SCR = 1.  StemP has 0.95 as an average for best prediction and 0.92 for top prediction on this test. 
For StemP, Table \ref{T:trna parameters} parameters are used with $3 \leq SL \leq 5.4$ uniformly for all testing sequences. 

\begin{table}
\small
\centering
\begin{tabular}{rl|l|llr|rl|l|llr}
\hline
\multicolumn{1}{l}{} &
  \multicolumn{1}{c}{\multirow{2}{*}{Accn}} &
  \multicolumn{1}{c}{\multirow{2}{*}{\cite{poznanovic2020challenge}}} &
  \multicolumn{3}{|c|}{StemP} &
  \multicolumn{1}{l}{} &
  \multicolumn{1}{c}{\multirow{2}{*}{Accn}} &
  \multicolumn{1}{c}{\multirow{2}{*}{\cite{poznanovic2020challenge}}} &
  \multicolumn{3}{|c}{StemP} \\
  \cline{4-6} \cline{10-12}
 &
  \multicolumn{1}{c}{} &
  \multicolumn{1}{c|}{} &
  Top &
  Best &
  SCR/DR &
   &
  \multicolumn{1}{c}{} &
  \multicolumn{1}{c|}{} &
  Top &
  Best &
  SCR/DR \\
  \hline
1  & AY934184 & 0.63 & 0.98 & 0.98 & 1/1 & 25 & AE014184         & 0.73 & 0.98          & 0.98          & 1/1  \\
2  & AE013169 & 0.59 & 0.98 & 0.98 & 1/1 & 26 & CP000473         & 0.62 & 0.98          & 0.98          & 1/1  \\
3  & AY934018 & 0.60 & 0.73 & 0.75 & 6/2 & 27 & CP000492         & 0.58 & 0.77          & 0.98          & 2/2  \\
4  & AJ010592 & 0.47 & 0.98 & 0.98 & 1/1 & 28 & CR382129         & 0.55 & 0.68          & 0.73          & 79/4 \\
5  & U41549   & 0.58 & 0.93 & 0.93 & 1/1 & 29 & X03154           & 0.56 & 0.98          & 0.98          & 1/1  \\
6  & AE000520 & 0.70 & 0.57 & 0.78 & 3/2 & 30 & CP000254         & 0.55 & 0.85          & 0.85          & 1/1  \\
7  & CP000143 & 0.68 & 0.98 & 0.98 & 1/1 & 31 & CP000493         & 0.73 & 0.93          & 0.93          & 1/1  \\
8  & AY934254 & 0.67 & 0.98 & 0.98 & 1/1 & 32 & BA000021         & 0.50 & 0.98          & 0.98          & 1/1  \\
9  & AE009952 & 0.54 & 0.93 & 0.93 & 1/1 & 33 & CP000412         & 0.51 & 0.79          & 0.95          & 2/2  \\
10 & AE017159 & 0.48 & 1.00 & 1.00 & 1/1 & 34 & AP006618         & 0.56 & 0.98          & 0.98          & 1/1  \\
11 & AY934351 & 0.60 & 0.98 & 0.98 & 1/1 & 35 & CP000099         & 0.56 & 0.93          & 0.93          & 1/1  \\
12 & BA000023 & 0.70 & 0.95 & 0.95 & 1/1 & 36 & CP000141         & 0.68 & 0.98          & 0.98          & 1/1  \\
13 & AY934387 & 0.55 & 0.98 & 0.98 & 1/1 & 37 & AC006340         & 0.62 & 0.93          & 0.93          & 1/1  \\
14 & AY933864 & 0.53 & 0.98 & 0.98 & 1/1 & 38 & BX569691         & 0.63 & 0.98          & 0.98          & 1/1  \\
15 & AJ248288 & 0.68 & 0.95 & 0.95 & 1/1 & 39 & AF137379         & 0.51 & 0.98          & 0.98          & 1/1  \\
16 & CP000471 & 0.64 & 0.98 & 0.98 & 1/1 & 40 & DQ396875         & 0.59 & 0.98          & 0.98          & 1/1  \\
17 & BA000011 & 0.64 & 0.98 & 0.98 & 1/1 & 41 & CP000142         & 0.69 & 1.00          & 1.00          & 1/1  \\
18 & BX321863 & 0.61 & 0.98 & 0.98 & 1/1 & 42 & BX640433         & 0.58 & 0.73          & 0.98          & 2/2  \\
19 & CP000423 & 0.55 & 0.55 & 0.98 & 2/2 & 43 & AJ294725         & 0.43 & 0.98          & 0.98          & 1/1  \\
20 & CR936257 & 0.65 & 0.93 & 0.93 & 1/1 & 44 & X04779           & 0.57 & 0.98          & 0.98          & 1/1  \\
21 & AC004932 & 0.53 & 0.93 & 0.93 & 1/1 & 45 & AY934393         & 0.49 & 0.98          & 0.98          & 1/1  \\
22 & AY933788 & 0.58 & 0.98 & 0.98 & 1/1 & 46 & AE008623         & 0.61 & 0.98          & 0.98          & 1/1  \\
23 & X04465   & 0.41 & 0.93 & 0.93 & 1/1 & 47 & AE000657         & 0.73 & 0.98          & 0.98          & 1/1  \\
24 & DQ093144 & 0.62 & 0.98 & 0.98 & 1/1 &    &                  &      &               &               &      \\
\hline
   &          &      &      &      &     &    & \textbf{Average} & 0.59 & \textbf{0.92} & \textbf{0.95} &     \\
   \hline
\end{tabular}

\caption{Comparison of tRNA prediction between StemP and \cite{poznanovic2020challenge}.  The 47 sequences is in  \cite{poznanovic2020challenge} from Gutell Lab \cite{cannone2002comparative}. Accn denotes the Accession number of the corresponding sequences.  Top is the best $F_1$-score among predictions with SCR = 1.  Best is the highest $F_1$-score of among all predictions of clique.}
\label{T:comparison tRNA 50 seqs}
\end{table}

\section{StemP for 5s rRNA folding prediction}
\label{sec: 5S}

5S rRNA plays a critical role in ribosomes of organisms.  The sequences typically contains about 120 nucleotides with a particular structure \cite{szymanski20025s} consisting of five helices, two hairpin loops, two internal loops  and a hinge region forming the three-helix junction. This structure has been recognized by comparative sequence analysis and  is useful in its application to the problems of molecular evolution.  The size and ubiquity of 5S rRNA, that enabled RNA sequencing using direct methods, made it an ideal candidate for a molecular phylogenetic marker\cite{szymanski20025s}. 

Similar to tRNA, different methods such as  Minimum Free Energy\cite{washietl2005fast}, constructing tree structure\cite{fera2004rag} and  sequence analysis\cite{leontis2002motif, parisien2008mc} are available for predicting 5s rRNA. 
TurboFold II \cite{tan2017turbofold} is  an extension of TurboFold \cite{harmanci2011turbofold}, a comparative method that provide multiple sequence alignments by iteratively estimating the probabilities for nucleotide positions between all pairs of input sequences. In \cite{bellaousov2010probknot},  309 sequences were tested by a probability based method ProbKnot where 69.2\% of known pairs were correctly predicted in average. 

5S rRNA has a particular structure, consisting of 5 stems (I-V) and 5 loops (A-E).   We use the notation of Domain $\alpha, \beta, \gamma$ to help illustrate the structure of 5s rRNA. Domain $\alpha$ is identical to Helix I. Domain $\beta$ and Domain $\gamma$ can be understood as a structure starting from a base stem leading up to one hairpin loop, including all internal loop and bulge in between.  Domain $\beta$ has Helix II enclosing Helix IV while Domain $\gamma$ has Helix III enclosing Helix V.   Fig. \ref{F:5s} (b) shows a typical example of full Stem-graph of a 5s rRNA sequence. 

\subsection{Parameters} 
\begin{figure}
\begin{center}
\begin{tabular}{cc}
\includegraphics[width = 2.2in]{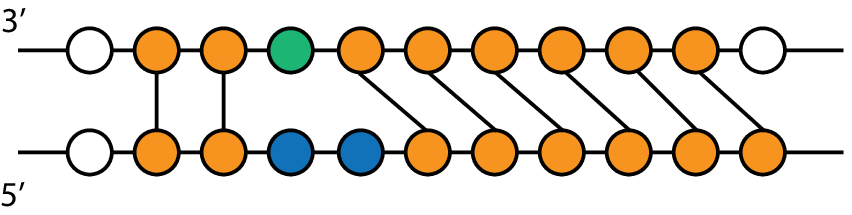} &
\includegraphics[width = 2.5in]{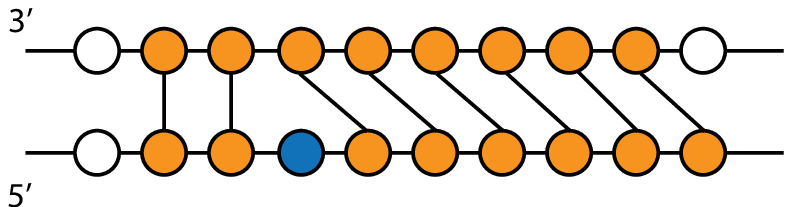} \\
(a) 2{\tiny[1/2]}6  & (b)  2{\tiny[0/1]}6 
\end{tabular} 
\end{center}
\caption{Two examples of 5s rRNA vertex variation.  Both has stem length $L=8$, but with gaps.  We consider such cases as one vertex for 5S rRNA prediction. }
\label{5s2-1}
\end{figure}

In 5S rRNA, often helix doesn't have consecutive basepair matches, but has one or two bases gap. We consider the stem variations to account for such cases, which counts a combination of shorter stems as one vertex.  The structure $l_1${\tiny $[n_1/n_2]$}$l_2$ or $ l_1${\tiny $[n_1/n_2]$}$l_2${\tiny $[n_3/n_4]$}$l_3$ denotes such variation, where  $l_i (i=1,2)$ the length of consecutive base pairs, and $n_i$s denote the gaps between the shorter stems, as illustrated in Fig. \ref{5s2-1}.  Notice that, not considering the gaps, this vertex have stem length to be $l_1 + l_2$. 
We use Stem-Loop score (\ref{E:sindex}) to identify each Helix (I-V) and Generalized-Stem-Loop score (\ref{E:gsl}) to identify the Domain $\beta, \gamma$ which contain more than one Helix.

For the computation, we add a step to use $GSL$ for more efficient computation.  From  the input data, 
\begin{enumerate}
\item{find five different sets of vertices $v_i$ with appropriate $SL$ score (each set of candidates for Helix I-V).}\vspace{-0.3cm}
\item{Use $GSL$ to find two (additional) sets of vertices $V_k$ to find Domain $\beta$ and $\gamma$.}\vspace{-0.3cm}
\item{Construct edges between each domain.}\vspace{-0.3cm}
\item{A maximal clique that represents the highest energy gives the prediction. }
\end{enumerate}

We summarize the parameters used for 5S rRNA folding prediction in Table \ref{T:5sRNA_Archaeal} for Archaelal (refined parameters), and general parameters in Table.  \ref{T:5s_parametersALL}.

\subsection{StemP Result 5s rRNA} 
\begin{figure}
\centering
\begin{tabular}{cc}
\includegraphics[width = 2in]{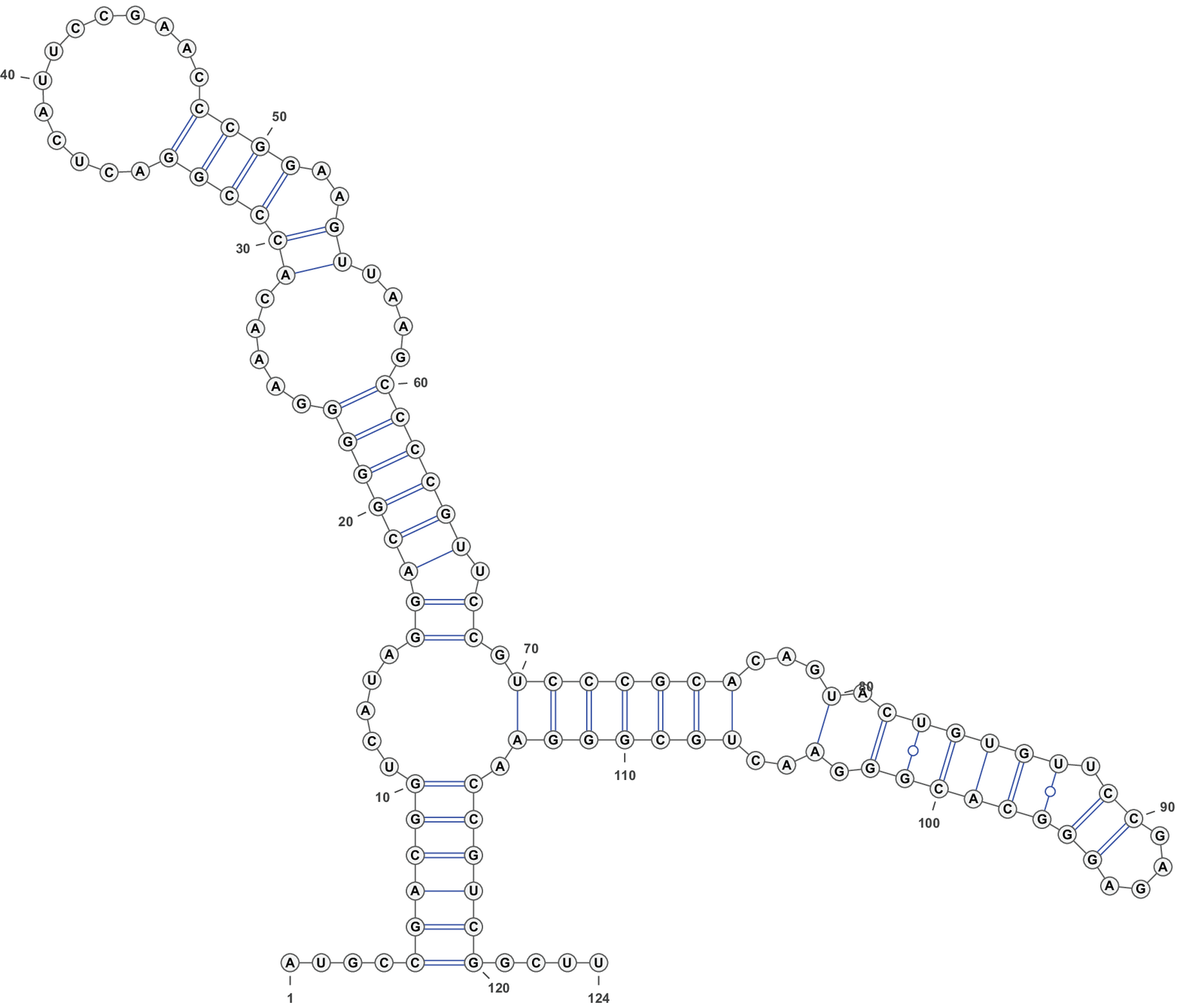} & 
\includegraphics[width = 2in]{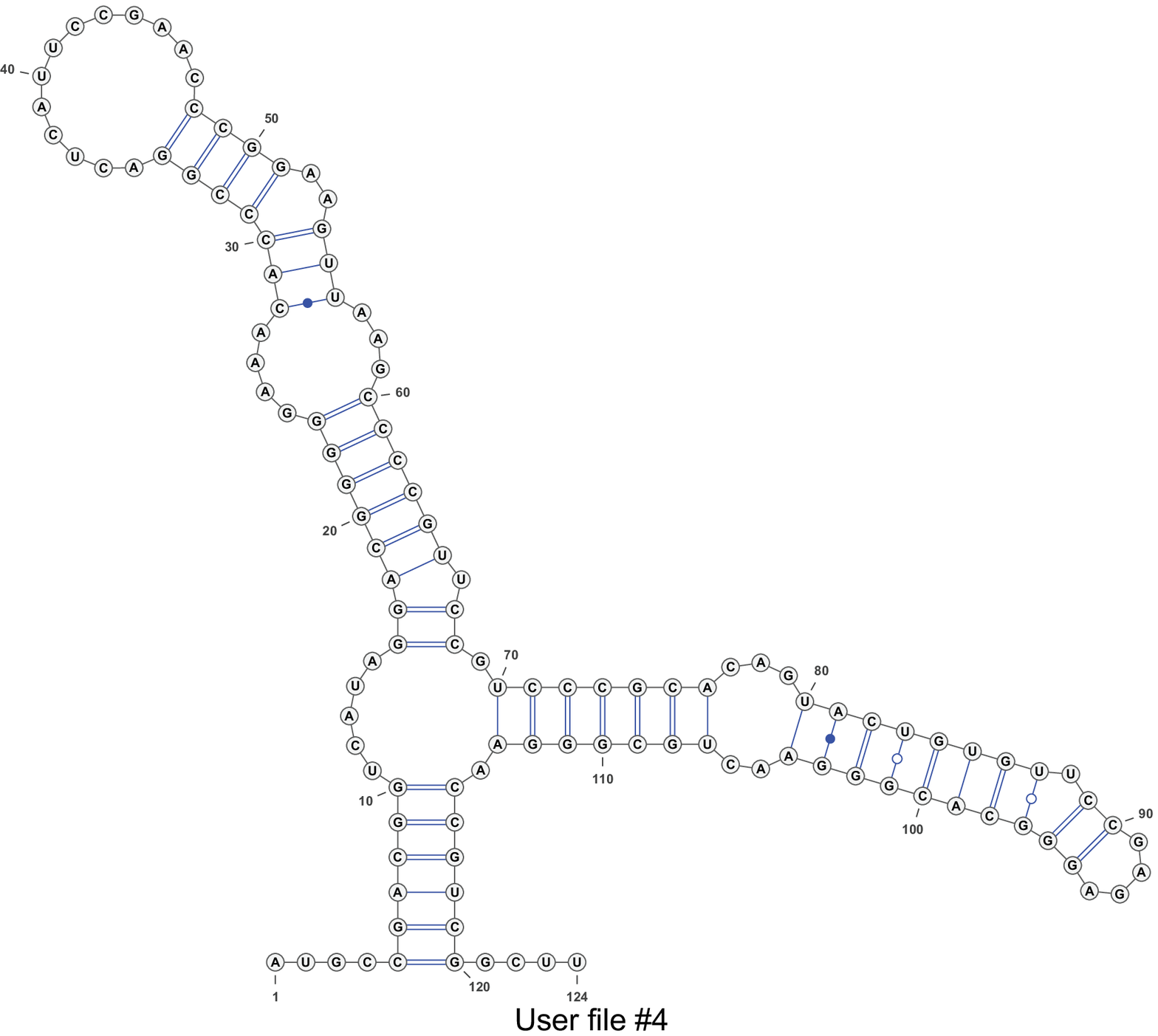}\\
(a) Prediction (MCC 0.97) &(b) True folding \\
\end{tabular}
\caption{ StemP for 5S rRNA (Accession number AE000782).   (a) StemP which has MCC 0.97 (not considering non-canonical pairs, MCC=1). (b) True folding. There are 154 vertices, and 8986 cliques. The best predicted is SCR=1(16). The average MCC for these 16 structures is 0.89 (not considering non-canonical pairs).} 
\label{F:5s} 
\end{figure}

Figure \ref{F:5s}  shows a typical result of StemP for a 5s rRNA sequence (Accession number AE000782).  StemP result in (a) gives MCC 0.97 accuracy with only 2 non-canonical base pairs \texttt{C(28)-U(56)} and \texttt{A(81)-G(103)} matching missing.  Not considering non-canonical pairs as positive pairs, StemP gives MCC 1 for this sequence.  

In Table \ref{T:5sRNA_results} and Table \ref{T: Archaeal sequences 53},  StemP is tested on 53 sequences in organism Archaeal based on the refined parameters in Table \ref{T:5sRNA_Archaeal}.  
For some sequences such as AE010349, X07545, the cpu time is longer than 1 minute because of the number of allowed possible structures is very large.  The average cpu time for each sequence is 15.5 seconds.
Table \ref{T:5sRNA_results} (a) SCR shows that 79.2\% of the sequences have rank 1, and 90.6\% of the sequences have rank 5 or higher. For 5S rRNA, maximum matching and maximal clique seems to be a good choice for the prediction.  Table \ref{T:5sRNA_results} (b) shows that 90.6\%  sequences have top MCC to be more than 0.92.  (c) shows that 66\% of the sequences have MCC higher than 0.95 and 100\% of the tested sequences has MCC higher than 0.92.  (d) shows the top and the best MCC values for each  Archaeal 5s rRNA sequences. 

\begin{table}
\begin{center}
\begin{tabular}{c|c|c}
\hline
Archaeal & Stem Length Variation &Stem Loop \\
\hline 
Helix I (Domain $\alpha$)  &  6, 5,  4{\tiny[1/0]}1, 4 &  $ 17.82 \leq SL \leq 27.5$  \\
 Helix II  & 8, 2{\tiny[0/1]}6, 2{\tiny[0/1]}5, 2{\tiny[1/2]}5,  1{\tiny[1/2]}6, 2{\tiny[0/0]}1{\tiny[2/1]}4 &  $ 6.37\leq SL \leq 7.72$  \\
 Helix III & 3{\tiny[0/2]}4, 2{\tiny[0/2]}4, 2{\tiny[2/4]}2 & $ 5.66 \leq SL \leq 10.76$  \\
 Helix IV & 7, 6, 5, 3{\tiny[1/1]}2, 3{\tiny[2/2]}1,2{\tiny[1/1]}2  &$ 3.9 \leq SL \leq 6.6$  \\
 Helix V  & 8, 1{\tiny[1/1]}5{\tiny[1/1]}2, 1{\tiny[1/1]}6{\tiny[1/0]}2,  8{\tiny[2/0]}2 \\
&2{\tiny[1/0]}2, 1{\tiny[1/1]}5, 1{\tiny[1/1]}8, 1{\tiny[1/1]}7 & \\ \hline
Domain $\beta$     & \multicolumn{2}{c}{ $ 3.46\leq GSL \leq 4.26$}\\
Domain  $\gamma$ & \multicolumn{2}{c}{ $ 2.52\leq GSL \leq 3.43$}\\
\hline
\end{tabular}
\end{center}
\caption{StemP parameters for 5S rRNA Archaeal.  Canonical and Wobble basepair matching is considered, and Partial Stems are included.  Helix I is Domain $\alpha$.}  \label{T:5sRNA_Archaeal}
\end{table}

\begin{table}
\centering
    \begin{tabular}{cccccc}
        \multicolumn{6}{c}{(a) SCR ranking of best MCC}\\
    \hline
 Organism  & \# & {$=$1}  & {$\leq$3}  & {$\leq $5} & {$>$5} \\
 Archaeal & 53  & 42 (79.2)  & +2 (83.0)  & +4 (90.6)  & 5 (9.4) \\
\hline
\\
    \multicolumn{6}{c}{(b) Top MCC values of StemP}\\
	\hline
 Organism  &  \# & {$\geq$0.97}  & {$\geq$0.95}  & {$\geq $0.92}  & { $<$0.92}  \\
 Archaeal  & 53  & 16 (30.2)   & +9(50.9)   & + 21 (90.6)  & 5 (9.4)  \\
\hline
\\
    \multicolumn{6}{c}{(c) Best MCC values of StemP}\\
    \hline
 Organism  &  \# & {$\geq$0.97}  & {$\geq$0.95}  & {$\geq $0.92}  & { $<$0.92}  \\

 Archaeal  & 53  & 16 (30.2)   & +19(66.0)   & + 18 (100.0)  & 0 (0.0)  \\
\hline
\end{tabular}               
\caption{StemP for 53 different 5S rRNA sequences.  
(a) The number and the percentage in parenthesis for SCR $\leq 1,3, 5$ or $>5$.  The true folding is mostly within top 5 ranking.  
(b) The number of the StemP results (with SCR = 1) of MCC $\geq 0.97, 0.95, or 0.92$. 
(c) The number of the best StemP results of MCC $\geq 0.97, 0.95, 0.92$ and $< 0.92$.  }  
\label{T:5sRNA_results}
\end{table}

\begin{table}
\small
\begin{tabular}{llccrr|llccrr}
\hline
   & \textbf{Accn} & Top  & Best & SCR/DR & CPU    &    & \textbf{Accn} & Top  & Best & SCR/DR & CPU   \\ \hline
1  & AE000782      & 0.97 & 0.97 & 1/1    & 14.66  & 28 & X62859        & 0.78 & 0.95 & 97/3   & 34.87 \\
2  & AE006649      & 0.97 & 0.97 & 1/1    & 1.28   & 29 & U67537        & 0.96 & 0.96 & 1/1    & 0.26  \\
3  & AE010349      & 0.89 & 0.96 & 209/2  & 237.57 & 30 & U67518        & 0.96 & 0.96 & 1/1    & 0.21  \\
4  & AM180088\_b   & 0.95 & 0.95 & 1/1    & 0.29   & 31 & M34911        & 0.94 & 0.94 & 1/1    & 0.19  \\
5  & AP000006      & 0.95 & 0.96 & 5/2    & 40.49  & 32 & X62860        & 0.96 & 0.96 & 1/1    & 0.39  \\
6  & AP000986      & 0.96 & 0.96 & 1/1    & 0.38   & 33 & X62861        & 0.95 & 0.95 & 1/1    & 0.25  \\
7  & AP006878      & 0.97 & 0.97 & 1/1    & 12.34  & 34 & M34910        & 0.97 & 0.97 & 1/1    & 1.51  \\
8  & Arc.fulgidus  & 0.97 & 0.97 & 1/1    & 16.09  & 35 & X62862        & 0.97 & 0.97 & 1/1    & 2.83  \\
9  & BA000001\_b   & 0.95 & 0.96 & 5/2    & 39.76  & 36 & X62864        & 0.44 & 0.94 & 3/2    & 0.37  \\
10 & BA000002      & 0.83 & 0.97 & 111/4  & 0.65   & 37 & M26976        & 0.97 & 0.97 & 1/1    & 14.56 \\
11 & BA000023\_b   & 0.96 & 0.96 & 1/1    & 0.37   & 38 & X15364        & 0.96 & 0.96 & 1/1    & 0.37  \\
12 & CNSPAX02      & 0.82 & 0.96 & 43/3   & 2.05   & 39 & X72495        & 0.96 & 0.96 & 1/1    & 0.97  \\
13 & CNSPAX03      & 0.94 & 0.95 & 3/2    & 4.60   & 40 & M21086        & 0.99 & 0.99 & 1/1    & 47.40 \\
14 & CP000254\_b   & 0.95 & 0.95 & 1/1    & 0.23   & 41 & X15329        & 0.96 & 0.96 & 1/1    & 3.41  \\
15 & CP000477\_b   & 0.95 & 0.95 & 1/1    & 0.19   & 42 & V01286        & 0.97 & 0.97 & 1/1    & 3.85  \\
16 & CP000493      & 0.99 & 0.99 & 1/1    & 73.02  & 43 & U05019        & 0.96 & 0.96 & 1/1    & 0.47  \\
17 & CP000575      & 0.97 & 0.97 & 1/1    & 7.52   & 44 & X01588        & 0.97 & 0.97 & 1/1    & 4.61  \\
18 & CR937011      & 0.95 & 0.95 & 1/1    & 0.41   & 45 & Y08257        & 0.97 & 0.97 & 1/1    & 4.55  \\
19 & DQ314493\_b   & 0.95 & 0.95 & 1/1    & 0.19   & 46 & X05870        & 0.97 & 0.97 & 1/1    & 1.99  \\
20 & DQ314494      & 0.95 & 0.95 & 1/1    & 0.19   & 47 & X07692        & 0.95 & 0.96 & 5/2    & 19.63 \\
21 & X07545        & 0.97 & 0.97 & 1/1    & 105.90 & 48 & M12711        & 0.96 & 0.96 & 1/1    & 15.46 \\
22 & E.coli.ref    & 0.93 & 0.96 & 44/3   & 0.22   & 49 & X02709        & 0.95 & 0.95 & 1/1    & 1.83  \\
23 & AF034620      & 0.95 & 0.95 & 1/1    & 0.43   & 50 & M32297        & 0.95 & 0.95 & 1/1    & 1.77  \\
24 & L27343        & 0.97 & 0.97 & 1/1    & 1.67   & 51 & AL445066      & 0.95 & 0.95 & 1/1    & 1.82  \\
25 & L27169        & 0.95 & 0.95 & 1/1    & 0.17   & 52 & NC\_002689    & 0.95 & 0.95 & 1/1    & 27.18 \\
26 & L27236        & 0.96 & 0.96 & 1/1    & 0.33   & 53 & BA000011      & 0.95 & 0.95 & 1/1    & 27.00 \\
27 & X15364        & 0.93 & 0.95 & 4/2    & 43.09  &    &               &      &      &        &       \\ \hline
   &               &      &      &        &        &    & Average       & 0.94 & 0.96 &        & 15.51 \\ \hline
\end{tabular}\caption{Experimental results on 53 Archaeal 5s rRNA sequences. Accn denotes the Accession number of each sequences.  Top represents the highest MCC score among all predictions that has SCR = 1. Best represents the highest MCC score among all predictions. For each sequence, we provide the Standard Competition Ranking (SCR), Dense Ranking(DR) associated to the best prediction, and the cpu time.  StemP has an average of 0.94 MCC for these 53 sequences. In addition, 42 among 53 sequences reaches the highest MCC score when SCR = 1. }
\label{T: Archaeal sequences 53}
\end{table}

\subsection{Comparison on 5s rRNA}

\begin{figure}
\centering
\begin{tabular}{ccc}
\includegraphics[width = 0.23 \textwidth]{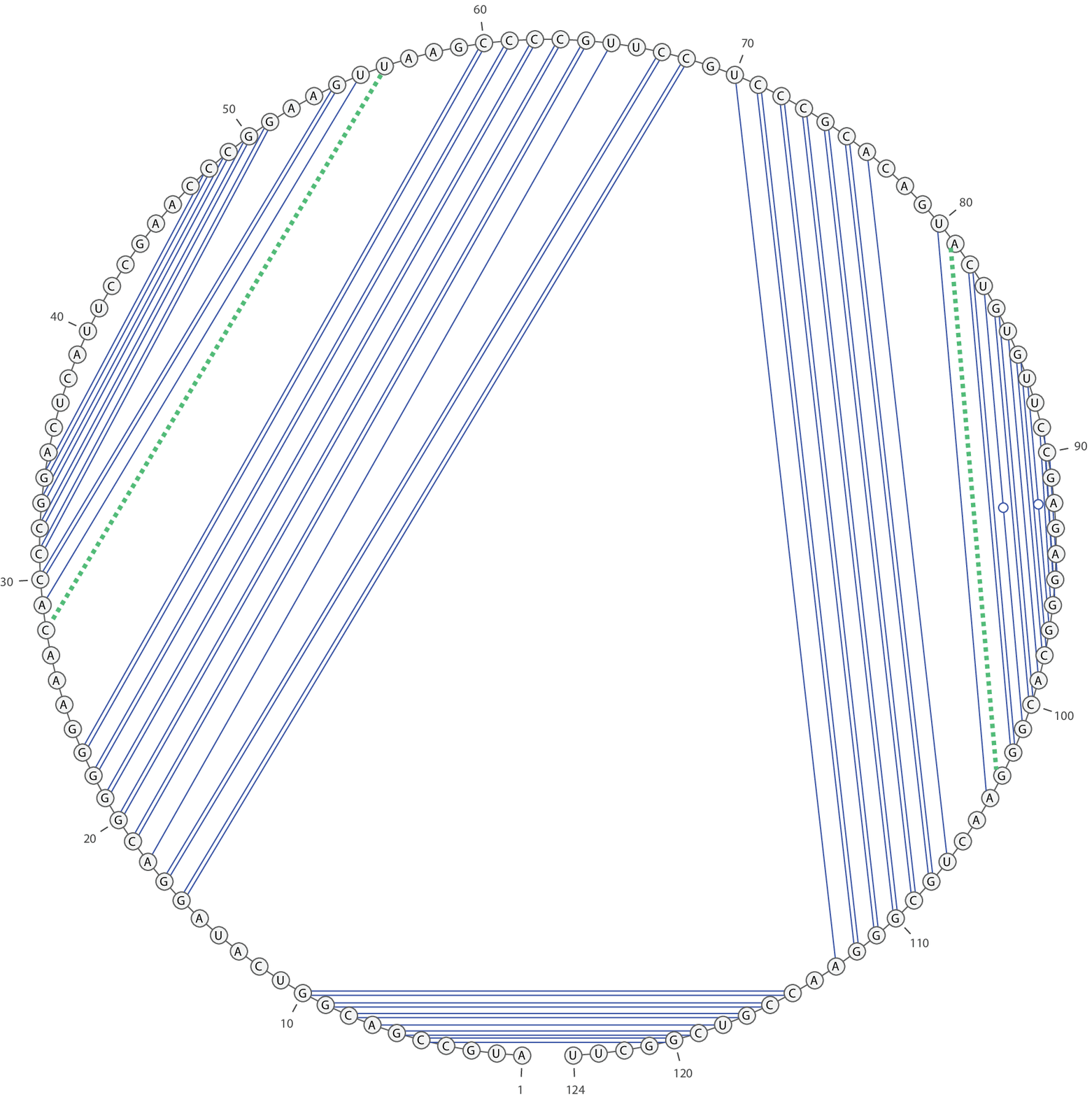} &
\includegraphics[width = 0.23 \textwidth]{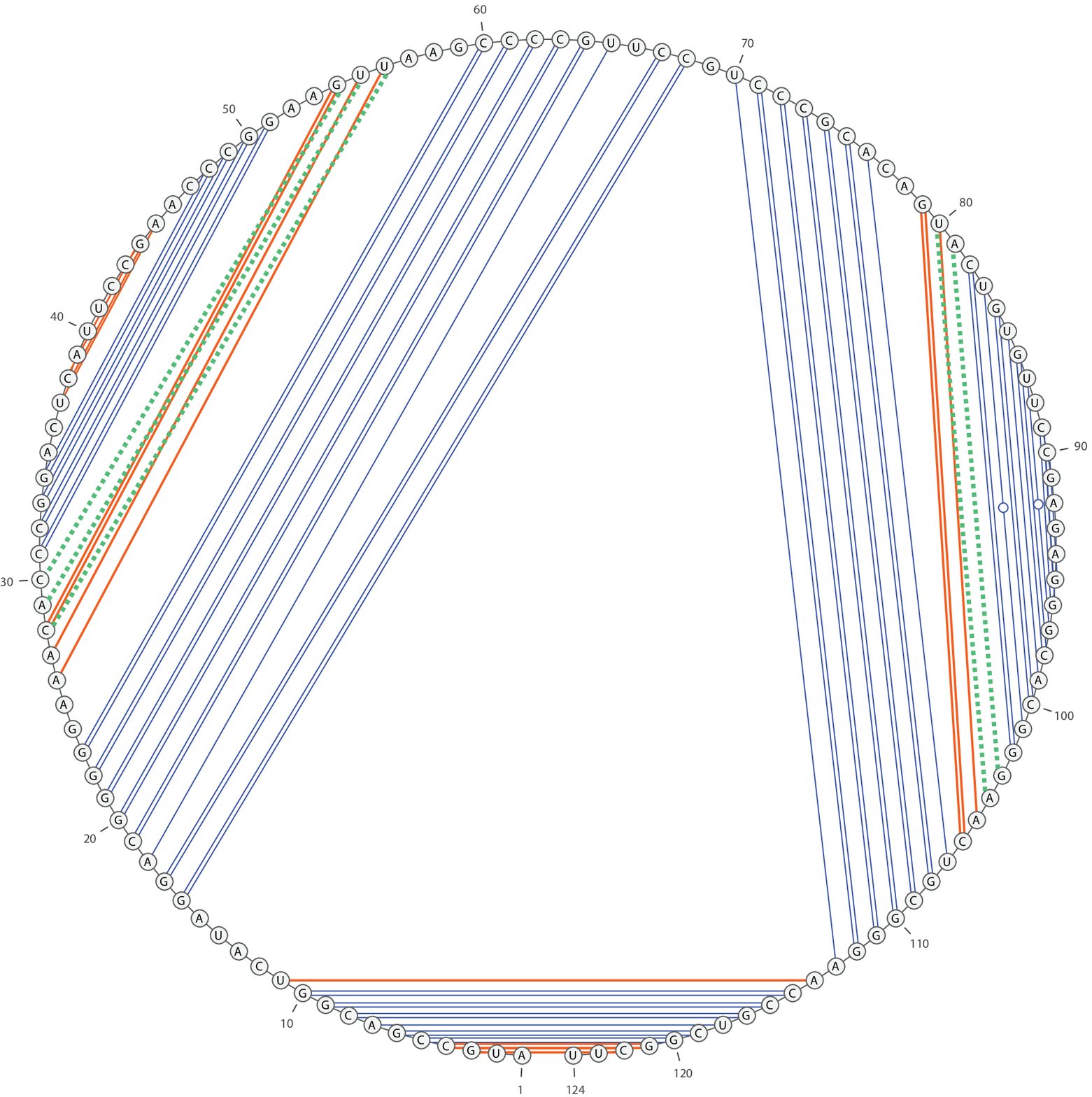} & 
\includegraphics[width = 0.23 \textwidth]{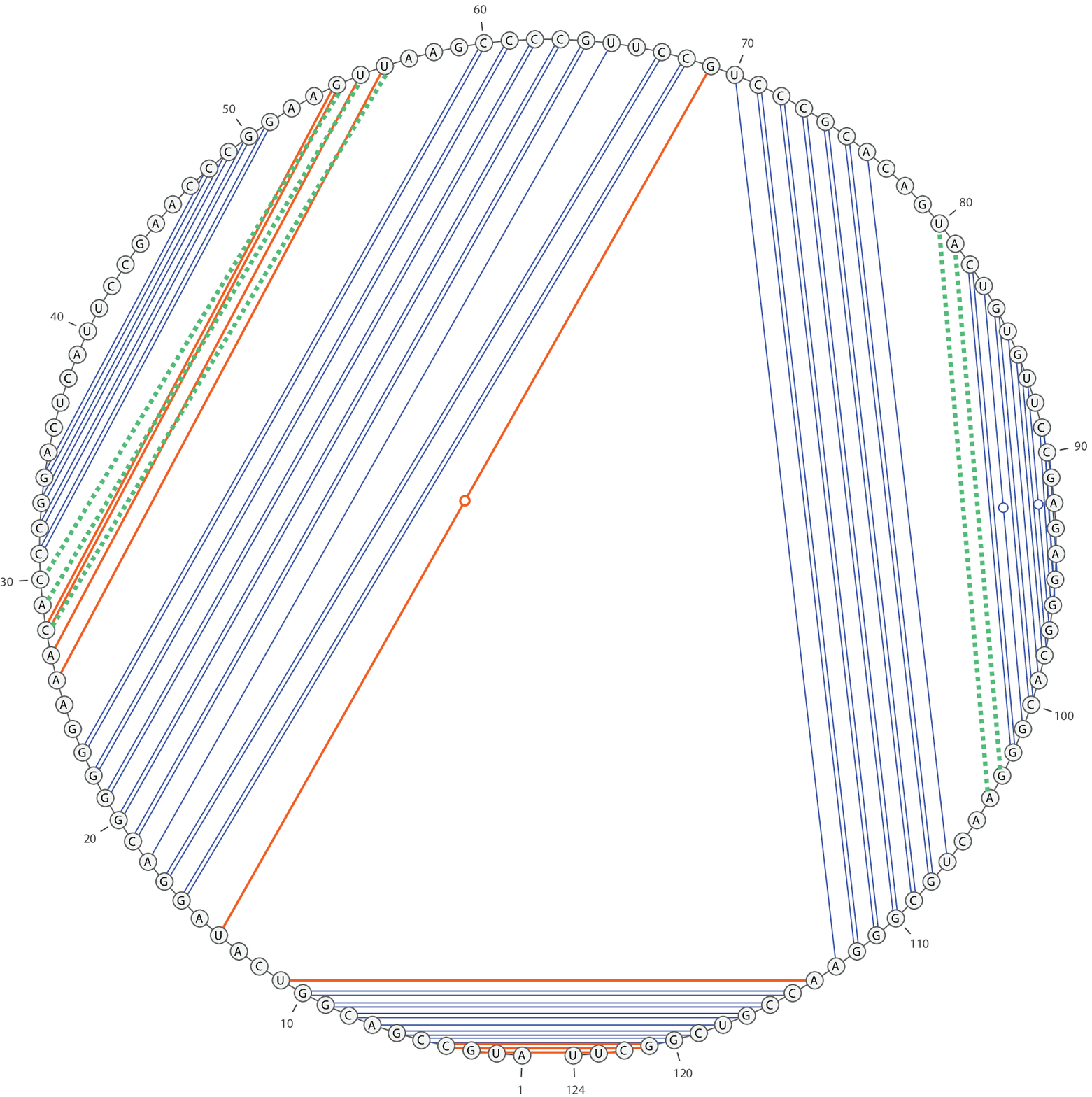} \\
(a) StemP (MCC 0.97) &  (b) MaxExpect (MCC 0.82) & (c) ProbKnot (MCC 0.85) \\
&
\includegraphics[width = 0.23 \textwidth]{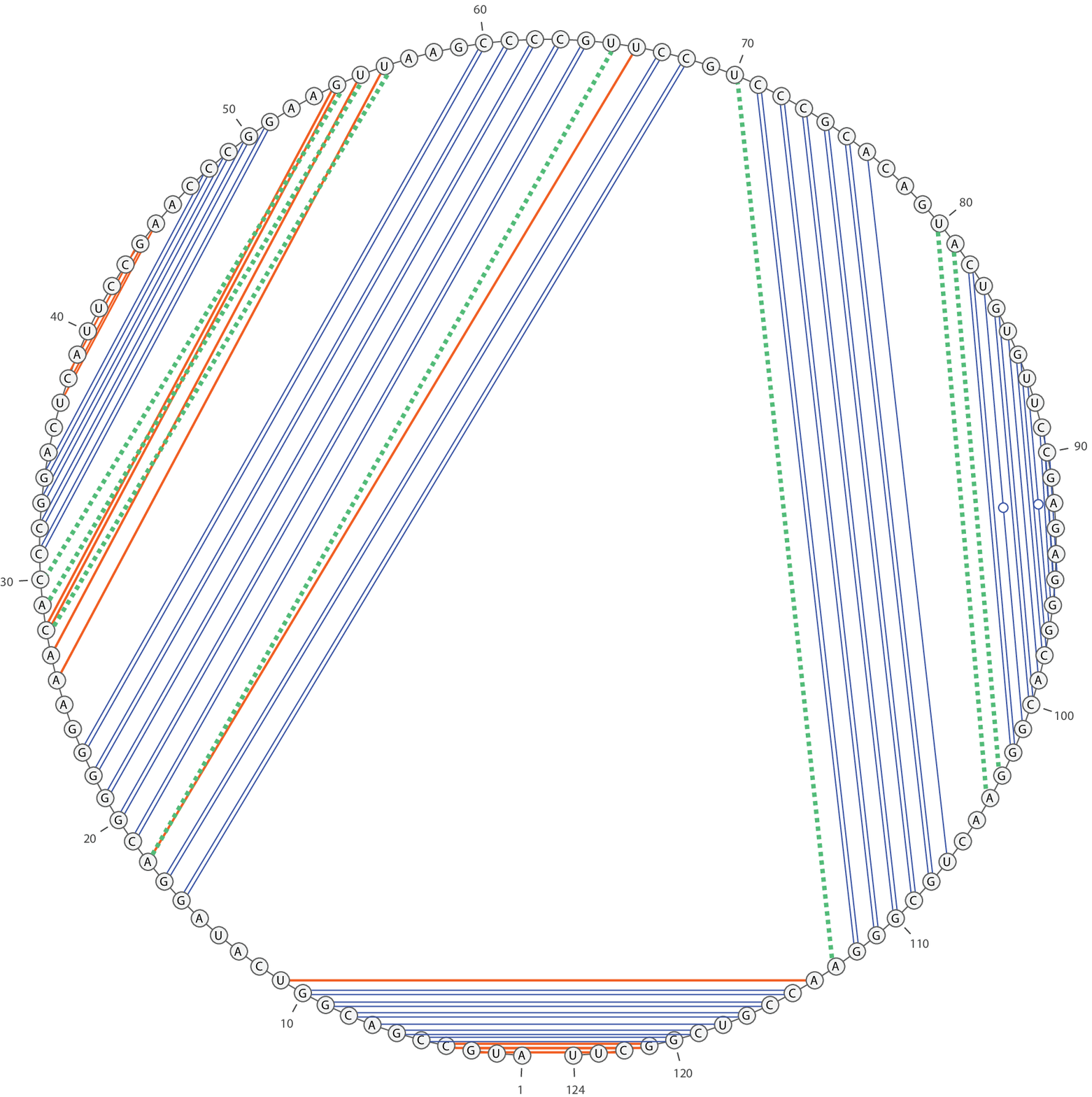}  &
\includegraphics[width = 0.23 \textwidth]{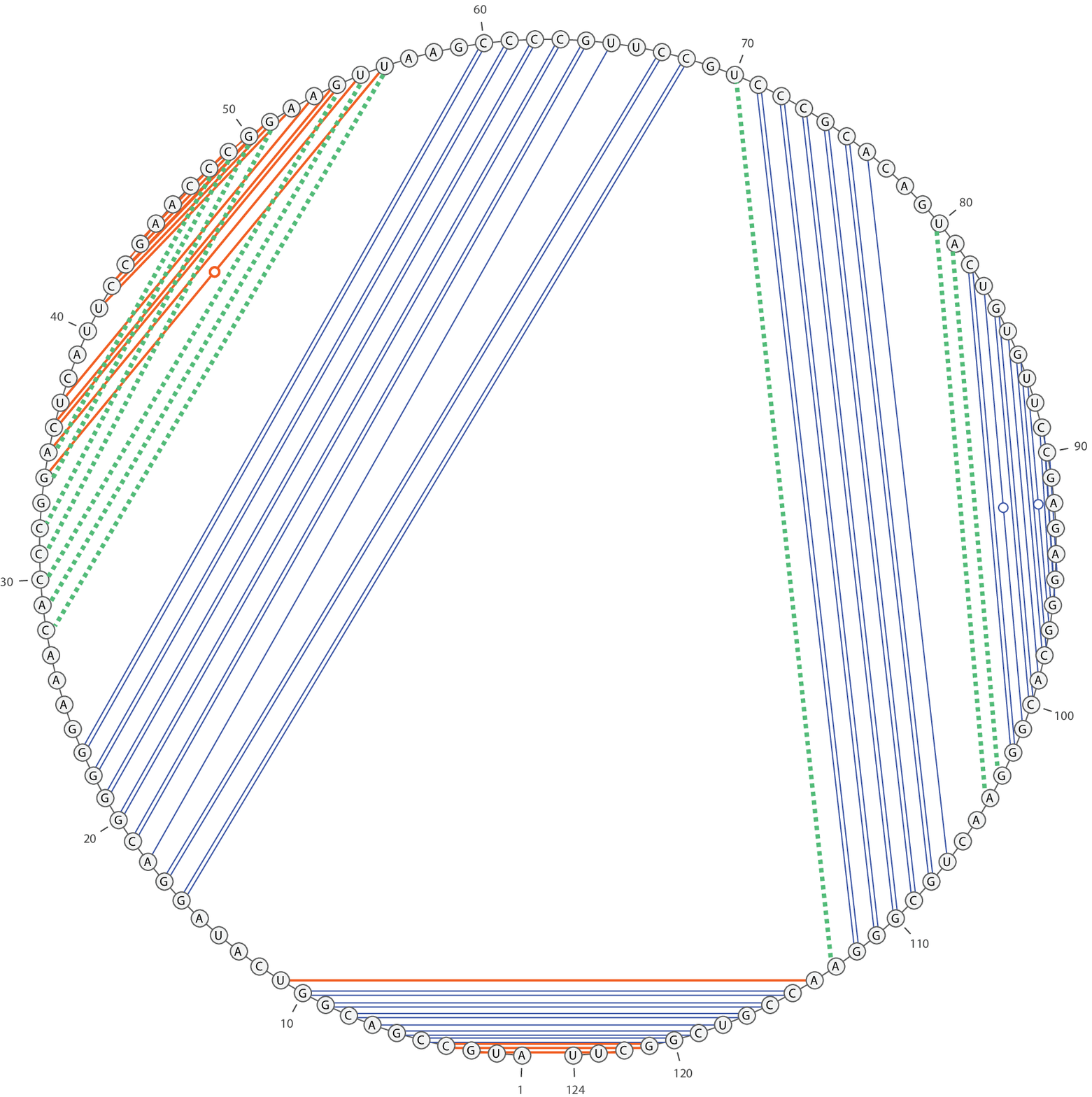} \\
& (d) FOLD1 (0.80) & (e) FOLD2 (0.73) \\
\end{tabular}
\caption{5S rRNA(AE000782) prediction.   (a) StemP (MCC 0.97), (b) MaxExpect  (MCC 0.82), (c) ProbKnot (MCC 0.85), (d)-(e) FOLD (MCC 0.73 - 0.80). False-Positive(Green),  and  False-Negative (Red). }
\label{F:5srRNAcompare1}
\end{figure}

Figure \ref{F:5srRNAcompare1}  shows comparison on sequences AE000782 using StemP,  MaxExpect, ProbKnot and Fold. Using the parameters in Table \ref{T:5sRNA_Archaeal}, StemP correctly found all base pairs except for two non-canonical base pairs \texttt{C(28)-U(56)} and \texttt{A(81)-G(103)}.    
MaxExpect, ProbKnot and FOLD all mismatched base pairs in the first branch. 
For Helix II, (a), (b), and (d) successfully identified the branch while (c) has one missing base pair (False negative) \texttt{U(14)-G(69)} and (d) has a missing  base pair (False Negative) \texttt{A(18)-U(65)} as well as a False Positive base pair \texttt{A(18)-U(66)}. 
Noticing that Helix IV has structure 3{\tiny [0/2]}4 with one non-canonical base pair \texttt{C(28)-U(56)}, (b)-(e) all failed to identify at least 3 base pairs in Helix IV completely. 
For Helix III, (a),(b),(c) successfully recognized this branch while (d),(e) both lost one pair \texttt{U(70)-A(113)} as the first pair of the vertex. 
The special structure of Helix V with one non-canonical base pair \texttt{A(81)-G(103)} in the middle made it difficult for all methods to identify the complete structure. FOLD, MaxExpect, ProbKnot all lost two pairs:  \texttt{U(80)-A(104)} and \texttt{A(81)-G(103)} and MaxExpect found two extra pairs in (b).  StemP is robust in predicting 5s rRNA sequences.

\begin{table}
\small
\centering
\begin{tabular}{@{}llrrrrrrrrr@{}}
\hline
Accn       & Method                                  & SCR / DR & TP & FN & FP & Spe \%         & Sen \%        & F1\%          & CPU(s)                         \\
\hline
{X67579}   & StemP                                   & 1   / 1  & 35 & 2  & 0  & \textbf{100.0} & \textbf{94.6}   & \textbf{97.2}  & \textbf{0.16}                  \\
           & RNAPredict\cite{wiese2008rnapredict}      &         & 33 & 4  & 6  & 84.6           & 89.2          & 86.8          & 101.89                         \\
           & SA \cite{eddy2004rna}                     &         & 33 & 4  & 4  & 89.1           & 89.1          & 89.1           & 401.34                         \\
           & TL-PSOfold\cite{lalwani2016efficient}      &         & 33 & 4 & 5  &86.8           & 89.2          & 88.0                                          \\
           & RNAfold\cite{lorenz2011viennarna} &  &   33& 4 & 7 & 82.5 & 89.2  & 85.7
           \\
           & 
            SP\cite{tsang2008sarna}                    &         & 33 & 4  & 6  & 84.6           & 89.2          & 86.8            &                                \\
           & Mfold\cite{zuker2003mfold}                 &        & 33 & 5  & 5  & {80.5}         & 89.2          & 84.6             &                                \\
           & COIN\cite{srikamdee2016rna}                &         & 33 & 4  & 1  & 97.1           & 89.2          & 93              &                                \\

           \hline
{AF03462}  & StemP                                    & 1   / 1  & 34 & 4  & 0  &\textbf{100.0} &  \textbf{89.5}           & \textbf{94.4}             & \textbf{1.97}                  \\

           & RNAPredict                                 &         & 27 & 11 & 3  & 90             & 71.1          & 79.4            & 102.66                         \\
           & SA                                         &         & 27 & 11 & 3  & 90             & 71            & 79.4            & 479.08                         \\
           & TL-PSOfold                                 &         & 31 & 7  & 5  & 81.6           & 86.1          & 83.8           &                                \\
          & RNAfold &&& 31 & 7  & 5  & 81.6           & 86.1          & 83.8           &                                \\
           & SP                                         &         & 27 & 11 & 3  & 90             & 71.1          & 79.4           &                                \\
           & Mfold                                      &         & 31 & 7  & 7  & 85.3           & 76.3          & 80.6           &                                \\
           & COIN                                       &         & 33 & 5  & 0  & \textbf{100}            & 86.8          & 93          &                                \\
        \hline
{X01590}   & StemP (Top $1(1)$)                          & 1  / 1   &  32  &  {8}  &  6       & 84.2             & 80.0          & 85.9           &  \multirow{2}{*}{\textbf{1.18}} \\
           & StemP (Best $2(14)$)                       & 2   / 2  & 37 & 3  & 0  & \textbf{100.0}   & \textbf{92.5}          & \textbf{96.1}         &                                \\
           & RNAPredict                                 &         & 33 & 7  & 3  & 91.7           & 82.5          & 86.8          &  120.45                         \\
           & SA                                         &         & 26 & 14 & 23 & 53             & 65            & 58.4            & 481.98                         \\
           & TL-PSOfold & &   36 & 4 & 3 & 92.3 & 90.0 & 91.1\\
           & RNAfold && 15 & 25 & 32 &  34.5 & 37.5 & 31.9\\
           & Mfold                                           &    & 29 & 12 & 15 & 65.9           & 70.7          & 68.4                                    \\
           
       \hline
AJ251080   & StemP (Top $1(8)$)                         & 1 / 1  &  33  &  5     & 2   & 94.3           & 86.8          & 90.4       &          \textbf{0.14}                       \\
        & RNAPredict                                     &    & 23 & 15 & 10 & 69.7           & 60.5          & 64.8          & 98.07                     \\
           & SA                                             &    & 22 & 16 & 20 & 52.3           & 57.9          & 55.0          & 398.23                         \\
    & TL-PSOfold && 27 & 11 & 7 & 79.4 & 71.1 & 75.0 \\
    & RNAfold && 23 & 15 & 18 & 56.1 & 60.5 & 58.2\\
      \hline
{V00336}   & StemP                                    & 1   / 1  & 37 & 3  & 0  & \textbf{100.0} & \textbf{92.5}          & \textbf{96.1}         & \textbf{0.27}                  \\
        & RNAPredict                                   &    & 10 & 4  & 6  & 25.6           & 25            & 25.3       & 99.09                          \\
           & SA                                      &    & 33 & 4  & 4  & 43.5           & 50            & 46.5       & 397.56                         \\
        \hline
{AE002087} & StemP                                    & 1   / 1  & 35 & 5  & 0  & \textbf{100.0} & \textbf{87.5}          & \textbf{93.3}  &  \textbf{0.12}                  \\
           & RNAPredict                                &    & 25 & 15 & 8  & 75.8           & 62.5          & 68.5          & 123.55                         \\
           & SA                                         &    & 18 & 22 & 21 & 46.2           & 45            & 45.6        & 490.00                            \\
           \hline
\end{tabular}
\caption{Comparison of 5s rRNA sequences prediction between StemP,  RNAPredict\cite{wiese2008rnapredict},  SA \cite{eddy2004rna}, TL-PSOfold\cite{lalwani2016efficient}, RNAfold\cite{lorenz2011viennarna},  SP\cite{tsang2008sarna}, Mfold\cite{zuker2003mfold}, and  COIN\cite{srikamdee2016rna}.  }
\label{T: comparison 6 5s seq}
\end{table}

\begin{figure}
    \centering
    \begin{tabular}{ccc}
    \includegraphics[width = 0.23 \textwidth]{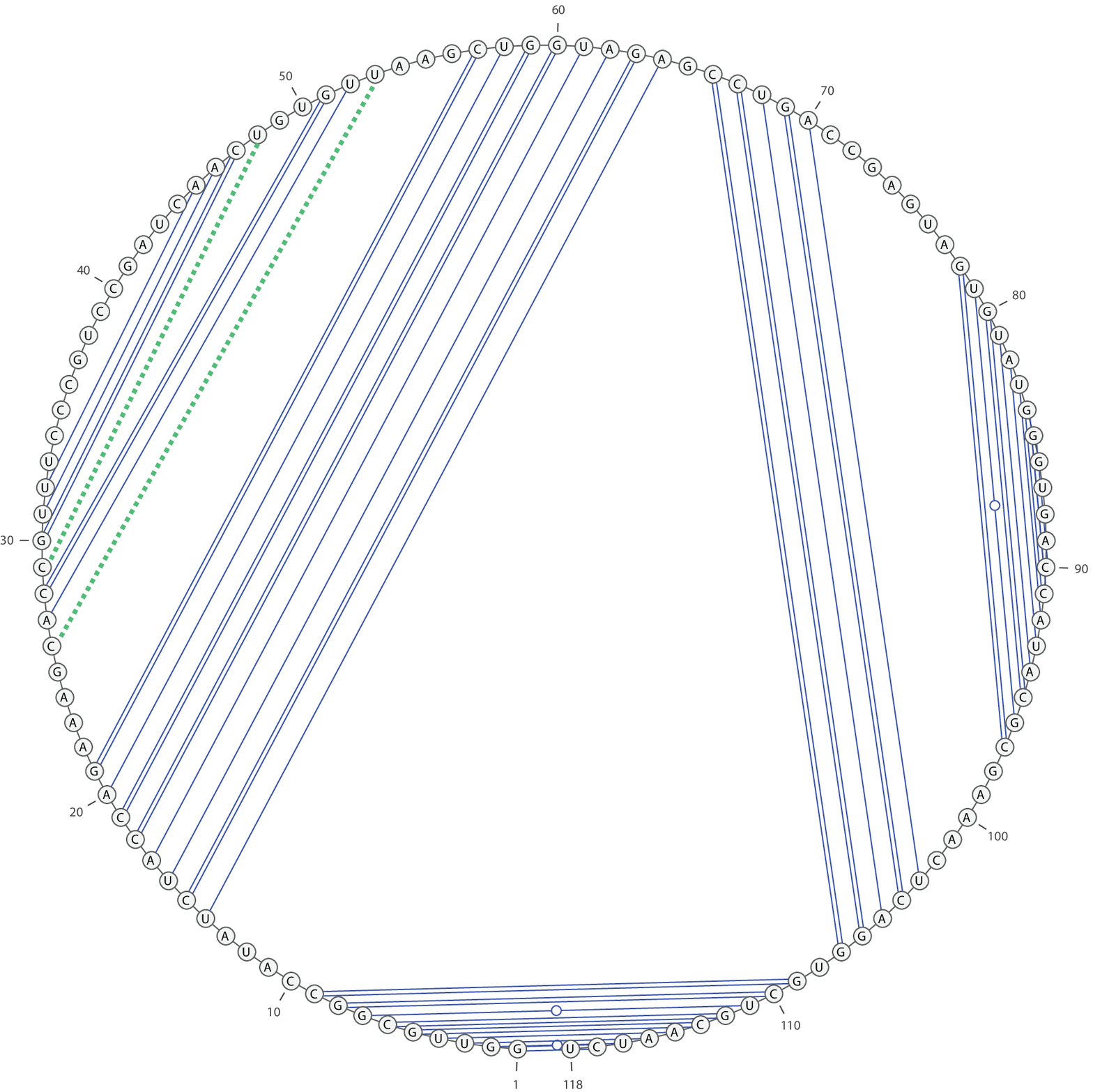} & 
    \includegraphics[width = 0.23 \textwidth]{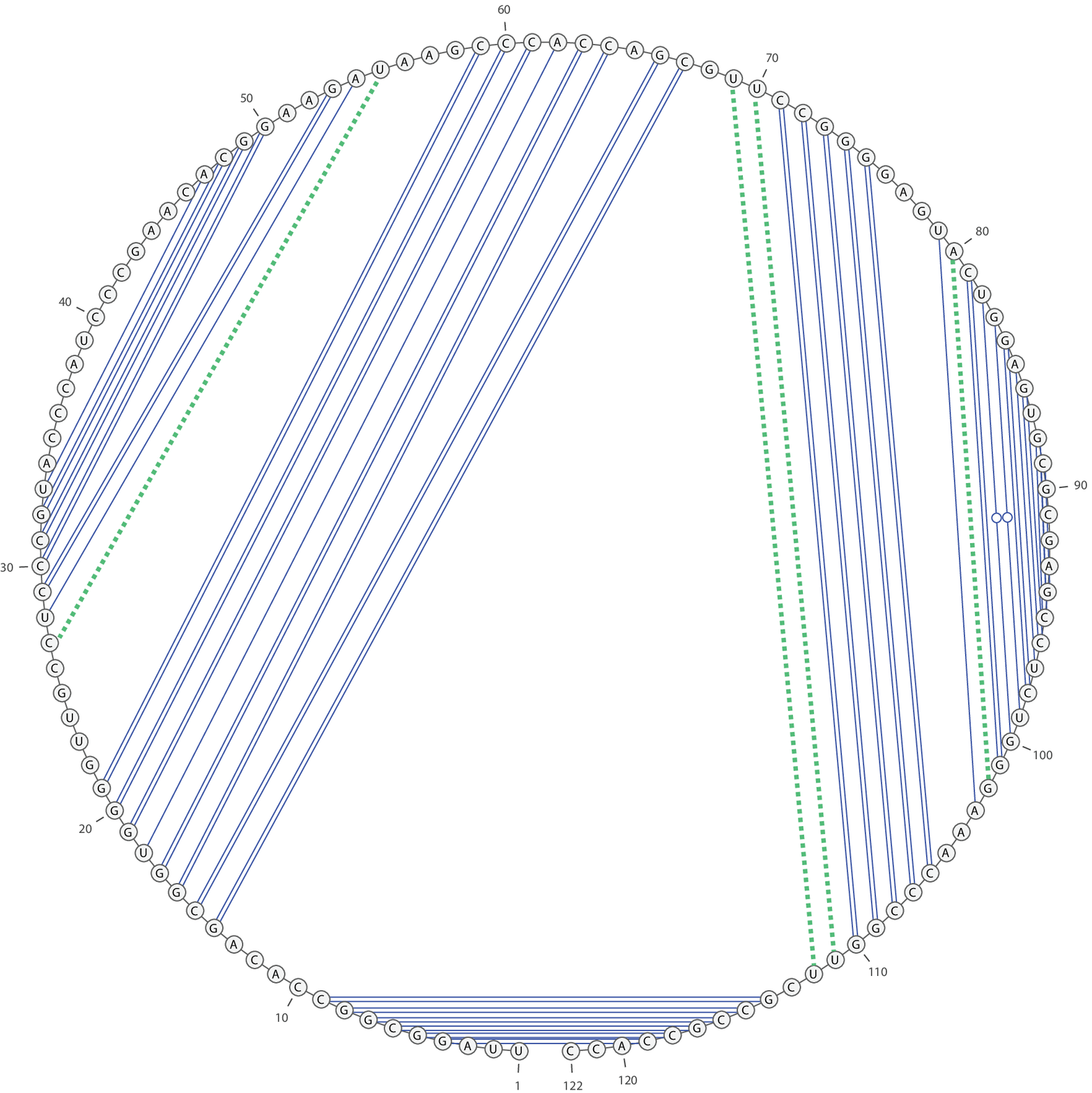}& 
    \includegraphics[width = 0.23 \textwidth]{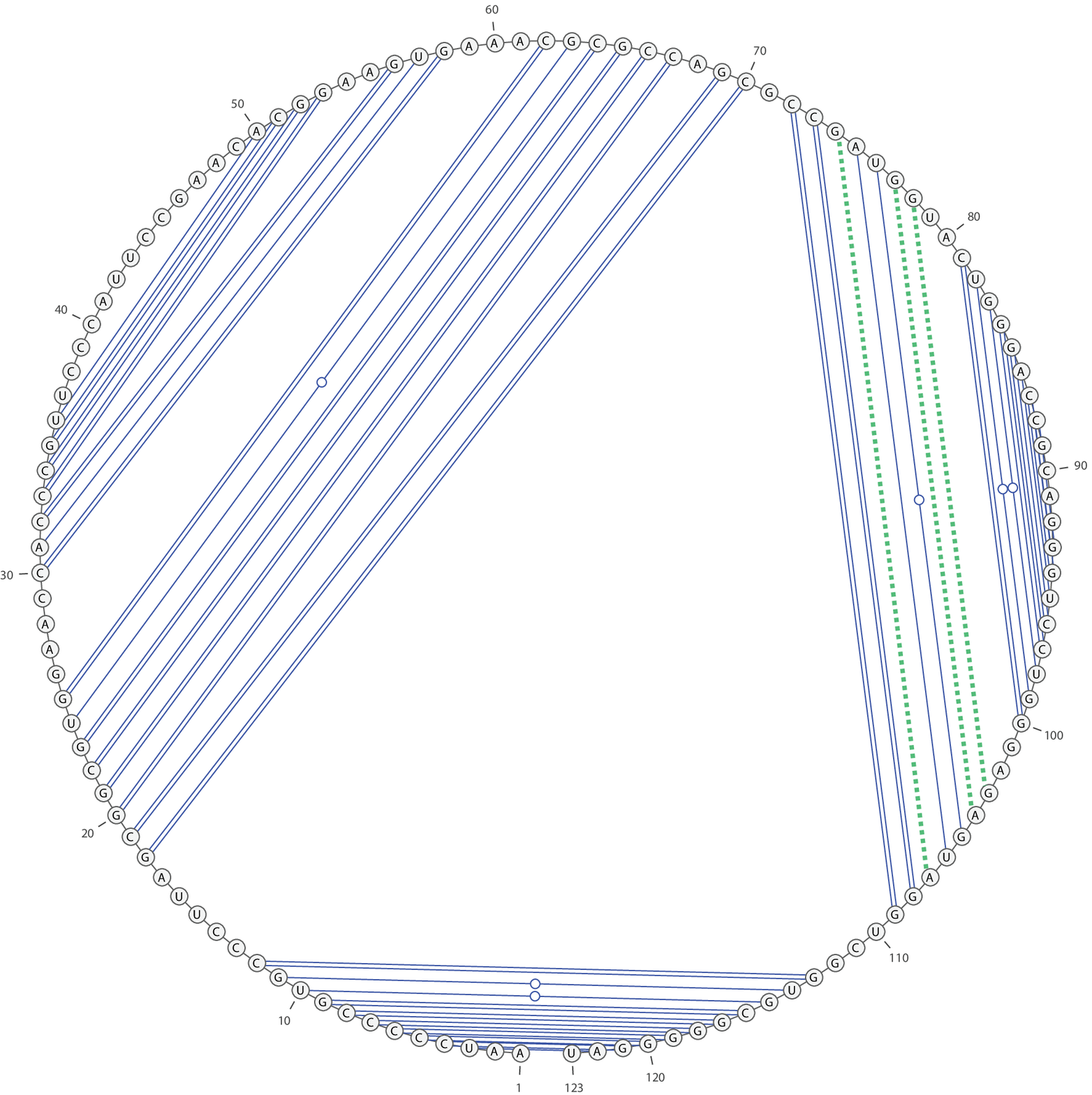}\\
   (a) X67579 & (b) AF034620  & (c) X01590  \\
    \includegraphics[width = 0.23 \textwidth]{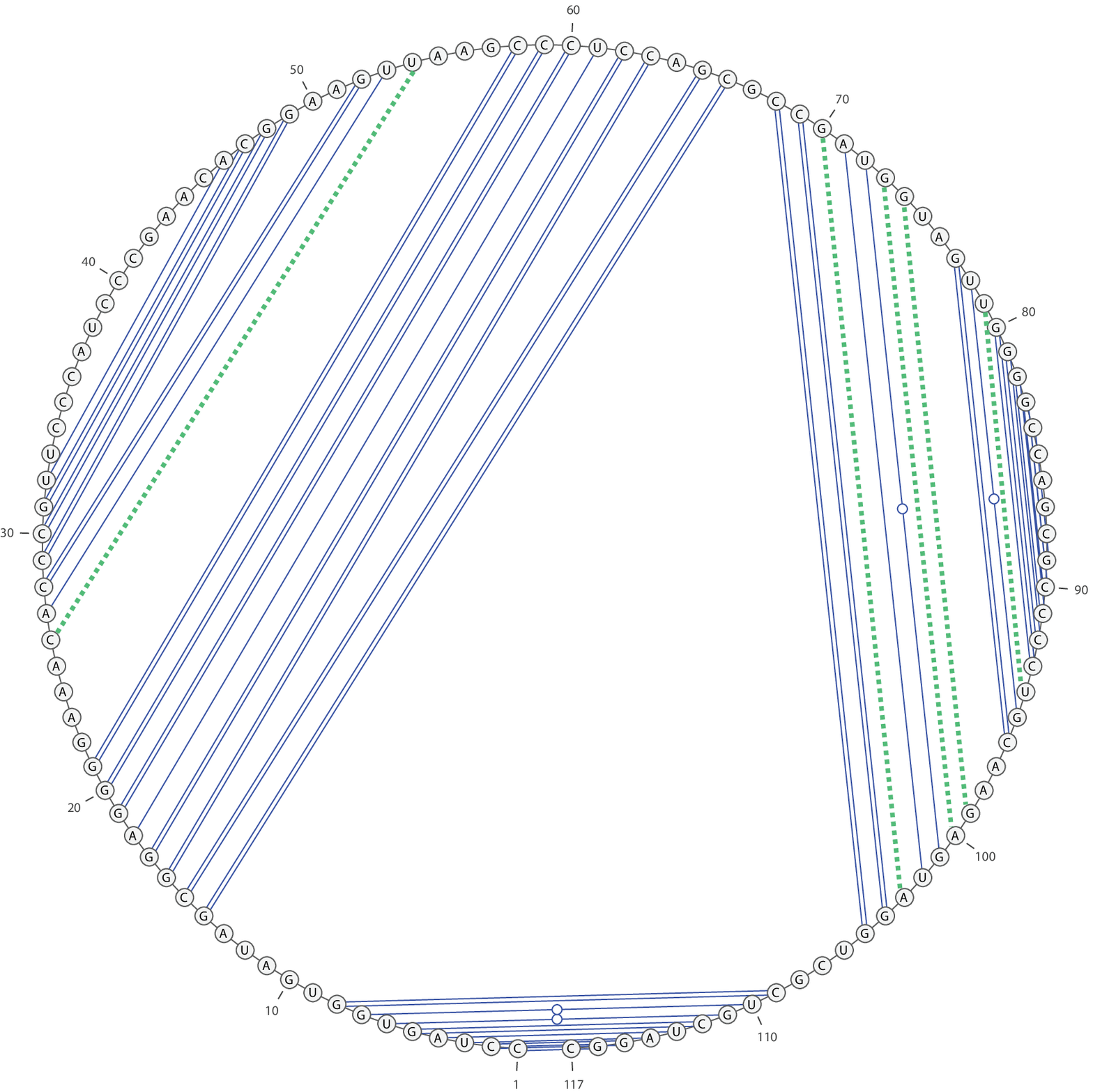}& 
    \includegraphics[width = 0.23 \textwidth]{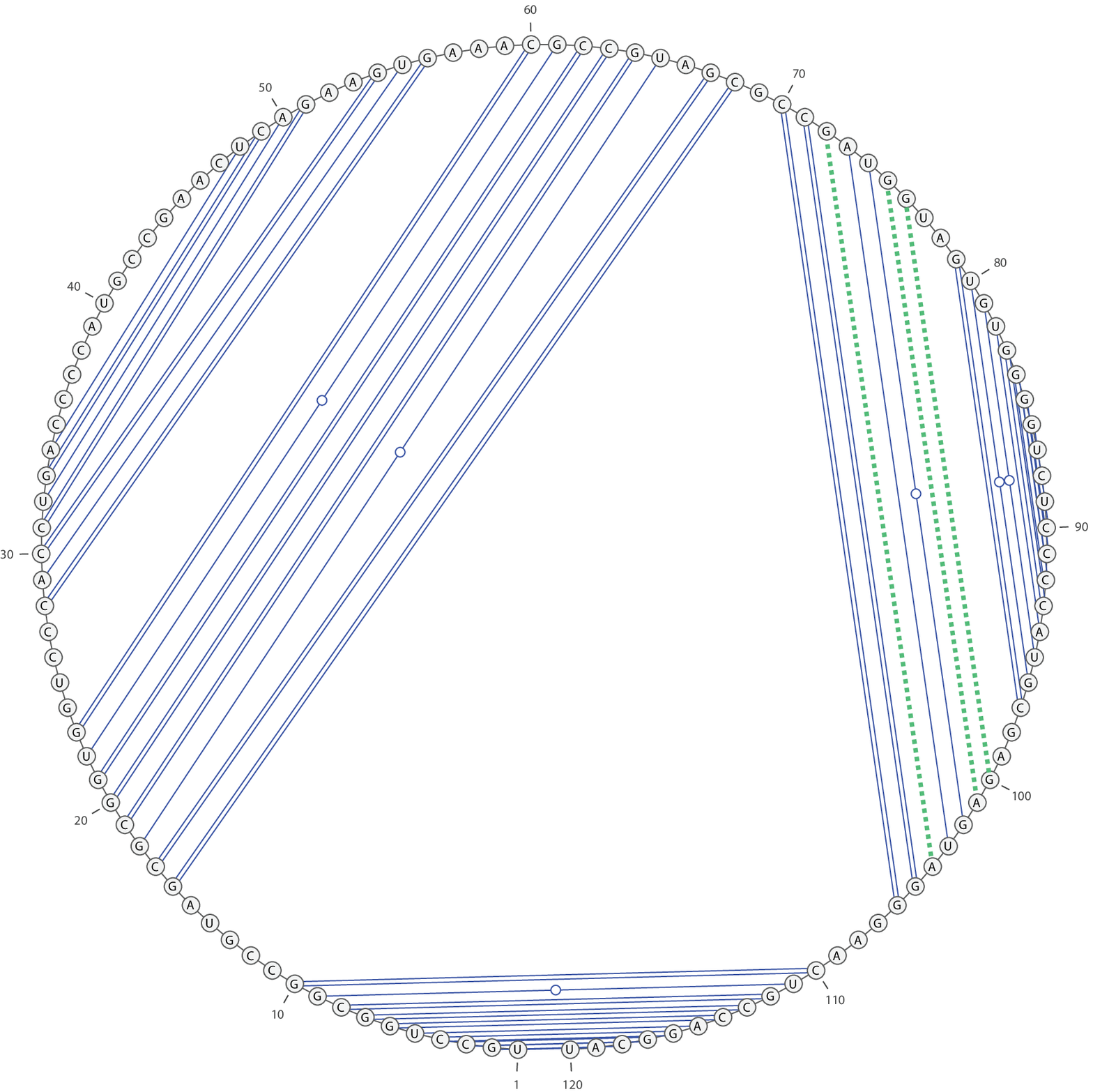}& 
    \includegraphics[width = 0.23 \textwidth]{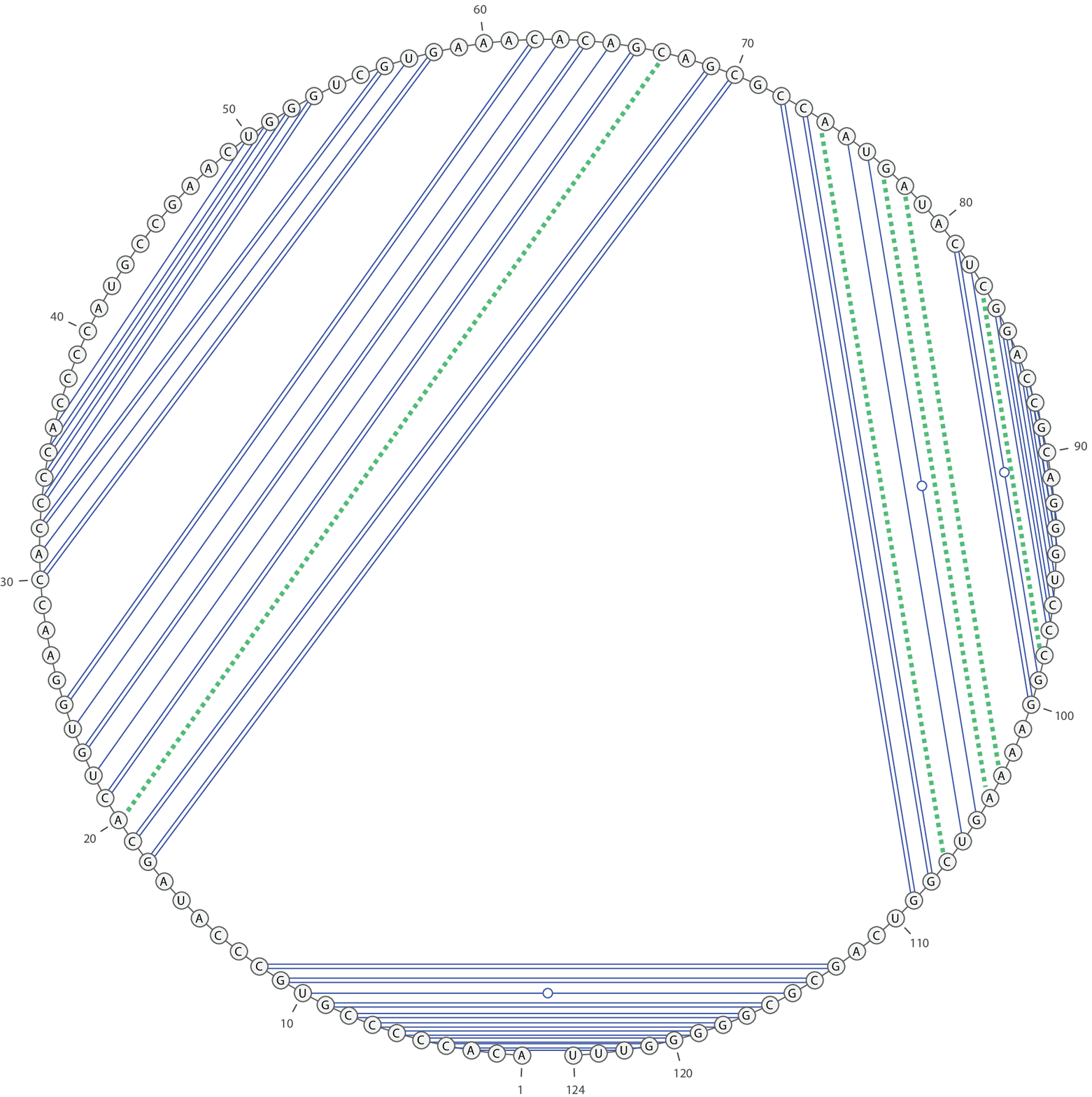}\\
     (d) AJ251080 & (e) V00336  & (f) AE002087 \\
    \end{tabular}
    \caption{Results of SPA on 6 5s rRNA sequences in Table \ref{T: comparison 6 5s seq}. (a)-(f) show the best predicted folding structures by StemP on sequences with Accession Number X67579, AF034620, X01590, AJ251080, V00336 and AE002087. Notice that there is no False Positive pairs found by StemP. The green dash line indicates the False Negative pairs. }
    \label{fig: 6 5s sequence}
\end{figure}

In Figure \ref{fig: 6 5s sequence} and Table \ref{T: comparison 6 5s seq}, we present the comparison of StemP on 6 different 5s rRNA sequences with  Genetic Algorithm (RNAPredict) \cite{wiese2008rnapredict},  SA \cite{eddy2004rna}, two-level particle swarm optimization algorithm (TL-PSOfold)\cite{lalwani2016efficient}, 
RNAfold\cite{lorenz2011viennarna},  Sarna-predict (SP) \cite{tsang2008sarna}, Mfold\cite{zuker2003mfold}, and Two-level particle swarm optimization algorithm(COIN)\cite{srikamdee2016rna}. The six sequences X67579(S.cerevisiae), AF03462(H.marismortui), X01590(T.aquaticus), AJ251080(G. stearothermophilus), V00336(E. coli), AE002087(D. radiodurans) are obtained from Gutell Lab \cite{cannone2002comparative}. X67579 belongs to Archaeal organism, AF03462 belongs to Eukaryotic organism and the rest belong to Bacterial organism. 
The result of  TL-PSOfold is from \cite{lalwani2016efficient},  RNAPredict and Mfold  from \cite{wiese2008rnapredict}.

For each sequence, we provide the top prediction with Sensitivity (\ref{eq_sens}), Specificity (\ref{eq_ppv}), and $F_1$-score (\ref{eq_f1})  listed across different methods.  
We use the general parameters in Table \ref{T:5s_parametersALL}.
For sequences X01590 and AJ251080, where the best prediction has SCR larger than 1, we show the highest measures among the unique or multiple sequences with SCR = 1.
All sequences except for X01590, StemP's top prediction has highest Specificity, Sensitivity, and $F_1$  among all methods. For X01590, the best prediction, which has SCR = 2, has the highest top prediction among all other methods.
In addition, in the best prediction of all 6 sequences, there is no incorrect base-pairs (False Positive) found by StemP,  all the False Negative base pairs associated with the best prediction are non-canonical base pairs including \texttt{U-C, G-A, G-G, A-A, A-C, U-U, C-C}.  Compared to other methods, there is significant improvement in cpu time, where the maximum of 2 seconds is needed for StemP while typically more than a minute is needed for other methods such as RNAPredict and SA.

\begin{table}
\centering
\small
\begin{tabular}{c|lc|lc|lc}
\hline
\hline
&  \multicolumn{2}{c}{  Archaeal }&   \multicolumn{2}{c}{Bacterial}  &   \multicolumn{2}{c}{Eukaryotic}\\
\hline
& StemLength & $SL$ & StemLength &  $SL$ & StemLength&  $SL$\\
\hline
Helix I  & 6,  &17 $\leq$SL$\leq$ 19
& 8,9,10,2{\tiny[1/0]}6  &11 $\leq$SL$\leq$ 21 
& 7,8,9, & 12.5 $\leq$SL$\leq$ 13.5 \\
Helix II & 2{\tiny[0/1]}6, 2{\tiny[1/2]}5, & 6 $\leq$ SL $\leq$ 7
& 2{\tiny[0/1]}6, 2{\tiny[0/1]}5  &6 $\leq$ SL $\leq$ 7.5
& 2{\tiny[0/1]}6, 2{\tiny[0/1]}5,  &6 $\leq$ SL $\leq$ 7 \\
& 1{\tiny[1/2]}6,  8 & &8,2{\tiny[0/1]}8 & & 2{\tiny[0/1]}2{\tiny[1/1]}3, 8 \\
Helix III & 2{\tiny[0/2]}4, 3{\tiny[0/2]}4 
& $3 \leq SL \leq 5 $
& 3{\tiny[0/2]}4, 2{\tiny[0/2]}4, 
& $3 \leq SL \leq 5 $
& 2{\tiny[0/2]}4,2{\tiny[0/3]}4, & $3 \leq SL \leq 5 $\\
& & & 3{\tiny[1/3]}3, 7 & & 3{\tiny[0/2]}4, 2{\tiny[1/3]}3 \\
Helix IV & 7,6,5 & $6\leq SL \leq 7$
& 2{\tiny[1/1]}2, 5 & $8 \leq SL \leq 10$
& 4,5 & $8 \leq SL \leq 11$\\
Helix V &   6{\tiny[1/0]}2, 8{\tiny[1/0]}2, 
& $2 \leq SL \leq 4$
& 8, 7, 6,  
& $2 \leq SL \leq 4$
& 2{\tiny[1/1]}2{\tiny[1/0]}3, 
& $2 \leq SL \leq 4$\\
&5{\tiny[2/1]}2  &&2{\tiny[1/1]}4,  3{\tiny[1/1]}3, && 5{\tiny[2/1]}2,5{\tiny[1/0]}3,\\
&&&2{\tiny[1/1]}5, 2{\tiny[2/2]}5&&2{\tiny[1/1]}2{\tiny[1/0]}2,7 \\
\hline
Domain $\beta$ & \multicolumn{6}{c}{$2 \leq GSL \leq 4$}\\
Domain $\gamma$ & \multicolumn{6}{c}{$2 \leq GSL \leq 4$}\\
\hline
\end{tabular} 
\caption{A general bound of parameters for 5S rRNA. 
This table is based on the true folding sequences in Gutell Lab \cite{cannone2002comparative}. (Canonical and Wobble base-pair matching is considered, and Partial Stems are included.)}
\label{T:5s_parametersALL}
\end{table}


In Table \ref{T: comparison 12 5s seqs}, we present  comparison results of 5s rRNA sequences with PMmulti\cite{hofacker2004alignment}, RNAalifold\cite{hofacker2002secondary} and Profile-Dynalign \cite{bellamy2007can}. We implement StemP on 12 different 5s rRNA Bacterial sequences  from Gutell Lab \cite{cannone2002comparative} in \cite{bellamy2007can}. 
The general parameters of Bacterial Sequences in Table \ref{T:5s_parametersALL} is used. In Table \ref{T: comparison 12 5s seqs}, we show the Sensitivity in (\ref{eq_sens}), PPV in (\ref{eq_ppv}), and MCC in (\ref{eq_mcc}) of the top and the best prediction results of StemP.  StemP has an average of MCC 0.922 for top predictions and an average of MCC 0.936 for the best predictions.  
PMmulti + RNAalifold has the highest Sensitivity, which means that fewer False Negative pairs while more incorrect pairs (False Positive pairs) were found. Overall, StemP has higher performance over the other methods when considering both False Negative rate and False Positive rate. 

\begin{table}
\begin{center}
(a)StemP on 12 5s rRNA bacterial sequences.
\begin{tabular}{l|rrr|rrrr|r}
\hline
&
  \multicolumn{3}{c|}{Top Results} &
  \multicolumn{3}{c}{Best results} &
  \multicolumn{1}{c|}{Ranking} &
   \\ \hline
Accn &
  \multicolumn{1}{l}{Sens} &
  \multicolumn{1}{l}{PPV} &
  \multicolumn{1}{l|}{MCC} &
  \multicolumn{1}{l}{Sens} &
  \multicolumn{1}{l}{PPV} &
  \multicolumn{1}{l}{MCC} &
  SCR /
  DR &
  CPU \\ \hline
X08002   & 91.7 & 100.0 & 0.920 & 84.6 & 100 & 0.920 & 1  / 1 & 0.107 \\
X08000   & 91.7 & 100.0 & 0.920 & 84.6 & 100 & 0.920 & 1  / 1 & 0.107 \\
X02627   & 93.3 & 97.2  & 0.934 & 89.7 & 100 & 0.947 & 3  / 2 & 0.223 \\
AJ131594 & 93.2 & 97.1  & 0.932 & 89.5 & 100 & 0.946 & 2  / 2 & 0.142 \\
V00336   & 96.1 & 100.0 & 0.962 & 92.5 & 100 & 0.962 & 1  / 1 & 0.224 \\
M10816   & 91.7 & 97.1  & 0.918 & 86.8 & 100 & 0.932 & 5  / 2 & 0.113 \\
AJ251080 & 90.4 & 94.3  & 0.905 & 86.8 & 100 & 0.932 & 31 / 3 & 0.128 \\
M25591   & 90.4 & 94.3  & 0.905 & 86.8 & 100 & 0.932 & 23 / 3 & 0.113 \\
K02682   & 93.3 & 97.2  & 0.934 & 89.7 & 100 & 0.947 & 2  / 2 & 0.133 \\
X04585   & 90.4 & 94.3  & 0.905 & 86.8 & 100 & 0.932 & 7  / 3 & 0.122 \\
X02024   & 90.4 & 94.3  & 0.905 & 86.8 & 100 & 0.932 & 23 / 3 & 0.151 \\
M16532   & 91.9 & 97.1  & 0.920 & 87.2 & 100 & 0.934 & 2  / 2 & 0.151 \\ \hline
Average  & 92.0 & 96.9  & 0.922 & 87.7 & 100 & 0.936 &       & 0.143 \\ \hline
\end{tabular}\\
\vspace{0.5cm}
(b) Statistical comparison on 12 5s rRNA bacterial sequences\\
\begin{tabular}{l|ccc}
\hline
Method                                    & {\% Sens} & {\% PPV} & {MCC}   \\
\hline
{StemP(Top)}  &   92.0 & \textbf{96.9} & \textbf{0.922}      \\
{StemP(Best)}  &  87.7 & \textbf{100.0} & \textbf{0.936}      \\
PMmulti                                        & {36.8}          & {88.9}          & {0.572}           \\
{Profile-Dynalign}              & {35.9}          & {94.7}          & {0.583}           \\
{Clustal W +   RNAalifold}      & {86.5}          & {80.3}          & {0.833}           \\
{PMmulti +   RNAalifold}        & {96.6}          & {85.3}          & {0.908}           \\
{Profile$-$ynalign   + RNAalifold}  & {66.1}          & {80.5}          & {0.729}          \\
\hline
\end{tabular}
\end{center}
\caption{ Comparison of 5s rRNA sequences of StemP, PMmulti\cite{hofacker2004alignment}, RNAalifold\cite{hofacker2002secondary} and Profile-Dynalign \cite{bellamy2007can}. We implement StemP on the test set containing 12 5s rRNA Bacterial sequence obtained from Gutell Lab \cite{cannone2002comparative} in \cite{bellamy2007can}. These sequences are from Bacterial Organism and we use the general parameters of Bacterial Sequences in Table \ref{T:5s_parametersALL}.  Both the top and the best prediction from StemP is comparable with highest MCC and PPV among all the methods.}
\label{T: comparison 12 5s seqs}
\end{table}

In Table \ref{T: compare 50 5s sequences}, we present  comparison results on 50 different 5s rRNA sequences in  \cite{poznanovic2020challenge}.  For StemP, we used the parameters in Table \ref{T:5s_parametersALL} (for Archaeal, Bacterial and Eukaryotic) on sequences 1-15, 16-21 and 22-50 respectively. 
Here, for sequences 1-15, we present results not using any GSL for domain $\beta$ and $\gamma$. 
This is due to the absent of similar sequences in the learning set of StemP, which is where the parameter bounds are learned.  
For these sequences 1-15,  not using GSL (only using the top Helix I-V parameter) was enough to find the folding prediction, some even giving higher accuracy.  This allows more possible helix in each domain to construct maximal cliques.  
It is shown that 33 out of 50 sequences, StemP's top prediction (with or without GSL) with SCR=1 has higher $F_1$-score.  Overall, StemP has 0.77 as an average of best prediction and 0.73 as highest prediction on this test set, which is higher than 0.635 in \cite{poznanovic2020challenge}.

\begin{table}
\small
\centering
\begin{tabular}{rl|l|llr|rl|l|llr}
\hline
\multicolumn{1}{l}{} &
  \multicolumn{1}{c}{\multirow{2}{*}{Accn}} &
  \multicolumn{1}{c}{\multirow{2}{*}{\cite{poznanovic2020challenge}}} &
  \multicolumn{3}{|c|}{StemP} &
  \multicolumn{1}{l}{} &
  \multicolumn{1}{c}{\multirow{2}{*}{Accn}} &
  \multicolumn{1}{c}{\multirow{2}{*}{\cite{poznanovic2020challenge}}} &
  \multicolumn{3}{|c}{StemP} \\
  \cline{4-6} \cline{10-12}
 &
  \multicolumn{1}{c}{} &
  \multicolumn{1}{c|}{} &
  Top &
  Best &
  SCR/DR &
   &
  \multicolumn{1}{c}{} &
  \multicolumn{1}{c|}{} &
  Top &
  Best &
  SCR/DR \\
 \hline
1  & X07545   & 0.85 & 0.90                                            & 0.85 & 1/1    & 26 & K02343   & 0.86 & 0.85                                            & 0.86 & 1/1  \\
2  & X14441   & 0.68 & 0.19                                            & 0.84 & 1/1    & 27 & AB015590 & 0.97 & 0.67                                            & 0.97 & 1/1  \\
3  & X72588   & 0.66 & 0.20                                            & 0.84 & 7/4    & 28 & X06102   & 0.86 & 0.74                                            & 0.86 & 1/1  \\
4  & M10691   & 0.69 & 0.47                                            & 0.69 & 1/1    & 29 & M25016   & 0.86 & 0.72                                            & 0.86 & 2/2  \\
5  & M36188   & 0.00 & 0.77                                            & 0.00 & 1/1    & 30 & X13718   & 0.41 & 0.70                                            & 0.56 & 2/2  \\
6  & M26976   & 0.85 & 0.73                                            & 0.86 & 11/2   & 31 & X06996   & 0.94 & 0.86                                            & 0.96 & 2/2  \\
7  & X62859   & 0.60 & 0.63                                            & 0.66 & 19/4   & 32 & U31855   & 0.93 & 0.49                                            & 0.94 & 6/3  \\
8  & U67518   & 0.67 & 0.76                                            & 0.67 & 1/1    & 33 & M74438   & 0.31 & 0.84                                            & 0.70 & 1/1  \\
9  & M34911   & 0.26 & 0.86                                            & 0.26 & 1/1    & 34 & Z75742   & 0.86 & 0.36                                            & 0.86 & 1/1  \\
10 & X62864   & 0.41 & 0.55                                            & 0.45 & 49/5   & 35 & X01004   & 0.33 & 0.81                                            & 0.33 & 1/1  \\
11 & X72495   & 0.94 & 0.94                                            & 1    & 1/1    & 36 & X00993   & 0.86 & 0.61                                            & 0.86 & 1/1  \\
12 & AE009942 & 0.62 & 0.89                                            & 0.62 & 1/1    & 37 & D00076   & 0.59 & 0.82                                            & 0.59 & 1/1  \\
13 & M21086   & 0.89 & 0.88                                            & 0.89 & 1/1    & 38 & Z93433   & 0.86 & 0.38                                            & 0.86 & 2/2  \\
14 & X05870   & 0.90 & 0.88                                            & 0.90 & 1/1    & 39 & V00647   & 0.93 & 0.15                                            & 0.94 & 1/1  \\
15 & X07692   & 0.89 & 0.87                                            & 0.89 & 1/1    & 40 & M10432   & 0.86 & 0.31                                            & 0.86 & 2/2  \\
16 & X02627   & 0.93 & 0.33                                            & 0.95 & 3/2    & 41 & L49397   & 0.93 & 0.29                                            & 0.94 & 1/1  \\
17 & V00336   & 0.96 & 0.27                                            & 0.96 & 1/1    & 42 & M18170   & 0.86 & 0.58                                            & 0.86 & 2/2  \\
18 & AJ251080 & 0.90 & 0.75                                            & 0.93 & 41/3   & 43 & X00996   & 0.69 & 0.24                                            & 0.70 & 1/1  \\
19 & M24839   & 0.50 & 0.25                                            & 0.67 & 62/4   & 44 & Y14281   & 0.86 & 0.68                                            & 0.86 & 12/4 \\
21 & M25591   & 0.90 & 0.79                                            & 0.93 & 23/3   & 45 & Z33604   & 0.30 & 0.75                                            & 0.59 & 4/2  \\
23 & U39694   & 0.80 & 0.72                                            & 0.80 & 1/1    & 46 & AJ242949 & 0.69 & 0.83                                            & 0.70 & 2/2  \\
22 & X99087   & 0.43 & 0.89                                            & 0.75 & 20/3   & 47 & M24954   & 0.93 & 0.17                                            & 0.94 & 1/1  \\
23 & X13035   & 0.44 & 0.74                                            & 0.70 & 2/2    & 48 & X13037   & 0.70 & 0.77                                            & 0.70 & 1/1  \\
24 & Y00128   & 0.70 & 0.79                                            & 0.70 & 1/1    & 49 & K00570   & 0.86 & 0.87                                            & 0.86 & 1/1  \\
25 & AB015591 & 0.91 & 0.60                                            & 0.91 & 1/1    & 50 & X06094   & 0.99 & 0.69                                            & 0.99 & 1/1 \\
\hline
                      &                                         &                       &      &     &           & & \textbf{Average} & 0.73                 & 0.64                                    & 0.77   \\ \hline

\end{tabular}
\caption{ Comparison of 5s rRNA sequences between StemP and \cite{poznanovic2020challenge}. The 50 sequences are from  \cite{poznanovic2020challenge} obtained from Gutell Lab \cite{cannone2002comparative}. This table follows the format in Table \ref{T:comparison tRNA 50 seqs}. We adopted the general bound of parameters for 5S rRNA for Archaeal, Bacterial and Eukaryotic in Table \ref{T:5s_parametersALL} on sequences 1-15, 16-21 and 22-50 respectively.   This table shows that overall, StemP has higher $F_1$-score in both the top and the best prediction compared to \cite{poznanovic2020challenge}.}
\label{T: compare 50 5s sequences}
\end{table}


We present additional StemP results on 15 number of 5s rRNA sequences, with or without using the $GSL$ which helps to reduce computation by identifying domain $\beta$ and $\gamma$ separately. 
The 15 sequences are from \cite{poznanovic2020challenge} and obtained from Gutell Lab \cite{cannone2002comparative}.  We adopt the general parameters for 5S rRNA for Archaeal  in Table \ref{T:5s_parametersALL}.  Among 15 sequences, the best MCC value for 9 sequences are improved without GSL, and StemP  has higher MCC compared to \cite{poznanovic2020challenge} for 11 out of 15 sequences. For sequence M36188, both top and best MCC has increased from 0 to a positive rate. 
However, both CPU and SCR increased when GSL is not used. When computer power is not limited, the true folding can be close to some clique structure, while $GSL$ helps to reduce the candidate set in general.
\begin{table}
\centering
\begin{tabular}{cc|ccrc|cccrc|c}
\hline 
   &    &        \multicolumn{4}{c|}{StemP with $GSL$}  &  & \multicolumn{4}{c|}{without $GSL$}            & \multicolumn{1}{c}{\multirow{2}{*}{\cite{poznanovic2020challenge}}}                      \\               
   & Accn     & Top  & Best & SCR/DR  & CPU  &  & Top  & Best          & SCR/DR & CPU  &                                                                \\ \hline
1  & X07545   & 0.85 & 0.85 & 1   / 1  & 0.11 &  & 0.85 & 0.85          & 1   / 1  & 2.41 & 0.90                                                                \\
2  & X14441   & 0.68 & 0.68 & 1   / 1  & 0.07 &  & 0.58 & \textbf{0.84} & 7   / 3  & 0.18 & 0.19                                                                \\
3  & X72588   & 0.66 & 0.68 & 7   / 4  & 0.07 &  & 0.84 & \textbf{0.84} & 1   / 1  & 0.14 & 0.20                                                                \\
4  & M10691   & 0.69 & 0.69 & 1   / 1  & 0.07 &  & 0.82 & \textbf{0.82} & 1   / 1  & 0.68 & 0.47                                                                \\
5  & M36188   & 0.00 & 0.00 & 1   / 1  & 0.08 &  & 0.24 & \textbf{0.44} & 77  / 5  & 0.24 & 0.77                                                                \\
6  & M26976   & 0.85 & 0.86 & 11  / 2  & 0.10 &  & 0.85 & 0.86          & 11  / 2  & 1.50 & 0.73                                                                \\
7  & X62859   & 0.60 & 0.66 & 19  / 4  & 0.09 &  & 0.59 & \textbf{0.77} & 2   / 2  & 8.52 & 0.63                                                                \\
8  & U67518   & 0.67 & 0.67 & 1   / 1  & 0.07 &  & 0.77 & \textbf{0.77} & 1   / 1  & 0.16 & 0.76                                                                \\
9  & M34911   & 0.26 & 0.26 & 1   / 1  & 0.07 &  & 0.00 & \textbf{0.60} & 93  / 6  & 0.16 & 0.86                                                                \\
10 & X62864   & 0.41 & 0.45 & 49  / 5  & 0.09 &  & 0.41 & \textbf{0.62} & 177 / 7  & 0.30 & 0.55                                                                \\
11 & X72495   & 0.94 & 0.94 & 1   / 1  & 0.08 &  & 0.94 & 0.94          & 1   / 1  & 0.23 & 0.94                                                                \\
12 & AE009942 & 0.62 & 0.62 & 1   / 1  & 0.09 &  & 0.77 & \textbf{0.77} & 1   / 1  & 1.11 & 0.89                                                                \\
13 & M21086   & 0.89 & 0.89 & 1   / 1  & 0.10 &  & 0.89 & 0.89          & 1   / 1  & 1.32 & 0.88                                                                \\
14 & X05870   & 0.90 & 0.90 & 1   / 1  & 0.09 &  & 0.90 & 0.90          & 1   / 1  & 0.34 & 0.88                                                                \\
15 & X07692   & 0.89 & 0.89 & 1   / 1  & 0.10 &  & 0.89 & 0.89          & 1   / 1  & 1.86 & 0.87    \\
\hline
\end{tabular}
\caption{StemP results  of 15 different 5s rRNA sequences in \cite{poznanovic2020challenge} from Gutell Lab \cite{cannone2002comparative}.  We use the general parameter in Table \ref{T:5s_parametersALL} and $F_1$-score as validation measurement.  
We present StemP with and without $GSL$ restriction for comparison.  Among 15 sequences, the best MCC value for 9 sequences improved not using the GSL, while the CPU time increased.}
\label{T: compare 15 5s sequences relaxed}
\end{table}


\section{Concluding Remarks and discussion}\label{sec:conclu}
We proposed a simple deterministic Stem-based Prediction algorithm for the RNA secondary structures prediction.  By using mainly the Stem Length and Stem-Loop score, we explore the question whether the Stem is enough for folding prediction.   We experimented across different type of sequences: Protein, tRNA, and 5s rRNA. While there were small variations to true folding structure, StemP gives a good general form within a few seconds of CPU time.  

In the step of constructing the vertices, biological environment or features of interest can be further added for better prediction.  Different learning based algorithm or Motif based idea can be incorporated to refining what is allowed for vertices.  
A sophisticated  hierarchical structure clique models as well as chemical reactions can be further considered.   
StemP is stable and deterministic, and this makes it easier to study folding energy more concretely.

\section*{Acknowledgment}
The third author would like to thank a colleague, Prof. Christine E. Heitsch, for her passion for
the topic of secondary structure prediction and introducing this topic. This work would not have
been possible without submerging in the topic via various seminars she organized many years ago.

\begin{appendices}

\section{StemP Algorithm}

We present the outline of the StemP algorithm and extended algorithms for tRNA and 5s rRNA.
One of the best part of this algorithm is its simplicity: a simple code can be easily written and experiments can be done on a regular laptop for the sequence of length up to 150.  

In Algorithm \ref{alg:StemP} [Step 1] constructs vertex and edges respectively. Then we employ Bron-Kerbosch algorithm with pivoting \cite{Tomita2006} to find the cliques of the full Stem-graph. 
Along with the bound of \textit{Stem-Loop score}, Length Threshold $L$ of a stem is introduced to narrow down the choice of possible vertex. Stems of length shorter than $C$ is not considered as effective vertices in [Step 1-2]. This gives priority to overall skeleton of the folding structure, which is driven from big binding forces, i.e. by longer stems.  The longer the sequence is, the longer this threshold can be.

Algorithm \ref{alg:trRNA} illustrates the details in finding partial stems when predicting tRNA sequences.  An example of predicting Archaeal  5s rRNA sequence can be found in Algorithm \ref{alg:5S rRNA archaeal}. This algorithm is based on the results of vertex construction for 5 helix $H_1,H_2,H_3,H_4,H_5$ with minimum Stem Length $L$.  Example of finding particular structure $l_1${\tiny[$n_1/n_2]$}$l_2$ vertex  can be found in Algorithm \ref{alg:5s rRNA vertex 2+6}.

\begin{algorithm} 
\caption{Structure Predicting Algorithm (StemP)}
\begin{algorithmic}
\STATE \textbf{Input:} The sequence $r$ of size $n$, 
the stem length threshold $L$, and bounds of  $SL$ : $SL_{\min}$ and $SL_{\max}$. \\
\underline{Step 1-1: full Stem-graph   construction (Vertex Construction)}  \\
\STATE $k \leftarrow  1$\\
\FOR {$i=1 : n$} 
\FOR{ $ j  = i+3 : n$}
\IF{  IsBasePair$(r_i,r_j)$}
\STATE $k \leftarrow k+1$, $l \leftarrow 1$
\WHILE{ $j-l>j+l$ and IsBasePair$(r_{i+l},r_{j-l})$ }
\STATE $l = l+1$
\ENDWHILE
\STATE $l \leftarrow l-1$,$d \leftarrow j-i$, $SL \leftarrow \frac{d}{l}$
\IF{$l>L$ and $SL_{\min}\leq SL \leq SL_{\max}$}
\STATE $v_k \leftarrow (i,j,l,d,SL)$,$k \leftarrow k+1$ 
\ENDIF
\ENDIF
\ENDFOR
\ENDFOR\\
\underline{Step 1-2: full Stem-graph   construction (Edge Construction)}
\FOR{$m = 1: length(V)$}
\FOR{$n = m+1: length(V)$}
\IF{(i) $j_m < i_n$ or (ii) $j_n < i_m$ or (iii) $i_m + l_m -1 < i_n$ and $ j_n < j_m-l_m +1 $ or (iv) $i_m+l_m -1 < i_n $ and $ i_n+l_n-1 < j_m - l_m+1 $ and $ j_m <j_n-l_n+1$}
\STATE $e_{mn}=1$
\ENDIF
\ENDFOR
\ENDFOR\\
\underline{Step 2: Choose a Subgraph}
\begin{quote} \vspace*{+0.2cm} 
Find all the cliques, using \cite{Tomita2006}. \\
Compute the total matching energy for each cliques.\\
Choose the maximum matching and/or maximal clique as the folding prediction.
\end{quote}
\end{algorithmic}
\footnotesize{
Remark: Along with the bound of \textit{Stem-Loop score}, Length Threshold $L$ of a stem is introduced to narrow down the choice of possible vertex. Stems of length shorter than $C$ is not considered as effective vertices in [Step 1-2]. This gives priority to overall skeleton of the folding structure, which is driven from big binding forces, i.e. by longer stems.  The longer the sequence is, the longer this threshold can be. }
\label{alg:StemP}
\end{algorithm}

\begin{algorithm}
\caption{tRNA algorithm}
\begin{algorithmic}
\STATE \textbf{Input:}  (i)tRNA Sequence  $r$ of size $\hat{l}$, (ii)lower bound $SL_{\min, Acceptor}$ of $SL$ for Acceptor stem, (iii)the stem length threshold $L$, (iv)bounds of  $SL$ : $SL_{\min}$ and $SL_{\max}$, and (v)  bounds of distance $d_{\min},d_{\max}$.
\ENSURE secondary structure of tRNA
\STATE \underline{Step 1}: vertex construction
\STATE Find all possible vertices $v_k$ of $l_k \geq L$ and store them in $V_{temp}$\\
V $ \leftarrow \phi$\\
\FOR {$v_k \in  V_{temp}$} 
\IF{$d_k > \hat{l}/2$}
\STATE $SL_k \leftarrow (\hat{l}- d_k + 2l_k - 2)/l_k$
\IF{$SL_k \leq SL_{\min, Acceptor}$}
\STATE $V \leftarrow V \cup v_k$
\ENDIF
\ELSE
\STATE $SL_k \leftarrow d_k / l_k$
\WHILE{$SL_k \leq SL_{\min}$}
\STATE $l_i \leftarrow l_k-1, SL \leftarrow d_k / l_k$
\ENDWHILE
\IF{ $SL_{\min} \leq SL \leq SL_{max}$and$ l_k \geq L$ and $ d_{\min} \leq d_k \leq d_{\max}$}
\STATE $V \leftarrow V \cup v_k$
\ENDIF

\ENDIF
\ENDFOR
\STATE \underline{Step 2}: go to step 2 of Algorithm \ref{alg:StemP}.
\end{algorithmic}
\label{alg:trRNA}
\end{algorithm}

\begin{algorithm} 
\caption{5S rRNA algorithm for Archaeal}
\begin{algorithmic}
\STATE \textbf{Input:} (i) five sets of possibles vertex in each Helix: $H_1, H_2, H_3, H_4, H_5$, (ii) lower/upper bounds of \textit{Generalized-Stem-Loop score}   in $\alpha$ domain and $\beta$ domain:  $GSL_{\min}^{\alpha}, GSL_{\max}^{\alpha}, GSL_{\min}^{\beta}, GSL_{\max}^{\beta}$.
\STATE \textbf{Output:} secondary structure of 5S rRNA\\
 $V \leftarrow \phi, V = V \cup  H_1$\\
\FOR{$v_m \in H_2$} 
\FOR{$v_n  \in H_4$}
\IF{ \text{ExistEdge}($v_m,v_n$) and $ GSL_{\min}^{\alpha} \leq GSL(v_m,v_n) \leq GSL_{\max}^{\alpha}$}
\STATE $V \leftarrow V \cup ( \min(i_m,i_n), \max(j_m,j_n), l_m+l_n, d_m, GSL(v_m,v_n))$ 
\ENDIF
\ENDFOR
\ENDFOR

\FOR{$v_m  \in H_3$}
\FOR{$v_n \in H_5$}
\IF{ExistEdge($v_m,v_n$) and $ GSL_{\min}^{\beta} \leq GSL(v_m,v_n) \leq GSL_{\max}^{\beta}$}
\STATE $V \leftarrow V \cup ( \min(i_m,i_n), \max(j_m,j_n), l_m+l_n, d_m, GSL(v_m,v_n))$ 
\ENDIF
\ENDFOR
\ENDFOR
\STATE \underline{Step2}: go to step 2 of Algorithm \ref{alg:StemP}.
\end{algorithmic}
\footnotesize{
Remark: ExistEdge can be found in Algorithm \ref{alg:StemP} Step 1-2. each vertex $v$ in $V$ has 5 attributes: $(i,j,l,d,SL)$  }
\label{alg:5S rRNA archaeal}
\end{algorithm}




\begin{algorithm} 
\caption{algorithm to construct $l_1${\tiny[$n_1/n_2]$}$l_2$ vertex in 5S rRNA}
\begin{algorithmic}
\REQUIRE (i)5S rRNA Sequence  $r$ of size $n$, (ii)bound of $SL$:  $SL_{\min}, SL_{max}$, (iii) subvertex size: $l_1,l_2$, and (iv) gap size $n_1,n_2 $
\ENSURE All possible vertex of structure $l_1${\tiny[$n_1/n_2]$}$l_2$ with stem-loop score $SL$ satisfying $SL_{\min} \leq SL \leq SL_{\max}$
\STATE $l \leftarrow 0, V \leftarrow \phi, k \leftarrow 0$
\FOR{$i=1:n$}
\FOR{$j=i+2:n$}
\STATE $d \leftarrow j-i$
\STATE idx $\leftarrow 0$, $l \leftarrow 0$
\WHILE{IsBasePair$(r_{i},r_{i})$ and $i<j$ and $l< l_1 + l_2$}
\STATE index $\leftarrow$ idx+1, $l \leftarrow l + 1$
\IF{ idx $\neq 1_1$}
\STATE $i \leftarrow i +1, j \leftarrow j -1 $
\ELSE
\STATE $i \leftarrow i +1 + n_1, j \leftarrow j -1 - n_2$
\ENDIF
\IF{$i > n $ or $j \leq i$}
\STATE break
\ENDIF
\ENDWHILE
\STATE $LS \leftarrow \frac{d}{l}$
\IF{ $s_{\min} \leq SL \leq s_{\max}$}
\STATE $k \leftarrow k + 1, v_k \leftarrow (i,j,l,d,SL)$, $V \leftarrow V \cup v_k$
\ENDIF

\ENDFOR
\ENDFOR

\end{algorithmic}
\label{alg:5s rRNA vertex 2+6}
\end{algorithm}

\section{PDB results for long sequences}

We present additional StemP results for PDB sequences with length larger than 50.  The sequences are obtained from the Protein Data Bank (PDB) \cite{bernstein1977protein}.  In Table \ref{T:pdb_ more than 50}, we compare the prediction with 
FOLD, MaxExpert, Probknot, MC and NAST with the same set up as mentioned in the main paper. 
Typically, these sequences have a large quantity of possible structures as well as stems of size as small as 1 or 2. 
For the case where the best prediction does not has SCR = 1, we provide the top prediction here, which is the highest MCC value among possible prediction with maximum number base pairs. These top MCC values are 2FK6 0.55, 3E5C 0.88, 1DK1 0.33, 1MMS 0.34, 3EGZ 0.72, 2QUS 0.00, 1KXK 0.81, 2DU3 0.49 and 2OIU 0.58.

The highest accuracy of StemP result MCC is comparable to FOLD, MaxExpect, ProbKnot, MC and NAST in the majority. 

\begin{table}
\small
\begin{center}	
\begin{tabular}{lclrllccccccc}
\hline
PDB & Size & StemP &SCR(m) & DR  &  CPU(s)   & FOLD & MaxExpect & ProbKnot  & MC & NAST  \\ \hline
2FK6 $^p$     & 53 & \textbf{0.83}    & 10065(1991) & 10   & 35.49 & 0.76            & 0.76              & 0.78            & O & X \\
3E5C          & 53 & \textbf{{0.97}$^L$ }         & 9(69)     & 2  & 6.43  &\textbf{0.94}$^m$             & \textbf{0.97}$^m$ & 0.81            & O & O \\
1MZP $^u$     & 55 & \textbf {0.89}      & 53(49)    & 4   & 0.12  & 0.36            & 0.36              & 0.36            & X & O\\
1DK1 $^{w}$ & 57 & \textbf{0.83$^S$} & 2(12)  & 2 & 10.61& 0.69 & 0.72 & {0.82}  & O & O \\
1MMS  &  58  & \textbf{0.87}$^{S, L}$ & 477(643) & 6 &  7.80  & 0.71 &0.73 &0.69  & O & X\\
3EGZ $^p$ & 65 & 0.80 & 11(36) & 3  & 0.60 & \textbf {0.84}& 0.81 & 0.81 & X & O\\ 
2QUS $^p$ & 69 & \textbf{0.95} & 465(250)  &6   &  2.00  &0.93 &0.93&0.93 & O & O\\ 
1KXK $^w$ & 70 & \textbf{0.96}$^S$ & 3(11) &2 &  11.16$^{N}$ &0.81 & 0.81& 0.79 & O & O \\ 
2DU3 $^p$ & 71 & \textbf{0.90}$^S$ & 745(917) & 6&   3.608 & 0.82 & 0.47 & 0.49 & X & O \\
2OIU $^w$ & 71 & \textbf{0.94}$^S$ & 7(25) &3  & 16.62  & 0.74 & 0.93 & 0.93  &  O & X\\
 \hline
\end{tabular}
\end{center}
\caption{StemP for Protein sequence of length over 50. In the protein list (first column), superscript $p$ indicates pseudo knots, superscript $w$ indicates including wobble base pairs, and $u$ includes \texttt{UU} base pairs.   Superscript $L$ indicates using $L=2$ otherwise $L=3$, and $S$ indicates when $SL$ is used.  Superscript $m$ shows when there are multiple folding with the same ranking.  For 1KXK,  the superscript $N$ indicate that if $SL$ is not imposed, it take 89s to get the same result.    In the FOLD column, $m$ represents picking the best MCC among multiple possible predictions: for 1DK1, the best MCC among (0.69, 0.21, 0.34), for 2QUS, among (0.93, 0.73), for 2DU3, among (0.42, 0.82), and for 2OIU, among (0.88, 0.50, 0.74, 0.71). In MaxExpect column for 3E5C, $m$ represents picking the best MCC among multiple possible predictions (0.65, 0.97).}   \label{T:pdb_ more than 50}
\end{table} 

\section{Learning structures of Archaeal sequences.}
In Table  \ref{5s:st1}, we present how the parameters are learned for Archaeal organism 5s RNA sequences, using the data from Gutell Lab \cite{cannone2002comparative}.  We count different variations of the stems. 


\begin{table}
\centering
\small
\begin{tabular}{lrrrrrrrr}
\hline
          & \multicolumn{8}{c}{\textbf{Stem Length Variation (Count)}}                                      \\ \hline
Helix I   & 6 (47)     & None (2)  & 5 (2)   & 4{\tiny[1/0]}1 (1) & 4 (1)    &         &          &        \\ 
Helix II  & 2{\tiny[0/1]}6 (26)   & 8 (20)    & 2{\tiny[1/2]}5 (3) & 1{\tiny[1/2]}6 (2) & 2 {\tiny[0/1]}1{\tiny[2/1]}4(1) &  2 {\tiny[0/1]}5 (1)       &          &        \\ 
Helix III & 7 (27)     & 6 (18)    & 5 (4)   & 2{\tiny[1/1]}2 (2) & 3{\tiny[1/2]}2 (1)  & 3{\tiny[2/2]}1 (1) &          &        \\ 
Helix IV  & 2{\tiny[0/2]}4 (44)   & 2{\tiny[2/4]}2 (5)   & 3{\tiny[0/2]}4 (4)  &         &          &         &          &        \\ 
Helix V   & 1{\tiny[1/1]}6{\tiny[1/0]}2 (39) & 1{\tiny[1/1]}5{\tiny[2/1]}2 (4) & 1{\tiny[1/1]}5 (3) & 1{\tiny[1/1]}8 (2)  & 8{\tiny[1/0]}2 (2) & 1 {\tiny[1/1]} 7(1)  \\ & 1{\tiny[1/1]}6{\tiny[2/0]}1(1) & 8(1) \\

 \hline
\end{tabular}
\caption{Learned structure in Archaeal organism. (None denotes that the corresponding Helix does not exist in a given structure)}
\label{5s:st1}
\end{table}

\end{appendices}
\bibliographystyle{plain}
\bibliography{cite_spa}  

\end{document}